\documentclass{aa}  

\usepackage{graphicx}
\usepackage{txfonts,textcomp}
\usepackage{natbib}
\usepackage{hyperref}
\hypersetup{
  colorlinks   = true, %Colours links instead of ugly boxes
  urlcolor     = blue, %Colour for external hyperlinks
  linkcolor    = blue, %Colour of internal links
  citecolor   = blue %Colour of citations
}
\usepackage{booktabs}

\newcommand\mc[1]{\multicolumn{1}{c}{#1}}
\usepackage{subfig}

\begin{document}

   \title{Search for radio halos in starburst galaxies}

   % \subtitle{I. Overviewing the $\kappa$-mechanism}

   \author{C.A. Galante\inst{1,2}
          \and J. Saponara\inst{1}
          \and G.E. Romero \inst{1,2} \and
          P. Benaglia \inst{1}
          }

   \institute{Instituto Argentino de Radioastronomía (CONICET--CIC--UNLP), C.C.5, (1984) Villa Elisa, Buenos Aires, Argentina\\
              \email{cgalante@iar.unlp.edu.ar}
         \and
             Facultad de Ciencias Astronómicas y Geofísicas, Universidad Nacional de La Plata, Paseo del Bosque s/n, 1900 La Plata, Argentina\\
             }

   \date{}

% \abstract{}{}{}{}{} 
% 5 {} token are mandatory
 
  \abstract
   {Starburst galaxies are undergoing intense episodes of star formation. In these galaxies, gas is ejected into the surrounding environment through winds created by the effect of hot stars and supernova explosions. When interacting with the intergalactic medium, these winds can produce strong shocks capable of accelerating cosmic rays. The radiation from these cosmic rays mainly occurs in radio and gamma rays. The radio halo can be characterized using the scale height, which is an important parameter for understanding cosmic ray acceleration and transport.}
   {We searched for the presence of radio halos in a sample of edge-on starburst galaxies gathered from the MeerKAT 1.28 GHz Atlas of Southern Sources in the IRAS Revised Bright Galaxy Sample. The investigation of how the radio halos relate to the global properties of the galaxies can shed light on the understanding of the halo origin and the underlying cosmic ray population.
}
   {We selected a sample of 25 galaxies with inclinations $i>80^{\circ}$ from the original sample and modeled their disk and halo contributions. We determined the scale heights and the radio luminosity of the halos when detected.
}
   {We have detected and characterized 11 radio halos from a sample of 25 edge-on galaxies. Seven of them are reported here for the first time. The average radio scale height is $\sim$1~kpc. We found that the halo scale heights increase linearly with the radio diameters and this relation does not depend on the star formation rate. All galaxies in our sample follow the radio-infrared relation with a $q$ parameter value of 2.5$\pm0.1$. The halo luminosity linearly increases with the infrared luminosity and star formation rate. }
   {The dependence of the halo luminosity on the star formation rate and the infrared luminosity supports the hypothesis that the radio halos are the result of synchrotron radiation produced by relativistic electrons and points toward the fact that the star formation activity plays a crucial role in halo creation. The average scale height of 1~kpc implies a dynamical range of 4~Myr, several orders of magnitude greater than the synchrotron losses for electrons of 10~TeV. This suggests that some process must exist to reaccelerate cosmic rays in the halo if gamma-ray emission of a leptonic origin is detected from the halo. According to the relation between the radio and gamma-ray luminosities, we found that NGC~4666 is a potential gamma-ray source for future observations.}


   \maketitle
%
%-------------------------------------------------------------------

\section{Introduction}
\label{sec:intro}

Starburst galaxies are characterized by undergoing an intense and rapid burst of star formation in their disks. This rate can be several times or even hundreds of times higher than that of normal galaxies \citep[e.g., Milky Way $\sim$1~M$_{\odot}$\,yr$^{-1}$,][]{Robitaille2010}. The high star formation rates (SFRs) are only possible if abundant gas feeds the process. Dust heated by the young stars reprocesses the radiation, which is then reemitted in the infrared (IR). This is why these galaxies are very luminous in the IR, reaching values larger than $10^{10}~\mathrm{L_{\odot}}$.

Massive young stars eject hot gas through their winds into the surrounding medium, creating a hot bubble around the star-forming region. Supernova explosions also contribute to heating the gas and increasing the pressure inside the bubble. Cold gas is pushed away, and the bubble eventually bursts. The hot gas, heated to temperatures of $T\sim 10^8$~K by supernova shocks, is unbound by the gravitational potential because its temperature is greater than the local escape temperature. The gas then expands adiabatically, becomes supersonic at the edge of the starburst region, and finally escapes from the disk, sweeping up cooler and denser gas. In the process, some of the cold gas and cosmic rays originally accelerated by the shocks within the starburst region are transported away from the disk, forming a galactic superwind \citep{Chevalier1985}. 

The superwind produces several effects. It depletes the disk of gas, quenching star formation and transporting metals into extragalactic space \citep{Bolatto2013}. The hot gas expands outside the disk, creating an X-ray emitting region surrounded by swept warm ($T\sim10^4$~K) gas that produces H$\alpha$ lines \citep[e.g.,][]{Heckman1990,Strickland2002}. The effects of the superwind can be detected up to distances of $\sim 10$ kpc from the disk of some edge-on galaxies. Most local starburst galaxies, such as M82 and NGC\,253, show clear evidence of superwinds \citep[e.g.,][]{Veilleux2005}. The superwind region in the halo is filled with a multiphase gas with hot, warm, cool, and relativistic components, and each are observable at different ranges of the electromagnetic spectrum. The relativistic component can be formed by particles that are accelerated in the disk and transported to the halo and/or by locally accelerated particles in the shocks that are produced by gas collisions in the outflows. 

Radio observations of nearby starbursts show extraplanar nonthermal emission on both sides of the galactic disks, which is produced by synchrotron radiation \citep[e.g.,][]{Beck1994,Heesen2009,Heesen2011}. From now on, we refer to this emission as radio halos. Such observations support the idea that relativistic particle acceleration and transport can be associated with the superwind. Radio halos require the coexistence of extraplanar cosmic rays and magnetic fields. As cosmic rays are injected and accelerated by shocks associated with supernova remnants and stellar winds in the star-forming region within the disk, they must be transported from the disk into the halo \citep{Heesen2009b,Peretti2019}. Due to adiabatic and radiative losses, the halos cannot extend very far unless there is some reacceleration of the particles. The natural place for such a process is the reverse shock at the terminal interaction region between the superwind and the intergalactic medium \citep[e.g.,][]{Anchordoqui1999,Romero2018,Peretti2022}. If some obstacles, such as large clouds or fragments from the disrupted disk, interact with the superwind, the bow shocks that formed around them can also be sites of particle acceleration \citep{Muller2020}. 

Identifying the radio halos to probe galactic winds and the cosmic ray population outside the disk of starburst galaxies is complex. The galaxy plane can be described by considering at least two exponential functions or a Gaussian and exponential function: one with a small scale height representing the disk itself and another with a larger scale height representing the halo \citep{Krause2018,Dahlem2001}.
Therefore, high-resolution imaging and very good sensitivity are required. Several surveys have been performed with different results. For example, \cite{Irwin1999} used the Very Large Array (VLA) in its C and D configurations to image 16 edge-on galaxies from the northern hemisphere at 20~cm. Only six of these galaxies were starbursts. They found that all but one galaxy (NGC\,4517) of their sample showed some evidence of extraplanar emission and concluded that such radiation appears to be common in star-forming galaxies.
\cite{Dahlem2001} made radio continuum observations with the VLA and the Australia Telescope Compact Array (ATCA) of a sample of 15 edge-on spiral galaxies, none of which had active galactic nuclei or nearby interacting partners. Radio halos were found in six of these galaxies. A trend was found that galaxies with radio halos have the highest IR 60~$\mu$m to 100~$\mu$m flux ratios, which is an indicator of galaxies with high SFRs. The measured exponential scale heights of these six detected radio halos range from about 1.4 to 3.1 kpc. 
More recently, \cite{Krause2018} determined the scale heights and radial scale lengths of the radio halos for a sample of 13 galaxies from the Continuum Halos in Nearby Galaxies -- an EVLA Survey (CHANG-ES) radio continuum survey \citep{Irwin2012} in two frequency bands. The average values they found for the radio scale heights of the halos are $1.1\pm 0.3$ kpc in the C band and $1.4\pm 0.7$ kpc in the L band. 

Radio and IR emission are observed to be correlated, both in local star-forming late-type galaxies and even in merging galaxies  \citep[e.g.,][]{Condon1993,Condon2002,Murphy2013}. The radio-IR correlation is characterized by the parameter $q$, defined as \citep{Helou1985}

\begin{equation}
    q = \log \left( \frac{F_{\mathrm{IR}}/(3.75\times 10^{12}~\mathrm{Hz})}{S_{1.4}} \right),
\label{eq:radio-IR-q}
\end{equation}

\noindent where $F_{\mathrm{IR}}$ is the IR flux and $S_{1.4}$ is the 1.4~GHz flux density. The typical values found for this parameter are $q=2.5$, $\sigma_{q}=0.1$ \citep{LI-shao2018}, $q=2.2$, $\sigma_{q}=0.1$ \citep{Irwin1999}, $q=2.3$, $\sigma_{q}=0.2$ \citep{Condon1992}, and $q=2.1$, $\sigma_{q}=0.16$ \citep{Helou1985}. Looking into this relationship might be key to better understanding the origin of the radio continuum halos.\\

In this paper, we present an investigation of the radio halos of all galaxies with inclinations $i>80^{\circ}$ in the recently published MeerKAT atlas of IRAS bright southern galaxies \citep{Condon2021}. The observations were carried out at 1.28~GHz with the South African radio interferometer MeerKAT, a precursor to the Square Kilometre Array (SKA). Our sample of 25 nearly edge-on galaxies is the largest one investigated to date.

In Section~\ref{sec:sample} we define our sample, while in Section~\ref{sec:method} we describe the preparation of the images, the modeling of the galactic disks, and the subtraction method used to separate the extraplanar emission. In Sections~\ref{sec:results-general} and \ref{sec:results-individual}, we explain how we applied this method to the sample and present the results obtained. Section~\ref{sec:correlations} presents an exploration of the correlations found in our results. In Section~\ref{sec:discussion} we discuss some implications of the results for other wavelengths, and finally we close with a summary and some conclusions in Section~\ref{sec:conclusions}.

%--------------------------------------------------------------------
\section{Galaxy sample}
\label{sec:sample}

\cite{Condon2021} published a MeerKAT atlas of 298 IRAS revised bright southern galaxies ($\delta<0$) \citep[IRAS, RBGS, ][]{Neugebauer1984,Sanders2003}. The observations were performed at $1.28~\mathrm{GHz}$ using the powerful MeerKAT radio interferometer. MeerKAT has 64 antennas with 13.5~m of diameter; 70\% of the antennas form a dense inner component distributed in two dimensions with a Gaussian $uv$ distribution with a dispersion of 300~m, a shortest baseline of 29~m and a longest baseline of 1~km. The maximum baseline of the array is 7.7~km.

We gathered the cropped MeerKAT images in FITS format available online\footnote{\url{https://doi.org/10.48479/dnt7-6q05}}. The final Stokes I images have an angular resolution of $7.5''$, and the rms is approximately 20~$\mu \rm Jy~beam^{-1}$.
The high angular resolution and low rms of these images make it possible to reveal faint extended emission, study its morphology, and look for the presence of galactic winds.

The edge-on galaxies are the best targets to separate the emission originating in the galactic plane from the extraplanar one in a galaxy. In the case of lower inclinations, the emission from the disk overlaps with the extraplanar emission, and a fraction of it will be challenging to distinguish. The methods commonly used to estimate the inclinations of the galaxies present high uncertainties. We decided to use INCLINET\footnote{\url{https://edd.ifa.hawaii.edu/inclinet/}}, an online application for determining the inclinations of spiral galaxies from their optical images, which uses neural networks to provide more reliable estimates of this parameter \citep{Kourkchi2020}. We used INCLINET to calculate the inclinations of all the galaxies in the MeerKAT Atlas and selected those with $i\gtrsim 80^{\circ}$. Besides, we selected those with radio angular sizes more significant than five times the angular resolution. We carefully analyzed each image's rms and quality (e.g., we discarded those images with reduction artifacts). Our final sample consists of 25 galaxies listed with their main properties in Table~\ref{table:general}.

\begin{table*}
    \caption{Galaxy sample.}
    \setlength\tabcolsep{4pt}
    \centering
    \begin{tabular}{l c c c c c c c c c c}
        \toprule
        Name & \mc{RA} & \mc{DEC} & \mc{$i$} & \mc{P.A.} & \mc{$D_{\mathrm{C}}$} & \mc{$\log(L_{\mathrm{IR}}/\mathrm{L_{\odot}})$} & \mc{SFR} & \mc{$S_{60}/S_{100}$} & \mc{$L_{1.4}$} & \mc{$S_{\mathrm{i}}/S$} \\
                 & (h m s) & ($^{\circ}\,'\,''$) & ($^{\circ}$) & ($^{\circ}$) & (Mpc) & & ($\mathrm{M_{\odot}\,yr^{-1}}$) & & ($\mathrm{10^{21}~W~Hz^{-1}}$) & (\%) \\
            (1) & \multicolumn{2}{c}{(2)} & \mc{(3)} & \mc{(4)} & \mc{(5)} & \mc{(6)} & \mc{(7)} & \mc{(8)} & \mc{(9)} & \mc{(10)} \\
                  
            \midrule
            ESO 005-G004 & 06 05 36.5 & -86 37 53 & 89 & -86.8 & 22.36 & 10.16 & 2.15 & 0.39 & 5.55 & 18.46 \\
            ESO 079-G003 & 00 32 02.0 & -64 15 14 & 87 & -48.5 & 35.31 & 10.51 & 4.84 & 0.40 & 8.81 & 38.09 \\
            ESO 163-G011 & 07 38 04.3 & -55 11 27 & 84 & 3.3 & 38.7 & 10.5 & 4.74 & 0.47 & 8.49 & -- \\
            ESO 209-G009 & 07 58 14.9 & -49 51 09 & 88 & -25.9 & 11.81 & 9.81 & 0.95 & 0.38 & 1.93 & 1.72 \\
            ESO 428-G028 & 07 23 38.3 & -30 03 08 & 89 & 60.1 & 31.08 & 10.35 & 3.32 & 0.41 & 5.68 & 11.46 \\
            IC 3908 & 12 56 41.1 & -07 33 46 & 82 & -8.3 (-10.0) & 20.28 & 10.07 & 1.74 & 0.51 & 3.35 & 11.85 \\
            IC 4595 & 16 20 44.4 & -70 08 35 & 88 & 59.7 & 45.75 & 10.78 & 8.87 & 0.38 & 19.24 & 8.26 \\
            NGC 134 & 00 30 21.6 & -33 14 38 & 81 & 49.6 & 15.9 & 10.37 & 3.46 & 0.38 & 6.52 & 0.79 \\
            NGC 1406 & 03 39 23.2 & -31 19 20 & 85 & 16.6 & 18.98 & 10.22 & 2.46 & 0.47 & 5.80 & 44.65 \\
            NGC 1421 & 03 42 27.7 & -13 29 23 & 82 & 0.9 & 21.2 & 10.25 & 2.62 & 0.48 & 5.54 & 21.11 \\
            NGC 1448 & 03 44 32.1 & -44 38 41 & 87 & 41.7 & 11.47 & 9.78 & 0.90 & 0.32 & 1.80 & 2.23 \\
            NGC 1532 & 04 12 01.2 & -32 53 11 & 83 & 34.2 & 15.51 & 10.01 & 1.51 & 0.31 & 3.32 & 0.98 \\
            NGC 2221 & 06 20 16.0 & -57 34 43 & 80 & -0.6 & 33.83 & 10.39 & 3.63 & 0.46 & 6.26 & 15.69 \\
            NGC 2613 & 08 33 23.2 & -22 58 28 & 82 & -72.0 & 20.18 & 10.18 & 2.27 & 0.29 & 3.05 & 0.59 \\
            NGC 2706 & 08 56 12.2 & -02 33 47 & 83 & -11.6 & 24.36 & 10.15 & 2.09 & 0.47 & 3.76 & 9.69 \\
            NGC 3175 & 10 14 42.9 & -28 52 25 & 81 & 55.1 & 13.45 & 9.96 & 1.37 & 0.44 & 1.80 & 39.36 \\
            NGC 3263 & 10 29 12.3 & -44 07 13 & 81 & -78.0 & 38.79 & 10.66 & 6.78 & 0.45 & 25.21 & 59.66 \\
            NGC 3717 & 11 31 31.2 & -30 18 24 & 85 & 32.0 & 21.39 & 10.29 & 2.90 & 0.46 & 4.17 & 61.75 \\
            NGC 4666 & 12 45 07.7 & -00 27 41 & 81 & 40.6 & 12.82 & 10.36 & 3.42 & 0.43 & 8.76 & 2.86 \\
            NGC 4835 & 12 58 08.2 & -46 15 56 & 81 & -30.5 & 23.96 & 10.57 & 5.48 & 0.44 & 13.30 & 2.96 \\
            NGC 5073 & 13 19 18.5 &-14 50 30 & 88 & -31.1 & 37.09 & 10.6 & 5.95 & 0.62 & 6.77 & 74.65 \\
            NGC 7090 & 21 36 28.4 & -54 33 30 & 88 & -51.8 & 7.65 & 9.22 & 0.25 & 0.41 & 0.36 & -- \\
            UGCA 150 & 09 10 49.5 & -08 53 36 & 87 & 29.5 & 26.85 & 10.3 & 2.96 & 0.29 & 4.01 & 3.07\\
            UGCA 394 & 14 47 24.1 & -17 26 49 & 88 & -3.9 (-10.0) & 30.02 & 10.26 & 2.67 & 0.55 & 3.71 & 48.75 \\
            UGCA 402 & 15 13 30.6 & -20 40 29 & 85 & 61.8 & 30.94 & 10.34 & 3.21 & 0.40 & 6.95 & 6.76 \\
            \bottomrule
        \end{tabular}
        \tablefoot{
            (1) Galaxy name. (2) J2000 coordinates of the IRAS source. (3) Inclination calculated using INCLINET. (4) Position angle, measured from the north, between 90$^{\circ}$ and -90$^{\circ}$. Obtained from HYPERLEDA\footnote{\url{http://leda.univ-lyon1.fr/}} \citep{Makarov2014}. For IC~3908 and UGCA~394 we had to use slightly different values, shown in parenthesis, to ensure that the major axis was horizontal after rotating them. (5) Comoving distance, taken from \cite{Condon2021}. (6) Logarithm of the absolute IR $\left(8<\lambda\left[\mathrm{\mu m}\right]<1000\right)$ luminosity in units of the solar bolometric luminosity $L_{\odot}=3.83\times 10^{26}~\mathrm{W}$, taken from \cite{Sanders1996}. (7) The SFR calculated as $\mathrm{SFR}=0.39\,L_{\mathrm{IR}}/10^{43}~\mathrm{erg\,s^{-1}}$ \citep{Heesen2018}. (8) $60~\mathrm{\mu m}$ to $100~\mathrm{\mu m}$ IR IRAS flux ratio \citep{Dahlem2001}. (9) Total $1.4~\mathrm{GHz}$ luminosity obtained from the MeerKAT 1.28~GHz flux densities. We assumed that $S_{1.4}\approx 0.94\,S_{1.28}$ given that most star-forming galaxies have spectral indices near $\alpha\approx-0.7$, with $S(\nu)\propto \nu^{\alpha}$ \citep{Condon1992}. (10) Contribution of the compact central source to the total flux density of the galaxy \citep{Condon2021}.}
    \label{table:general}
\end{table*}

\section{Method}
\label{sec:method}

\subsection{Preparation of the images}
\label{subsec:prep}

Before the analysis, a series of processes were applied to the selected MeerKAT images. To align the galaxies with the horizontal axis, we rotated the images using the position angle (PA) of each galaxy; PA values are shown in Table~\ref{table:general}, column five. We estimated the rms noise of the background of each image, using as an initial estimation the average value given by \cite{Condon2021} of $20~\mathrm{\mu Jy\,beam^{-1}}$. We fit a normal distribution to all the pixels for which the intensity is below three times the initial rms. The new estimation was obtained as 
\begin{equation}
 \sqrt{\sigma_{m}^{2}+\sigma_{s}^{2}},
 %\label{eq:rms}
\end{equation}
where $\sigma_{m}$ and $\sigma_{s}$ are the mean value and the standard deviation of the fitted normal distribution. Then, we selected only the pixels for which the intensity exceeds three times the new rms estimation. We estimated the radio diameter of each galaxy as the angular size of the radio intensity within three times the rms estimation. Since in the total image, there may be other sources than the galaxy in which we are interested, we cropped the images and kept only a rectangular section containing the emission from the galaxy. If, in that section, there were remaining non-related sources that could interfere with the analysis, they were removed by fitting a two-dimensional Gaussian function using the NOD-3 package \citep{Muller2017}. Finally, we analyzed the presence of strong nuclear sources that dominate the emission from the disk because they can create confusion when analyzing the presence of halos. The contribution $S_{\mathrm{i}}$ of these sources to the total flux density $S$ of the galaxies was calculated by \cite{Condon2021} by fitting a Gaussian function. We removed them in the cases in which the contribution $S_{\mathrm{i}}/S$ showed in Table~\ref{table:general}, column 10, was greater than $25\%$ of the total.

\subsection{Construction and modeling of the z profiles} \label{subsec:profiles}

The determination of the presence and size of the radio halos was performed using the $z$ profiles, with $z$ being the direction perpendicular to the galactic disk. The rectangular section of the image described in Sec. \ref{subsec:prep} was divided into rows of pixels in the $z$ direction. These strips were averaged in the direction of the galaxy's major axis, obtaining a single-row profile that describes the average intensity distribution in the $z$ direction. The width of the resulting profile is given by the combination of the inclination $i$ of the galactic disk with the line of sight, the intrinsic height of the disk, and the possible presence of extraplanar emission, that is, a radio halo, all convolved with the beam of the telescope. 

To determine the presence and size of a radio halo, we adopted an approach similar to \cite{Dahlem2001}. We consider that the disk is infinitesimally thin, and when we see it inclined and projected onto the sky, we assume that the corresponding $z$ profile can be described either by a Gaussian
\begin{equation}\label{eq:gauss}
    \omega_{g}(z)=\omega_{0}\exp(-z^{2}/z_{0,\mathrm{d}}^{2})
\end{equation}

\noindent or an exponential function
\begin{equation}\label{eq:exp}
    \omega_{e}(z)=\omega_{0}\exp(-z/z_{0,\mathrm{d}}),
\end{equation}

\noindent where $z_{0,\mathrm{d}}$ is the scale height of the disk. This parameter is usually considered to describe the halo size because it is a more physical quantity than the extent of a radio halo that simply depends on the sensitivity of the observations. To model the emission from the halo, a second exponential component with scale height $z_{o,\mathrm{h}}$ was added only when a single component was not enough to describe the shape of the profile. These components were then convolved with a Gaussian-shaped beam described by

\begin{equation}\label{eq:beam}
    g(z)=\frac{1}{\sqrt{2\pi \sigma^{2}}}\exp \left( -z^{2}/2\sigma^{2}\right),
\end{equation}

\noindent with $\mathrm{HPBW}=2\sqrt{2\ln 2}\,\sigma$. The convolutions can be obtained analytically, and the resulting convolved functions are
\begin{align}
    W_{\mathrm{exp}}(z) & =\frac{w_{0}}{2}\exp \left( -z^{2}/2\sigma^{2}\right) \nonumber \\
     & \times \Bigg[ \Bigg. \exp \left( \frac{\sigma^{2}+z\,z_{0}}{\sqrt{2}\,\sigma\,z_{0}} \right)^{2} \mathrm{erfc}\left( \frac{\sigma^{2}+z\,z_{0}}{\sqrt{2}\,\sigma\,z_{0}} \right) \nonumber \\
     & + \exp \left( \frac{\sigma^{2}-z\,z_{0}}{\sqrt{2}\,\sigma\,z_{0}} \right)^{2} \mathrm{erfc}\left( \frac{\sigma^{2}-z\,z_{0}}{\sqrt{2}\,\sigma\,z_{0}} \right) \Bigg. \Bigg]
\end{align}

\noindent for the exponential, where erfc is the complementary error function defined as
\begin{equation}
    \mathrm{erfc}(x) = \frac{2}{\sqrt{\pi}} \int_{x}^{\infty} \exp (-r^{2})\,\mathrm{d}r,
\end{equation}

\noindent and
\begin{equation}
    W_{\mathrm{Gauss}}(z)=\frac{w_{0}\,z_{0}}{\sqrt{2\sigma^{2}+z_{0}^{2}}} \exp (-z^{2}/(2\sigma^{2}+z_{0}^{2}))
\end{equation}

\noindent for the Gaussian. As the $z$ profiles can be asymmetric, we allow the scale heights on both sides of the disk to be different, with a common maximum. 

When a radio halo is detected, we also calculate how much it contributes to the total flux density of the galaxy (hereafter, the halo contribution). To do this, we take the beam-convolved component of the fit corresponding to the disk and match its maximum with the maximum of the $z$ profile. The halo contribution is then calculated as the area between the disk component and the total profile on each side of the disk separately. We assume the emission around $z=0$ is from the disk, as we cannot separate the two components. For this reason, the contribution is only calculated for values of $|z|$ greater than the FWHM of the beam-convolved component that describes the disk.

\section{Results: General}
\label{sec:results-general}

In general, the disk component of the profiles was better fitted with exponential functions than with Gaussian functions. The disk scale heights $z_{0}^{\mathrm{d}}$ range between $0.6''$ and $15''$, except for the galaxies NGC~134 and NGC~2613, where the disk scale heights resulted larger, possibly due to their low inclination. These parameters are shown in Tables~\ref{table:results-halo} and \ref{table:results-nohalo} together with the disk scale heights in kpc, that range between $0.1$ and $2.5$ kpc. We note that this value is the apparent height of the disk, which could be produced by an inclination $i<90^{\circ}$, intrinsic thickness, or a composition of both effects. Thus, the $z_{0}^{\mathrm{d}}$ are not, in all cases, a direct measure of the intrinsic disk thickness. The errors of the scale heights of the disk were generally less than $20\%$, except for NGC~7090, which had an error of $\sim 24\%$. It is important to note that the magnitude of these errors is expected, given that the disk scale heights are sometimes much smaller than the angular resolution, which is $\sim 8''$. 

For the 11 galaxies listed in Sections~\ref{subsec:first-time-detected} and \ref{subsec:prev-detected}, a single-component function was not enough to fully describe the shape of their $z$ profiles and a second component was added, what we interpret as the detection of a radio halo. We averaged the halo scale heights and halo contributions at the left and right sides of the disk to obtain the mean values. 
All these parameters, the rms obtained for each radio image, and the diameter of the radio emission $d_{\mathrm{r}}$, which were calculated according to the method described in Section~\ref{sec:method}, are shown in Table~\ref{table:results-halo}. The average halo scale heights range between $4''$ and $20''$, with a mean value of $11.5''$ and errors lower than $19\%$. This is equivalent to scale heights between $0.7$ and $1.5~\mathrm{kpc}$, with a mean value of $1~\mathrm{kpc}$. We found that sometimes, on one side of the $z$ profile, the halo scale height resulted in the same order or lower than the disk scale height. This means the halo component was necessary only on one profile side. In these cases, we consider a positive detection of a halo only on the other side. Also, we found in some cases that the angular sizes of the halo scale heights are smaller than the angular resolution of the image. Thus, these results are preliminary and higher resolution data is needed to ensure a positive halo detection. For seven of these galaxies, listed in Section~\ref{subsec:first-time-detected}, extended radio emission had never been studied in the past. Therefore, in the present work, we report their radio halos for the first time. The radio halos of the four galaxies listed in Section~\ref{subsec:prev-detected} were already detected in the past by other authors. In this work, we confirm their results and obtain more accurate estimations of the halo scale heights.

For the remaining 14 galaxies in Section~\ref{subsec:non-detections}, a single-component function was enough to describe the $z$ profiles fully. In some cases, a barely excess emission above the fitted function is observed, but it is insufficient for adding a second component to the fit. Thus, we conclude that no halo is detected. The disk scale heights $z_{0}^{\mathrm{d}}$ obtained for their fits are shown in Table~\ref{table:results-nohalo}, together with the obtained rms of the radio images and the diameter of the radio emission $d_{\mathrm{r}}$. Four of these galaxies --NGC~2613, NGC~3175, NGC~3717 and NGC~5073-- were already studied in the past by other authors, who also did not detect a radio halo except for NGC~3175.

\begin{figure*}
\centering
    \subfloat{
        \centering
        \includegraphics[width=0.45\textwidth]{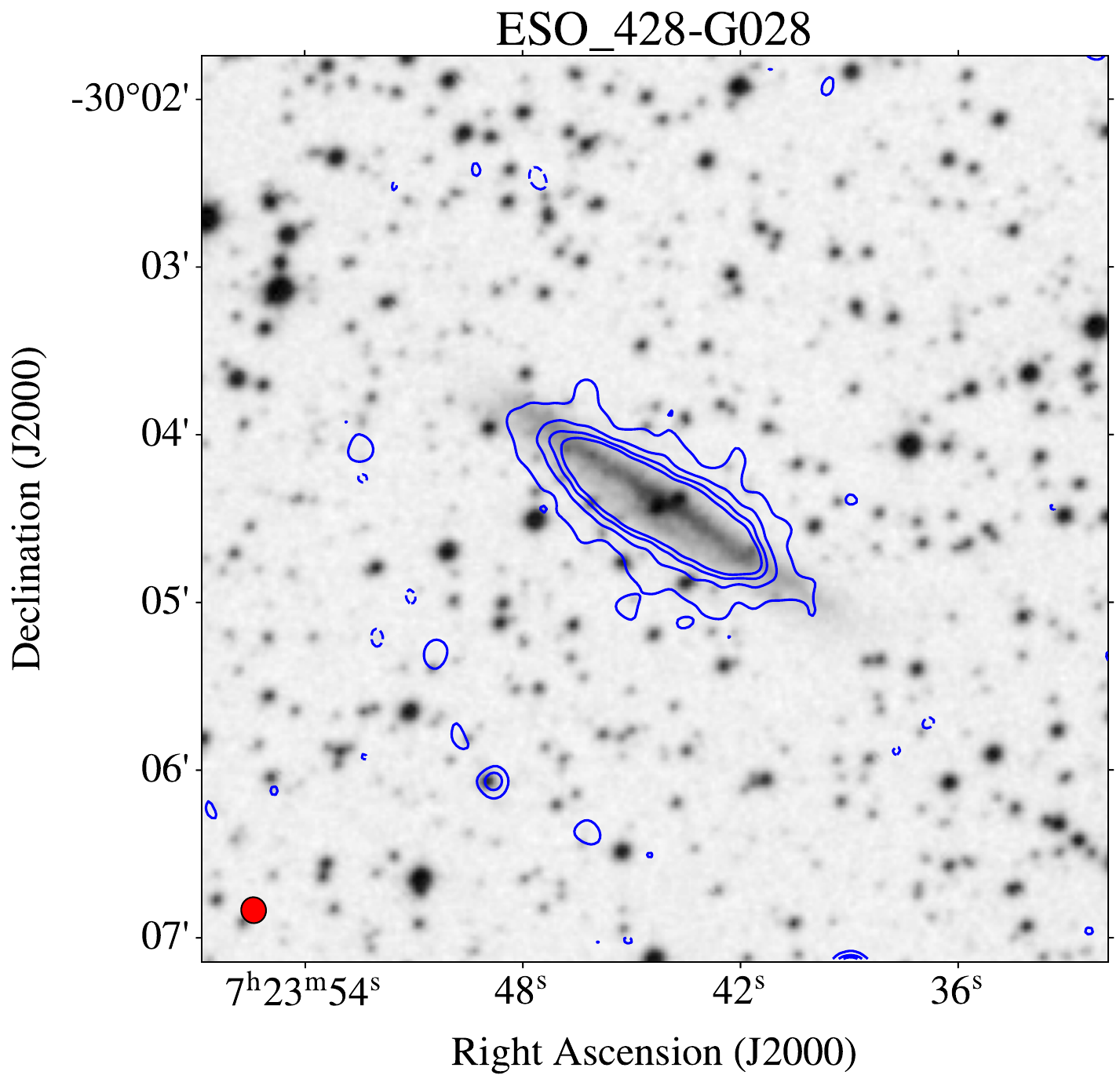}
        \hspace{0.3cm}
        \includegraphics[width=0.45\textwidth]{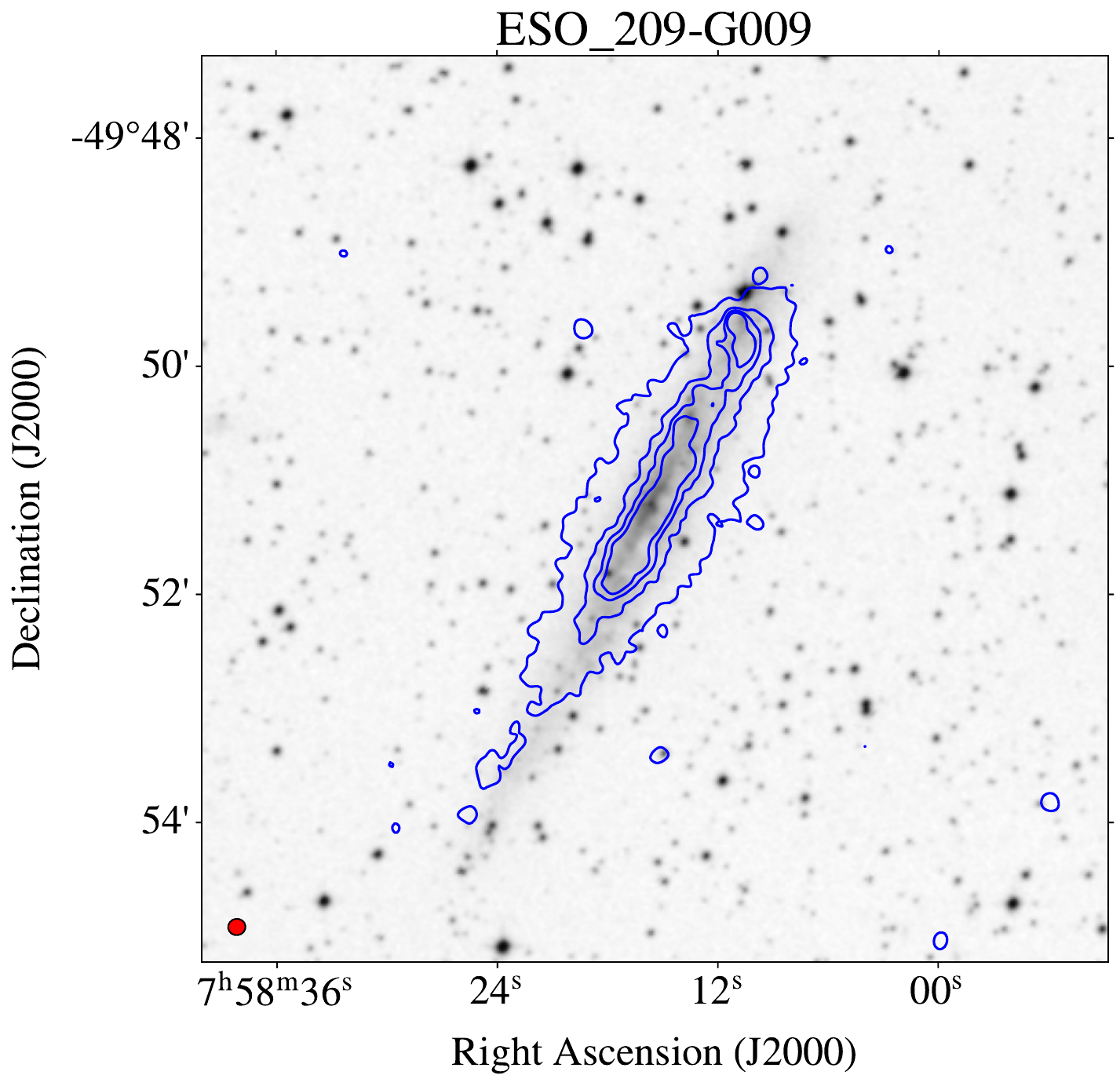}
    }\\
    \vspace{0.6cm}
    \subfloat{
        \centering
        \includegraphics[width=0.45\textwidth]{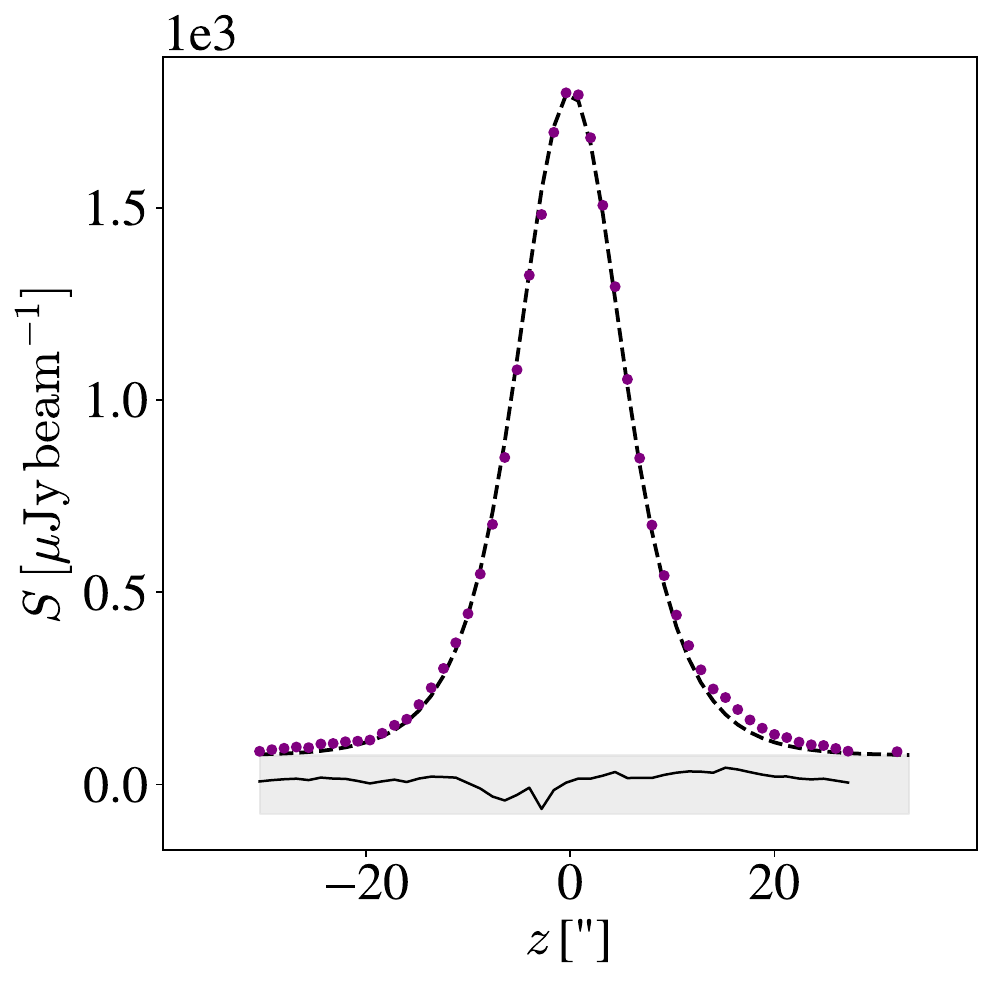}
        \hspace{0.5cm}
        \includegraphics[width=0.43\textwidth]{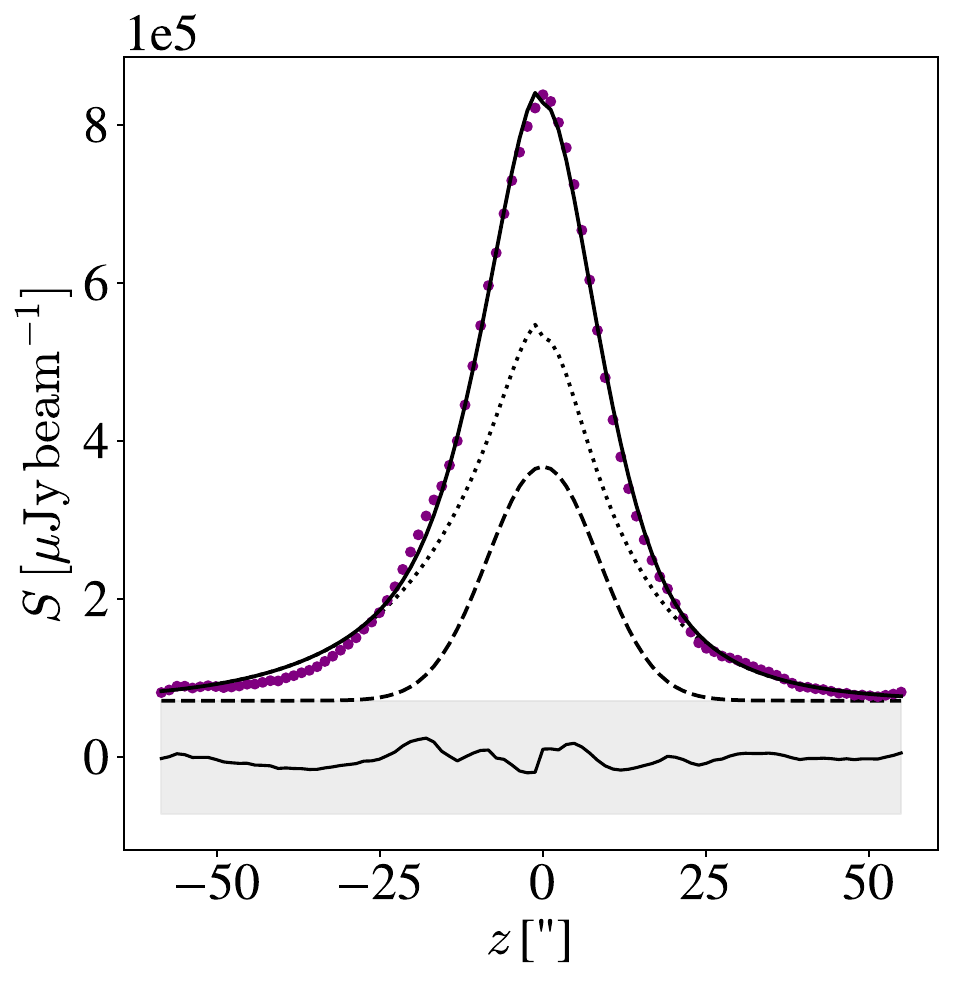} 
    }
\caption{Radio contour maps superimposed on a DSS image (\textit{top}) and $z$ profile (\textit{bottom}). The horizontal axis of the profiles is the offset in arc-seconds from the galactic plane in $z$ direction, and the vertical axis the average flux density in $\mu \mathrm{Jy\,beam^{-1}}$. The shaded gray area extends from -3~rms to 3~rms, and the solid line within this area is the residual of the fit. \textit{Left panel:} ESO~428-G028. The contour levels are -0.07, 0.07, 0.22, 0.37 and 0.52 mJy\,beam$^{-1}$. The data is shown with dots and the exponential fitted function with dashed line. \textit{Right panel:} ESO~209-G009. The contour levels are -0.12, 0.12, 0.36, 0.60 and 0.84 mJy\,beam$^{-1}$. The synthesized beam of $7.5''$ is shown in the bottom left corner of each image. The data are shown with dots, the fitted disk component in dashed line, the halo component in dotted line and the total fit in solid line.}
\label{figure:galaxies-example}
\end{figure*}

\begin{figure}
    \centering
    \includegraphics[width=0.45\textwidth]{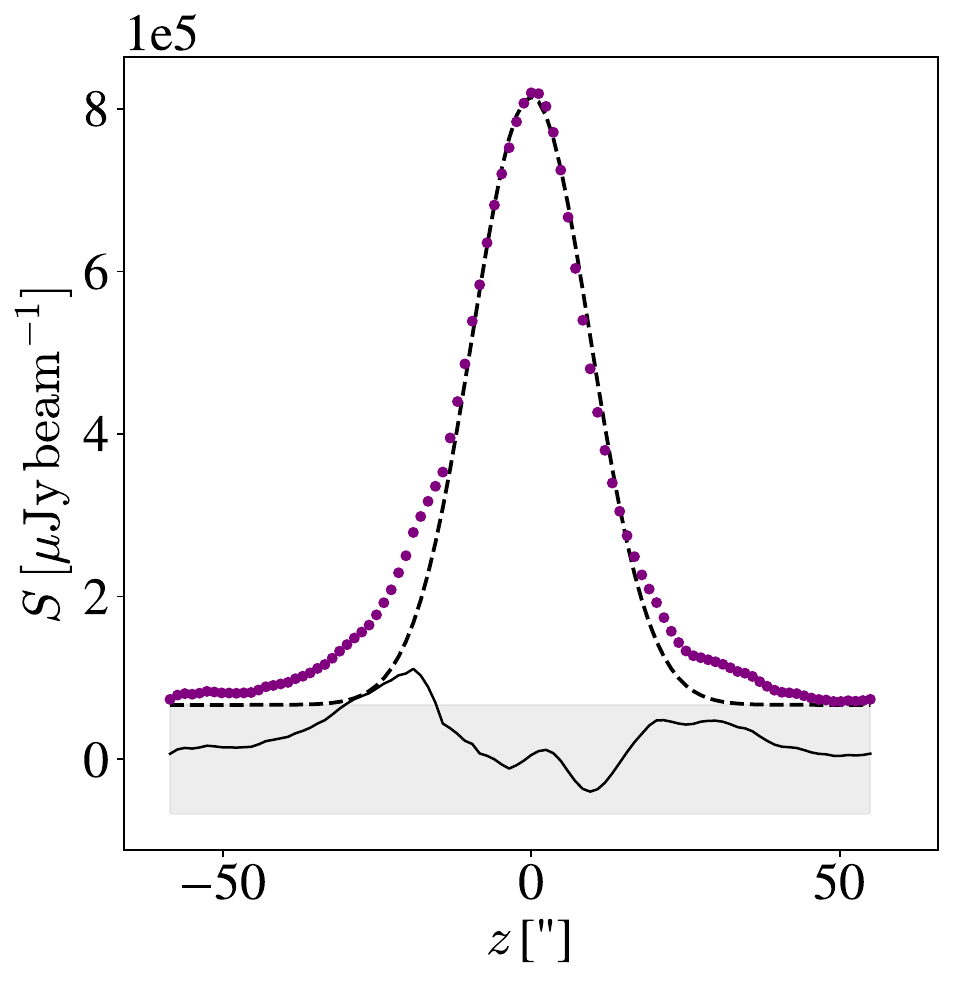}
    \caption{One-component fit for the galaxy ESO~209-G009. The horizontal axis is the offset in arc-seconds from the galactic plane in $z$ direction, and the vertical axis the average flux density in $\mu \mathrm{Jy\,beam^{-1}}$. The data are shown with dots and the Gaussian fitted function with dashed line. The shaded gray area extends from -3~rms to 3~rms, and the solid line within this area is the residual of the fit. The ``wings'' of emission above the fit at both sides of the profile together with the residuals being greater than 3~rms suggest that a second component is needed to accurately fit the data.}
    \label{figure:example-badfit}
\end{figure}

As an example, we show in Fig.~\ref{figure:galaxies-example} the radio contour map over the optical Digital Sky Survey (DSS) image, the $z$ profile and the fit for two galaxies: ESO~209-G009, one of the galaxies for which we detected a radio halo for the first time, and ESO~428-G028, for which we did not detect a radio halo. In Fig. \ref{figure:example-badfit} we show for comparison the one-component fit we obtained for ESO~209-G009. It can be seen that there are ``wings'' of emission above the fit on both sides of the profile. In addition, the residuals reach values greater than 3 rms, suggesting that a second component needs to be added to the fit. For visual references of the rest of the sample, please refer to Appendix~A, where the corresponding images and fitted profiles can be found. In the following paragraphs and sections, we provide a description of the results, the main characteristics of the sample galaxies, and an overview of the analysis conducted.

\section{Results: Individual sources}
\label{sec:results-individual}

\subsection{First-time detected radio halos}
\label{subsec:first-time-detected}

\noindent \textsf{ESO~005-G004:} The radio map of this galaxy, which harbors a hidden Seyfert~2 nucleus \citep{Mckernan2010,Garcia-Bernete2016}, reveals an extended emission, centrally peaked, covering the entire optical disk. Two optical point sources are located nearly symmetrically on both sides of the disk center. Up to date, its is not determined if these sources are related or not with the galaxy.

We found that an exponential function provided the most accurate fit for the disk, and we also observed highly asymmetric extraplanar emission that could not be fully explained with a single component, suggesting the presence of a halo. No trace of a jet that might be associated with the Seyfert nucleus was found.\\

\noindent \textsf{ESO~079-G003:} The contour map of this galaxy shows that the radio emission is centrally peaked. It covers almost all the optical disk, thicker on one side and with several clumps and plumes along its extension. 

We determined that a Gaussian function offered a good fit for the disk and found the presence of extraplanar emission that required the addition of a second component to achieve an accurate fit. However, the disk and halo scale heights resulted in smaller values than the angular resolution, making this a preliminary detection and indicating the need for even higher-resolution data.\\

\noindent \textsf{ESO~163-G011:} This galaxy is paired with ESO~163-G010. The contour map and profile exhibit an evident asymmetry; the left side of the profile displays greater emission, which corresponds to the side of the disk closer to its companion galaxy. For this reason, it was necessary to fit the profile with a two-component function: an exponential for modeling the disk and a second component to account for the asymmetric emission, which we interpret as a radio halo. However, the angular value of the average halo scale height was smaller than the angular resolution, making it another preliminary detection. \\

\noindent \textsf{ESO~209-G009:} Despite it having a significantly higher rms value than the other galaxy images, we could still thoroughly analyze this galaxy. To construct the $z$ profile, we had to select a smaller rectangular region than the entire galaxy, excluding the disk's outer regions that appeared as a tail on one side. The radio emission extends through the optical disk and is extended out of the galactic plane in $z$ direction, showing a clumpy structure, more prominent near the tail. The $z$ profile revealed that the galaxy exhibited emission beyond its disk, and a two-component model with a Gaussian disk was required to describe the profile accurately. However, as the halo scale height on the right side of the profile results in the same order as the disk scale height, we conclude that the radio halo is only detected on the left side. In the past, \cite{Rossa2003} detected extraplanar ionized gas in this galaxy, while recently, \cite{Mosenkov2022} found extraplanar dust. However, the radio halo was studied and detected for the first time in the present work. \\

\noindent \textsf{IC~3908:} For this galaxy, we had to adjust the position angle provided by HYPERLEDA by a small margin. We observed that the major axis was not horizontal when rotating, indicating the need for a slightly different angle. Upon examining the radio contours, we observed that the galaxy exhibits emission extending far beyond the galactic disk on both sides, forming a somewhat rectangular shape. To obtain the best fit for the $z$ profile, we utilized a Gaussian disk as well as a second component to account for the extraplanar emission. However, the halo scale height on the right side of the profile resulted in the same order as the angular resolution, making it a preliminary detection. \\

\noindent \textsf{NGC~1532:} Because this galaxy is part of an interacting pair with NGC~1531, its $z$ profile  was particularly difficult to fit. Its disk is highly distorted, and large plumes of extraplanar emission can be seen from the radio contour map. After removing the emission from NGC~1531, the best fit was found for a two-component function with a Gaussian disk. This galaxy has one of the most prominent radio halos in our sample, with an average scale height of $1.5$ kpc.\\ 

\noindent \textsf{NGC~2221:} This is a dwarf irregular galaxy, part of an interacting pair together with NGC~2222. Some small sources in the vicinity had to be removed before the analysis. Despite being one of the sample galaxies with a smaller angular size and a low inclination angle, this galaxy exhibits prominent extraplanar emission, which requires fitting its profile using a two-component function. An exponential disk provided the best fit, but the disk and halo scale heights were found to be smaller than the angular resolution. Thus, higher-resolution data are needed to confirm these results.\\

\subsection{Previously detected halos}
\label{subsec:prev-detected}

\noindent \textsf{NGC~1406:} The strong nuclear source present in this galaxy had to be removed before the analysis. The radio contour map shows evidence of extraplanar emission, more intense at the central part on one side of the galactic disk. The emission covers almost the entire optical disk in the radial direction. 

The best fit resulted from using a Gaussian disk and a component to account for the halo. As the left halo scale height resulted in the same order as the disk scale height, we conclude that the halo is detected only on the right side. The radio halo of this galaxy was detected for the first time by \cite{Dahlem2001}, who found an average halo scale height of 1.66~kpc using $43''$ resolution VLA observations. We correct this value to $1.06\pm 0.08~\mathrm{kpc}$.\\

\noindent \textsf{NGC~1421:} This galaxy has a relatively low inclination, allowing for clear visibility of its spiral arms and disk structure around the central source. We observed evidence of extraplanar emission in both the $z$ profile and contour map after removing nearby sources. To accurately describe the profile, a two-component function with a Gaussian disk was necessary, but due to the high asymmetry of the profile, obtaining a good fit was challenging. The resulting disk scale height was found to be large, and the left halo scale height was near the size of the telescope beam, resulting in a negligible halo contribution to the flux density on that side. This suggests the possible presence of a radio halo on this side of the galactic plane, but it is masked by the disk emission. On the other side of the disk, the halo emission is more reliably detected, resulting in a highly asymmetric halo. The presence of a radio halo in this galaxy was first suggested by \cite{Irwin1999}. Later, \cite{Dahlem2001} analyzed the profile of this galaxy, but they needed only one component function to describe it. Then, the present work is the first detection of a radio halo for this galaxy, so confirming \cite{Irwin1999} original suggestion. Also, this galaxy was observed in X-rays by the Einstein instrument \citep{Fabbiano1992}, showing extended emission with a very particular elongated shape beyond the stellar disk. \\

\noindent \textsf{NGC~4666:} This galaxy is known to host both a starburst and an AGN \citep{Persic2004}. Despite its low inclination, it exhibits a prominent wind, which has been studied by several authors in X-rays \citep{Ehle2004}, H$_{\alpha}$ \citep{Voigtlander2013}, and also shows HI emission more extended than the optical disk, with spurs and tidal features which suggest that it is interacting with NGC~4668, located to its southeast \citep{Zheng2022}. This galaxy was studied in radio by \cite{Stein2019}, who found a halo scale height of 2.16~kpc, almost twice the value found in this work. However, these two results are consistent if we take into account that they consider a distance of 27.5~Mpc, while in this work, we adopted the updated value of 12.82~Mpc. 

Using our method, we initially attempted to fit the $z$ profile with a single exponential function, but this resulted in an extremely large disk scale height. One possible explanation is that the relative intensity of the halo emission with respect to the disk emission is so high that it dominates the shape of the profile, making it impossible to separate the disk component. This effect could be amplified by the low inclination of the galaxy, which causes the disk emission to be more widespread. However, since the disk scale height resulting from a single-component fit was excessively large, and the existence of a radio halo produced by star formation activity in this galaxy has been well established \citep{Heesen2018,Heesen2019}, we accurately fitted the $z$ profile using a two-component function with a Gaussian fit.\\

\noindent \textsf{NGC~7090:} Several authors have also examined this nearby galaxy; \cite{Dahlem2001} detected for the first time a radio halo with an average scale height of 1.77~kpc. They used VLA observations at 1.43~GHz, with an angular resolution of 34.5", and considered a galaxy distance of 11.7~Mpc. Later, \cite{Dahlem2005} found extraplanar HI emission. \cite{Rossa2003} observed evidence of extraplanar $\mathrm{H}_{\alpha}$ emission, and recently \cite{Jo2018} found extraplanar dust and $\mathrm{H}_{\alpha}$ emission. As evidenced by both the $z$ profile and the radio contours, this galaxy exhibits a prominent and highly asymmetric radio halo. To achieve the best fit, we utilized a two-component function with an exponential disk. We obtained an average halo scale height of 0.7~kpc, considering the distance listed in Table~\ref{table:general}. The data we used has an angular resolution of $\sim 4$ times better than the previously used in \cite{Dahlem2001}. 

\subsection{Non-detections}
\label{subsec:non-detections}

\noindent \textsf{ESO~428-G028:} The radio emission of this galaxy is centrally peaked and extends over the entire disk. Despite its apparent thickness in the $z$ direction, with some plumes extending from the galactic plane, a one-component exponential function adequately described the $z$ profile. \\

\noindent \textsf{IC~4595:} The radio emission of this galaxy extends to the outermost parts of the optical disk. Before analysis, a very bright, nearby source had to be removed to ensure accurate results. The most optimal fit for this particular galaxy was achieved using an exponential disk. \\

\noindent \textsf{NGC~134:} This galaxy has a relatively low inclination angle that allows the structure of its disk to be clearly visible. The radio contour map analysis indicates that the emission closely follows the optical disk without a central peak and no detectable signs of extraplanar emission. It is possible that, if present, any extraplanar emission is masked by the strong disk emission due to the galaxy's low inclination angle. 

After removing nearby sources, we were able to obtain the most optimal fit for this galaxy using a Gaussian disk without requiring a second component to describe the $z$ profile. Interestingly, the resulting disk scale height was significantly larger than that of the rest of the sample, which may be a consequence of the low inclination angle of the galaxy.\\

\noindent \textsf{NGC~1448:} Although the radio emission of this galaxy, which hosts a Compton-thick obscured AGN \citep{Annuar2017}, extends into the outer regions of the galactic disk, the irregularity of this emission requires us to focus on a smaller region when constructing the $z$ profile to obtain an accurate fit. We used a single-component exponential function to obtain the best fit. \\

\noindent \textsf{NGC~2613:} \cite{Irwin2019} recently found evidence for a possible AGN in this galaxy. \cite{Li2006} studied this galaxy in X-rays and suggested the presence of an obscured AGN and a bubble that may correspond to extraplanar emission. They suggest that it could be produced by a starburst or an AGN in case of being an outflow. \cite{Irwin1999} studied the radio emission of this galaxy and found evidence of extraplanar emission. \cite{Chaves2001}, on the other hand, found extraplanar HI features.

The disk of this galaxy has a relatively low inclination angle, allowing its structure to be clearly visible. Its ring shape shows two regions brighter than the rest and even brighter than the emission from the galaxy center. It can be seen that the radio emission follows the optical disk, with no evidence of extraplanar emission. The high asymmetry and the low inclination of the disk made this galaxy's profile very difficult to fit. After the removal of surrounding sources and once obtained the profile, we found the best fit using a single Gaussian function. The disk scale height obtained was significantly larger than that of the rest of the sample, perhaps as a consequence of its low inclination.\\

\noindent \textsf{NGC~2706:} This galaxy is one of the sample galaxies with a smaller angular size and has a relatively low inclination angle, which causes the optical disk to be thick. The radio emission extends through the galactic disk and also appears broad in $z$ direction. However, we determined that a single-component function was enough to accurately describe the shape of the $z$ profile.\\

\noindent \textsf{NGC~3175:} This galaxy was previously studied by \cite{Dahlem2001}, who found a preliminary detection of a radio halo. However, they performed their analysis without removing the intense nuclear source present in this galaxy. When fitting the $z$ profile, the nuclear emission could be misinterpreted as the disk emission, and in turn, the disk emission was interpreted as the presence of a halo. 

In our work, the nuclear source of this galaxy had to be removed before analysis. The emission appears broad on the radio contour maps, but this may be due to the low inclination. A group of blobs is present on one side of the disk; however, they were excluded from the region used to build the $z$ profile, as it was impossible to determine whether they were part of the galaxy or unrelated sources. Contrary to \cite{Dahlem2001}, we found the best fit using a single-component exponential function, suggesting the absence of a radio halo. \\

\noindent \textsf{NGC~3263:} This galaxy is interacting with the radio-quiet galaxy NGC~3262, and is a member of the NGC~3256 group. It shows a tidal tail located at one end of the disk. When constructing the $z$ profile, it was challenging to identify an area that did not include any of the emission from the tail. However, we were able to select an appropriate region where the contribution from the tail was negligible, and the $z$-profile analysis could be performed. Despite the presence of the tidal tail and the galaxy's low inclination, we determined that a single-component function was enough to accurately fit the profile. \\

\noindent \textsf{NGC~3717:} The radio emission in this galaxy, paired with IC~2913, is primarily concentrated in the central region, with no discernible signs of extraplanar emission as it follows the optical disk. After removing the strong nuclear source and nearby sources, we fit the $z$ profile and determined that a single exponential function provided the best fit. Our results agree with \cite{Dahlem2001}, who also studied this galaxy and did not find a radio halo.\\

\noindent \textsf{NGC~4835:} This galaxy is similar to NGC~4666, in that its inclination is low, but it still shows signs of apparent extraplanar emission from both the $z$ profile and the radio contour map. However, a single-component exponential function was enough to accurately describe the profile. A possible explanation is that the low inclination of the galaxy does not allow one to differentiate the two components from each other if a radio halo is indeed present.\\

\noindent \textsf{NGC~5073:} This is one of the more distant galaxies in our sample, and its emission is primarily concentrated in a strong nuclear source, removed before analysis. It was studied in the past by \cite{Dahlem2001}, who could not make any conclusions because the galaxy was practically unresolved in the images. They also suggest that it could host an AGN, but a recent study made by \cite{Koulouridis2006} in which they study the relation between starbursts and AGNs, this galaxy is classified as a starburst. The remaining emission consists of a faint disk. Both the radio contour map and the $z$ profile indicate the absence of a radio halo. To accurately describe the profile, a single exponential function was utilized. The resulting disk scale height was found to be particularly small, indicating that this galaxy is essentially unresolved, as well as in previous images.\\

\noindent \textsf{UGCA~150:} The radio emission of this galaxy follows the optical disk, as can be seen from the radio contour map. However, we observed a small excess on both sides when attempting to fit a single-component function to the profile. Nevertheless, the observed excess is not significant enough to include a second component. Consequently, the best fit was achieved using a single Gaussian function.\\

\noindent \textsf{UGCA~394:} The nuclear emission of this galaxy is strong and had to be removed before the analysis. The radio emission follows the optical disk, except near the galaxy center, where is a small excess. This emission, however, is not enough to add a second component and a single exponential function was found to adequately fit the profile. The profile clearly displays a deviation from the exponential fit on the right side, indicating the presence of the excess emission mentioned earlier.\\

\noindent \textsf{UGCA~402:} After removing nearby unrelated sources and constructing the $z$ profile, we observed that the profile followed an exponential shape. Although we noted a small excess of radio emission outside of the galactic disk, this was not reflected in the profile fit, and there was no need to add a second component.

\begin{table*}
    \caption{Parameters obtained from fitting a two-component function to the $z$ profile of the sample galaxies with halos.}
    \centering
    \setlength\tabcolsep{5.5pt}
        \begin{tabular}{l c c c c c c c c}
            \toprule
            Name & rms & $d_{\mathrm{r}}$ & $z_{0}^{\mathrm{d}}$ & $z_{0,\mathrm{l}}^{\mathrm{h}}$ & $z_{0,\mathrm{r}}^{\mathrm{h}}$ & $<z_{0}^{\mathrm{h}}>$ & $<S^{\mathrm{h}}>$ & $L_{1.4}^{\mathrm{h}}$ \\
              & ($\mu \mathrm{Jy\,beam^{-1}}$) & ($'$) & (") & (") & (") & (") & (\%) & $(\mathrm{10^{21}~W~Hz^{-1}})$ \\
              & & (kpc) & (kpc) & (kpc) & (kpc) & (kpc) &\\
             (1) & (2) & (3) & (4) & (5) & (6) & (7) & (8) & (9) \\
            \midrule
            ESO\,005-G004 & 17    & 4.22  &   $2.8 \pm 0.2$   &  $8.1 \pm 0.5$  & $10.0 \pm 0.8$  &  $9.1 \pm 0.9$  & 18.6 & 1.03 \\
                          &       & 27.45 &  $0.31 \pm 0.02$  & $0.88 \pm 0.05$ & $1.09 \pm 0.09$ &  $1.0 \pm 0.1$  &      &      \\   
            ESO\,079-G003 & 18    & 2.46  &   $3.9 \pm 0.4$   &  $6.0 \pm 0.8$  &  $5.6 \pm 0.6$  &    $6 \pm 1$    & 17.7 & 1.56 \\
                          &       & 25.28 &  $0.66 \pm 0.07$  &  $1.0 \pm 0.1$  &  $0.9 \pm 0.1$  &  $1.0 \pm 0.2$  &      &      \\
            ESO\,163-G011 & 19    & 2.21  &  $0.66 \pm 0.04$  &  $7.3 \pm 0.4$  &  $4.6 \pm 0.2$  &  $5.9 \pm 0.5$  & 35.5 & 3.02 \\
                          &       & 24.52 & $0.121 \pm 0.007$ & $1.34 \pm 0.07$ & $0.84 \pm 0.03$ &  $1.1 \pm 0.1$  &      &      \\
            % ESO\,209-G009 & 22$^{*}$ & 4.96  &  $11.9 \pm 0.4$   & $13.7 \pm 0.5$  & $11.1 \pm 0.3$  & $12.4 \pm 0.7$&13.7 & 0.28 \\
            ESO\,209-G009 & 22000 & 4.96  &  $11.9 \pm 0.4$   & $13.7 \pm 0.5$  & $11.1 \pm 0.3$  & $12.4 \pm 0.7$&13.7 & 0.26 \\
                          &       & 17.03 &  $0.68 \pm 0.02$  & $0.78 \pm 0.03$ & $0.63 \pm 0.02$ & $0.71 \pm 0.04$ &      &      \\
            IC\,3908      & 22    & 2.33  &   $3.1 \pm 0.6$   &  $9.4 \pm 0.4$  &  $7.3 \pm 0.3$  &  $8.3 \pm 0.6$  & 35.5 & 1.19 \\
                          &       & 13.76 &  $0.30 \pm 0.05$  & $0.92 \pm 0.04$ & $0.71 \pm 0.03$ & $0.82 \pm 0.05$ &      &      \\
            NGC\,1406     & 20    & 3.46  &  $10.9 \pm 0.6$   &  $9.7 \pm 0.4$  & $13.5 \pm 0.8$  & $11.6 \pm 0.8$  & 17.1 & 0.99 \\
                          &       & 19.12 &  $1.01 \pm 0.05$  & $0.89 \pm 0.04$ & $1.24 \pm 0.08$ & $1.06 \pm 0.08$ &      &      \\
            NGC\,1421     & 21    & 4.06  &  $15.0 \pm 0.5$   &    $8 \pm 2$    &   $16 \pm 3$    &   $12 \pm 3$    & 4.9  & 0.27 \\
                          &       & 25.05 &  $1.54 \pm 0.05$  &  $0.9 \pm 0.2$  &  $1.7 \pm 0.3$  &  $1.3 \pm 0.3$  &      &      \\
            NGC\,1532     & 23    & 6.89  &    $12 \pm 1$     &   $21 \pm 3$    &   $19 \pm 2$    &   $20 \pm 4$    & 17.0 & 0.57 \\
                          &       & 31.11 &  $0.88 \pm 0.08$  &  $1.6 \pm 0.2$  &  $1.4 \pm 0.2$  &  $1.5 \pm 0.3$  &      &      \\
            NGC\,2221     & 16    & 1.73  &   $2.0 \pm 0.3$   &  $6.8 \pm 0.4$  &  $8.0 \pm 0.6$  &  $7.4 \pm 0.7$  & 27.1 & 1.69 \\
                          &       & 17.05 &  $0.33 \pm 0.05$  & $1.12 \pm 0.06$ &  $1.3 \pm 0.1$  &  $1.2 \pm 0.1$  &      &      \\
            NGC\,4666     & 22    & 5.74  &  $11.7 \pm 0.5$   & $15.1 \pm 0.4$  & $14.6 \pm 0.3$  & $14.9 \pm 0.5$  & 28.1 & 2.46 \\
                          &       & 21.39 &  $0.72 \pm 0.03$  & $0.94 \pm 0.02$ & $0.91 \pm 0.02$ & $0.92 \pm 0.03$ &      &      \\
            NGC\,7090     & 20    & 4.86  &     $5 \pm 1$     &   $14 \pm 2$    &   $25 \pm 5$    &   $19 \pm 4$    & 22.2 & 0.08 \\
                          &       & 10.82 &  $0.18 \pm 0.04$  & $0.51 \pm 0.06$ &  $0.9 \pm 0.2$  &  $0.7 \pm 0.1$  &      &      \\
            \bottomrule
        \end{tabular}
        \tablefoot{
            (1) Galaxy common name. (2) Value of the rms noise estimated with the method described in Sec. \ref{subsec:prep}. (3) Diameter of the radio emission calculated from the 3-$\sigma$ contour in arc-minutes and in kpc, using the distances from Table~\ref{table:general}. (4) Scale height of the disk in arc-seconds and kpc. (5) Scale height of the halo (left side of the $z$ profile) in arc-seconds and kpc. (6) Scale height of the halo (right side of the $z$ profile) in arc-seconds and kpc. (7) Average halo scale height in arc-seconds and kpc. (8) Average halo contribution. (9) Luminosity of the halo obtained from the halo contribution.}
    \label{table:results-halo}
\end{table*}

\begin{table}
    \caption{Parameters obtained from fitting a single-component function to the $z$ profile of the sample galaxies without halos.}
    \centering
    \setlength\tabcolsep{5pt}
        \begin{tabular}{l c c c }
            \toprule
            Name & rms & $d_{\mathrm{r}}$ & $z_{0}^{\mathrm{d}}$ \\
              & ($\mu \mathrm{Jy\,beam^{-1}}$) & ($'$) & (") \\
              & & (kpc) & (kpc) \\
             (1) & (2) & (3) & (4) \\
            \midrule
            ESO\,428-G028 & 24    & 2.54  & $4.28 \pm 0.08$ \\
                          &       & 22.98 & $0.64 \pm 0.01$ \\
            IC\,4595      & 18    & 3.16  & $5.95 \pm 0.06$ \\
                          &       & 42.06 & $1.32 \pm 0.01$ \\
            NGC\,134      & 24    & 6.53  & $26.2 \pm 0.2$  \\
                          &       & 30.20 & $2.02 \pm 0.01$ \\
            NGC\,1448     & 24    & 5.44  & $12.7 \pm 0.4$ \\
                          &       & 18.15 & $0.71 \pm 0.02$ \\
            NGC\,2613     & 20    & 5.43  & $25.7 \pm 0.7$  \\
                          &       & 31.86 & $2.52 \pm 0.07$ \\
            NGC\,2706     & 20    & 1.92  &  $5.2 \pm 0.1$ \\
                          &       & 13.60 & $0.62 \pm 0.01$ \\
            NGC\,3175     & 20    & 3.01  &  $8.9 \pm 0.1$  \\
                          &       & 11.76 & $0.57 \pm 0.01$ \\
            NGC\,3263     & 19    & 3.47  &  $5.4 \pm 0.1$ \\
                          &       & 39.13 & $1.02 \pm 0.02$ \\
            NGC\,3717     & 19    & 3.56  &  $4.4 \pm 0.1$  \\
                          &       & 22.15 & $0.45 \pm 0.01$ \\
            NGC\,4835     & 19    & 4.32  & $11.2 \pm 0.1$ \\
                          &       & 30.11 & $1.30 \pm 0.02$ \\
            NGC\,5073     & 21    & 2.59  & $1.02 \pm 0.07$ \\
                          &       & 27.90 & $0.18 \pm 0.01$ \\
            UGCA\,150     & 18    & 4.20  & $11.1 \pm 0.2$  \\
                          &       & 32.82 & $1.44 \pm 0.02$ \\
            UGCA\,394     & 26    & 3.18  &  $5.8 \pm 0.4$  \\
                          &       & 27.77 & $0.85 \pm 0.05$ \\
            UGCA\,402     & 26    & 3.49  &  $8.0 \pm 0.2$ \\
                          &       & 31.41 & $1.19 \pm 0.03$ \\
            \bottomrule
        \end{tabular}
        \tablefoot{
            (1) Galaxy common name. (2) Value of the rms noise estimated with the method described in Sec. \ref{subsec:prep}. (3) Diameter of the radio emission calculated from the 3-$\sigma$ contour in arc-minutes and in kpc, using the distances from Table~\ref{table:general}. (4) Scale height of the disk in arc-seconds and kpc.}
    \label{table:results-nohalo}
\end{table}

\section{Correlations} \label{sec:correlations}

In this section, we aim to investigate the potential relationship between the existence and physical properties of galactic halos and the star formation activity of the galaxies. We analyze the results obtained in Sections~\ref{sec:results-general} and \ref{sec:results-individual} and examine how they relate to the global properties of the galaxies.\\

\subsection{The galaxy sample into context: Looking for biases} \label{subsec:biases}

We tested our method to see any dependence on the resulting parameters (halo scale heights, flux density contribution and asymmetries) with the inclination. We did not find any relation, meaning that for values larger than 80°, our method can detect halos with no difficulties. 

\begin{figure}
    \centering
    \includegraphics[width=\linewidth]{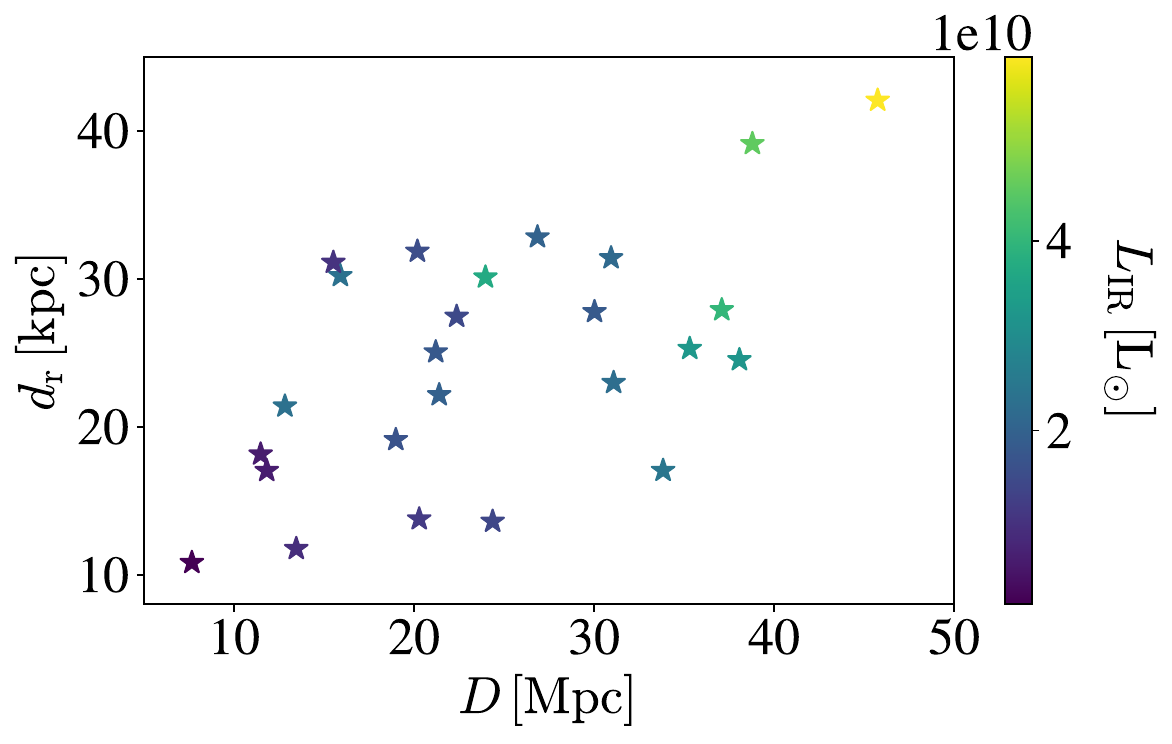}
    \caption{Radio diameter in kiloparsecs vs. distance in megaparsecs for the sample galaxies. The color scale corresponds to the IR luminosity in units of $10^{10}\,\mathrm{L_{\odot}}$. The direct relation between these quantities is a consequence of the selection criteria.}
    \label{figure:diameter-distance}
\end{figure}

We could detect halos in galaxies at all distances within our sample, which indicates that, in principle, our method does not prefer nearby galaxies. The galaxies in our sample with higher distances have larger IR luminosities. This is a consequence of the selection criteria: as the IRAS RBGS is flux-limited, distant galaxies with low IR luminosity are not part of the sample because of their low flux density. This effect can also be seen in Fig.~\ref{figure:diameter-distance} -although with large dispersion- as a direct relation between the radio diameter (measured in kpc) and the distance: larger and nearby galaxies are omitted for having an angular size too large, and distant and small galaxies are also omitted for having an angular size too small. The color scale in Fig.~\ref{figure:diameter-distance} corresponds with the increasing IR luminosity.

\begin{figure}
    \centering
    \includegraphics[width=\linewidth]{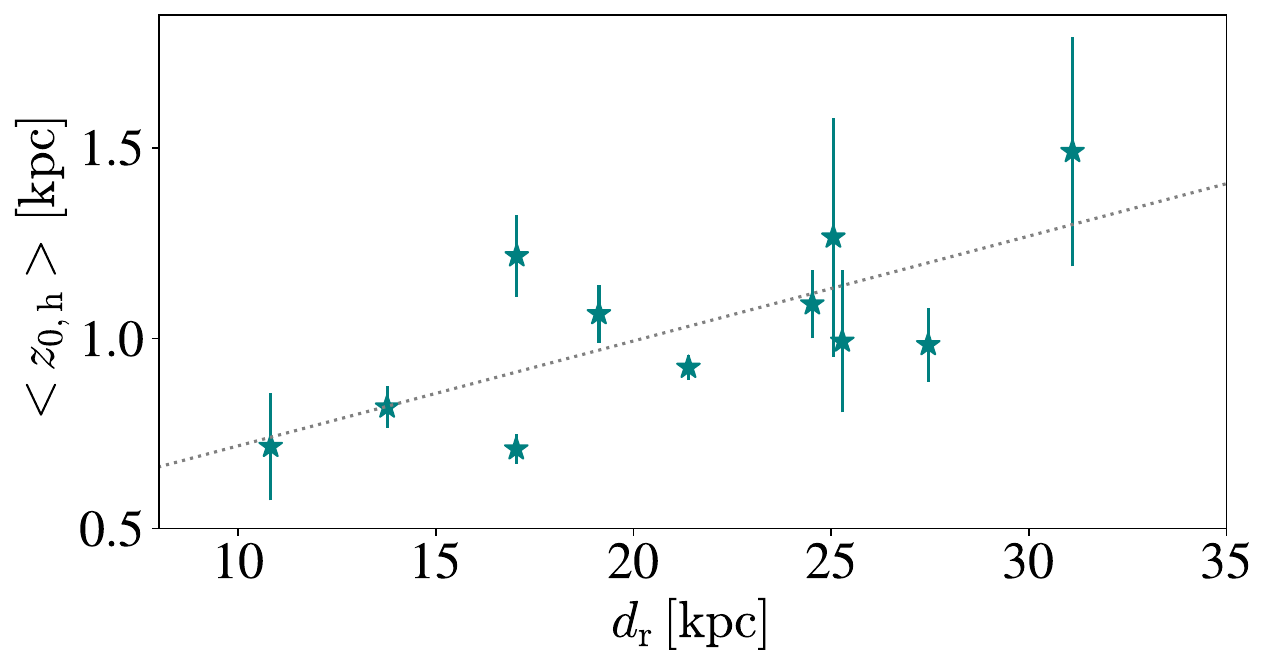}
    \caption{Halo scale heights vs. radio diameter, both in kiloparsecs. The sample galaxies are shown in stars, and the linear fit in dotted line, with a slope of $0.028 \pm 0.009$ and an intercept of $0.4 \pm 0.1$. The direct relation between these quantities cannot be explained by attributable to any selection criteria, and we consider its origin physical. }
    \label{figure:scale-height-diameter}
\end{figure}

Of all the halo parameters, only the halo scale heights correlate with the distance. However, comparing directly the halo scale heights with the radio diameters, we can see that the dependence on the distance is a consequence of a direct relation between the halo scale heights and the radio diameter. This relation, shown in Fig.~\ref{figure:scale-height-diameter}, cannot be attributed to any selection effect.

\subsection{Relations between the global properties of the galaxies and the presence of radio halos} \label{subsec:global-relations}

The presence of a radio halo is thought to be related to the star formation activity in the galactic disk. For this reason, we tested if the parameters that describe the radio halos depend on the global properties and how they relate with each other for our sample galaxies.

\begin{figure}
    \centering
    \includegraphics[width=\linewidth]{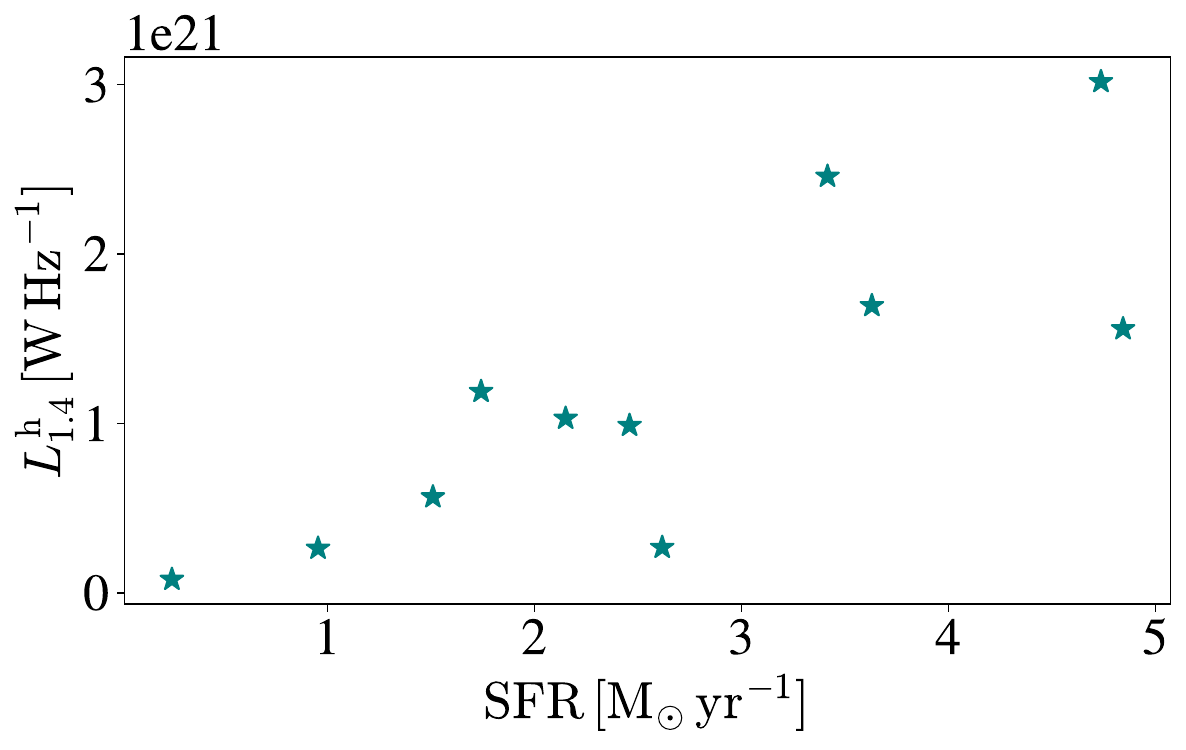}
    \caption{1.4 GHz luminosity of the halo in units of $10^{21}\,\mathrm{W~Hz^{-1}}$ vs. SFR in $\mathrm{M_{\odot}~yr^{-1}}$. A direct trend can be seen between these two quantities.}
    \label{figure:Lhalo-SFR}
\end{figure}

We found that the scale height does not follow a linear relation with the SFR. Still, it correlates with the galaxies' radio diameter, as previously mentioned in Section~\ref{subsec:biases}; similar results were reported by \cite{Krause2018}. This likely reflects the fact that age and environment are important in the determination of the scale heights. Besides, as shown in Fig.~\ref{figure:Lhalo-SFR}, the halo luminosity increases with the SFR.

\begin{figure}
    \centering
    \includegraphics[width=\linewidth]{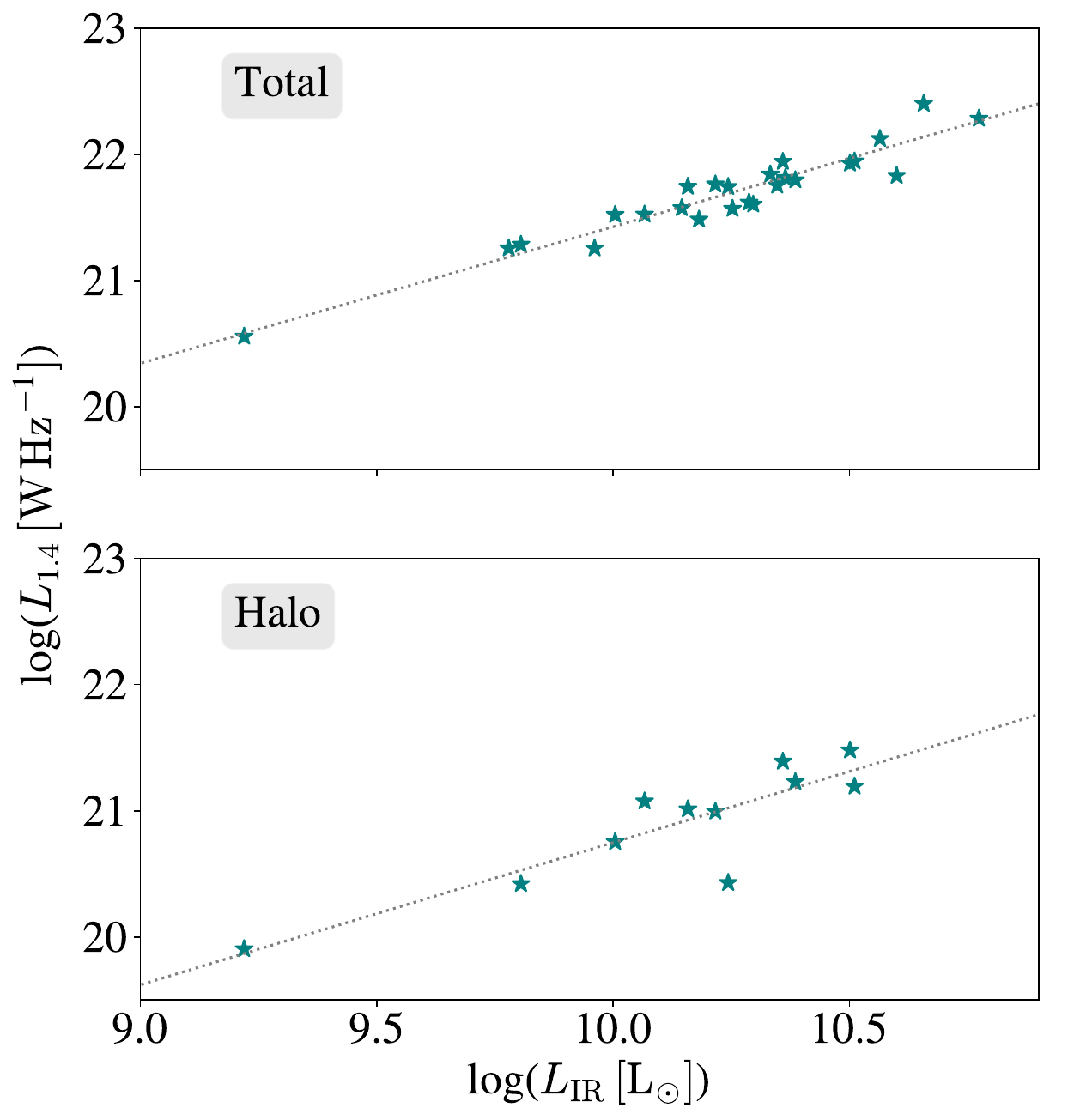}
    \caption{Radio-IR correlation for the total luminosity (\textit{upper panel}) and the halo luminosity (\textit{bottom panel}). The horizontal axis corresponds to the logarithm of the IR luminosity in $\mathrm{L_{\odot}}$ and the vertical axis corresponds to the logarithm of the $1.4\,\mathrm{GHz}$ luminosity in $\mathrm{W\,Hz^{-1}}$. Our sample galaxies are shown as stars and the dotted line corresponds to the linear fit of our data. The fit for the upper panel has a slope of $1.08 \pm 0.07$ and an intercept of $10.6 \pm 0.7$, while for the bottom panel, we obtained a slope of $1.1 \pm 0.2$ and an intercept of $9 \pm 2$.}
    \label{figure:radio-IR}
\end{figure}

All of our sample galaxies follow the radio-IR relation, shown in the left panel of Fig.~\ref{figure:radio-IR}. According to Equation~\ref{eq:radio-IR-q}, we obtained a mean value $q=2.5$ and a dispersion $\sigma_{q}=0.1$, in agreement with the values obtained by \cite{LI-shao2018}, \cite{Condon1992} and \cite{Helou1985}. Moreover, we obtained a similar relation --shown in the right panel of Fig.~\ref{figure:radio-IR}-- between the IR luminosity and the luminosity of the radio halos, calculated from the contribution to the total flux density of the galaxy listed in column 8 of Table~\ref{table:results-halo}. This result suggests that the emission coming from the radio halos is, as expected, produced by star formation activity.

\begin{figure}
    \centering
    \includegraphics[width=\linewidth]{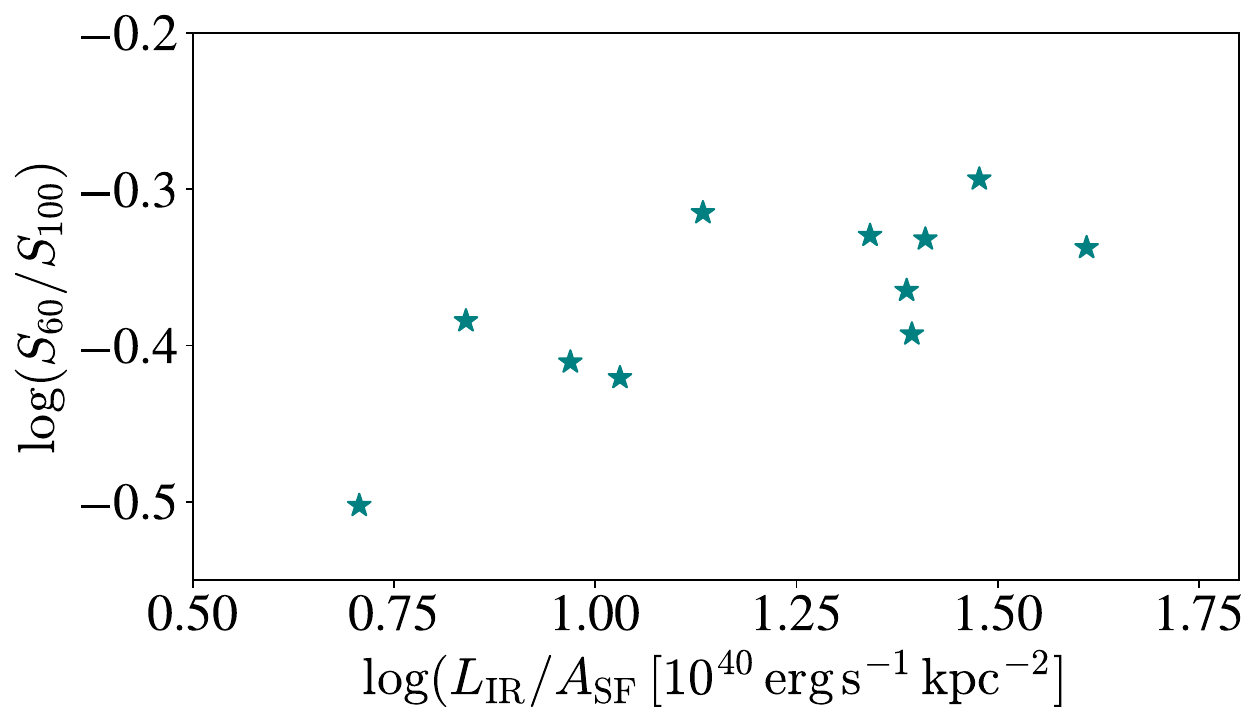}
    \caption{FIR flux ratio $S_{60}/S_{100}$ vs. IR luminosity divided by the star formation area $A_{\mathrm{SF}}$ for galaxies with detected radio halos. All the galaxies fall in the starburst region of the plot, i.e., they have $\log(S_{60}/S_{100})>-0.5$ and $\log(L_{\mathrm{IR}}/A_{\mathrm{SF}})>0.5$.}
    \label{figure:S60_S100-LIR_ASF}
\end{figure}

The relation of the far-infrared (FIR) flux ratio $S_{60}/S_{100}$ and the star formation activity per unit area with the presence of halos was studied by several authors in the past. \cite{Rossa2000} claim that all nearby, IR-luminous and IR-warm (i.e., with $S_{60}/S_{100}>0.4$) starburst galaxies have diffuse ionized halos. They show that the starburst and non-starburst galaxies are located in different places in the $S_{60}/S_{100}$ vs. $L_{\mathrm{IR}}/D_{25}^{2}$ plot, where $D_{25}$ is the diameter measured at the 25th blue magnitude. On the other hand, \cite{Dahlem2001} reproduce this same plot, but they use the radio diameter instead of the optical diameter to calculate the star-forming area $A_{\mathrm{SF}}$. We reproduced the plot for our sample galaxies with detected radio halos; see Fig.~\ref{figure:S60_S100-LIR_ASF}. We can see that all of them fall in the starburst region of the plot, that is, with $\log(S_{60}/S_{100})>-0.5$ and $\log(L_{\mathrm{IR}}/A_{\mathrm{SF}})>0.5$, except for one of them that has $\log(S_{60}/S_{100})\sim-0.5$ and falls in the transition of the two populations. It also appears to be a tendency that galaxies with larger $L_{\mathrm{IR}}/A_{\mathrm{SF}}$ also have larger $S_{60}/S_{100}$ ratios. Given the results obtained from this plot, from now on, we consider these galaxies as hosting starbursts.

We searched for any possible relations with the optical properties of the galaxies. \cite{Dahlem2006} found that for their sample galaxies, those with halos tend to have higher IR-to-blue luminosity ratios and more compact star formation distributions, given by the ratio $A_{\mathrm{SF}}/A_{25}$, where $A_{25}=\pi\;r_{25}^{2}$ is the optical disk area. We do not see the same trend within our sample. It is worth mentioning that the plot shown by \cite{Dahlem2006} combines a few galaxies from their work with several others from other works, each one of them using different methods to detect the radio halos and all of them using observations with low angular resolution (about $45''$) and high rms noise. We also looked for the relation between the $S_{60}/S_{100}$ ratio and the $A_{\mathrm{SF}}/A_{25}$ ratio, which is another plot made by \cite{Dahlem2006}. They claimed that galaxies with halos have more compact star-forming distributions and higher $S_{60}/S_{100}$ than those that do not show a halo. For our sample galaxies, however, the halos are present in all the $A_{\mathrm{SF}}/A_{25}$ range, and we did not find the same relation as these authors. On the other hand, \cite{Dahlem2001} studied a sample of galaxies and claimed that within their sample, galaxies physically smaller tend to have larger halos. However, the number of galaxies in their sample is too small to yield a conclusive result. We reproduced the same plot for our sample galaxies and, contrary to \cite{Dahlem2001}, did not obtain a relation between these quantities.

\section{Discussion}\label{sec:discussion}

 We have detected 11 radio halos around edge-on nearby galaxies from a sample of 25 candidates. Seven of the halos are reported here for the first time. The SFRs of the haloed galaxies range from 0.25 to 4.84~$\mathrm{M_{\odot}\,yr^{-1}}$, while the halos have average scale heights of 0.7 to 1.5~kpc. The halo scale heights cover a range of an order of magnitude with respect to the disk scale heights of the galaxies: $1 \lesssim \; <z_0^{\rm h}>/z_0^{\rm d}\; \lesssim 10$, that is, we have radio halos that are up to 10 times larger than the galactic disk. As expected for emission driven by a superwind produced by star formation activity, there is a clear correlation between the radio luminosity of the halos and the IR luminosity, which scales linearly with the SFR (see Fig.~\ref{figure:radio-IR}). The same is true for the radio luminosity of the halos and the SFR (see Fig.~\ref{figure:Lhalo-SFR}). All this supports the hypothesis that the radio halos result from synchrotron radiation produced by relativistic electrons either transported from the disk by the superwind, or locally accelerated in the halo by processes occurring in the interaction of the superwind with the external medium. The halo scale heights, on the other hand, may depend on other factors, such as the age of the system and the density of the surrounding medium, which could explain the absence of a relation with the SFR. A powerful but young wind, for example, will not have enough time to propagate to large distances, resulting in a bright but small radio halo. The same could happen if a wind propagates through a very dense medium.
 
The basic theory of superwinds in starburst galaxies was put forward by \cite{Chevalier1985} and later developed by \cite{Heckman1990} and \cite{Strickland2002}, among others. The superwind is produced when the ejecta from supernovae and stellar winds is efficiently thermalized in the nuclear region of the starburst. The result is a very hot ($T\sim10^8$~K), high-pressure bubble that expands and displaces the surrounding gas. As the bubble disrupts the disk, it expands adiabatically into the extragalactic medium, creating the multiphase region mentioned in the Introduction. The outflow quickly reaches the terminal velocity given by 
\begin{equation}
  v_{\infty}\sim\sqrt{2\dot{E}/\dot{M}}, \label{eq:vinfty} 
\end{equation}
where $\dot{E}$ and $\dot{M}$ are the total energy output and the total mass input, respectively. The wind is bipolar, with asymmetries reflecting differences in matter distribution and pressure on either side of the galaxy. The expanding gas forms a system of shocks, one sweeping matter outward and heating it, while the reverse shock moves through the hot medium and, being adiabatic, is suitable for particle acceleration \citep{Romero2018}. Since the external medium is most likely not uniform, the reverse shock is actually an ensemble of shocks, which can result in a very complex environment that would certainly affect the acceleration efficiency \citep[see, for example, the simulations performed by][]{Jana2020MNRAS.497.2623J}.

Both $\dot{E}$ and $\dot{M}$ are proportional to the total contribution of supernovae and stellar winds in the central starbursts. If we denote by $*$ the quantities corresponding to these contributions, we get the total injected energy \citep[e.g.,][]{Veilleux2005}
\begin{equation}
  \dot{E}=\epsilon\; \dot{E}_*,
\end{equation}
where $\epsilon$ is the thermalization efficiency (the fraction of the central SNe and stellar wind energy that goes into the outflow). This coefficient depends strongly on the local conditions. On the other hand, the total mass that goes into the outflow is made up of the mass supplied by the starbursts ($\dot{M}_*$) plus the gas loaded by the wind from the surrounding medium
\begin{equation}
  \dot{M}=\dot{M}_* + \dot{M}_{\textrm{ISM}}=\beta\; \dot{M}_*,
\end{equation} 
where $\beta$ is the mass loading parameter. 

Both $\dot{E}_*$ and $\dot{M}_*$ scale with the SFR as \citep{Veilleux2005}
\begin{equation}
 \dot{E}_{*}= 7\times 10^{41}\;(\mathrm{SFR/M_{\odot}\,yr^{-1}})\,\,\mathrm{erg\,s^{-1}},
\label{dotE}
\end{equation}
\begin{equation}
 \dot{M}_{*} = 0.26 \; (\mathrm{SFR/M_{\odot}\,yr^{-1}})\,\,\mathrm{M_{\odot}\,yr^{-1}}.
 \label{dotM}
\end{equation}

The final velocity of the superwind does not depend on the SFR, since this rate cancels out in the calculation. Therefore, the equation \ref{eq:vinfty} can be written as
\begin{equation}
  v_{\infty}= \sqrt{\frac{2\epsilon \dot{E}_*}{\beta \dot{M}_*}} \approx 3000 \sqrt{\frac{\epsilon}{\beta}}\; \textrm{km}\; \textrm{s}^{-1}.
\end{equation}  

The values of $\epsilon$ and $\beta$ are not well constrained. In the case of very local galaxies such as NGC~253 and M82, the thermalization efficiency seems to be very high \citep{Strickland2009} and the mass loading parameter has been calculated to be $\approx 12$ \citep{Bolatto2013,Romero2018}. Assuming that the galaxies in our sample have on average $\epsilon\approx0.75$ and $\beta\approx10$, we get $v_{\infty}\approx 820$~km s$^{-1}$. The actual velocity at which the cosmic ray populations are transported is much less than this value because of various effects, including diffusion, effective magnetic viscosity, gravitational pull, etc. \cite{Vijayan2020MNRAS.492.2924V} have found a dependence of the advection velocity of cosmic ray electrons produced in the disk on the surface density of star formation as $v_{\rm adv}\propto \Sigma_{\rm SFR}^{0.3}$. Even more recently, \cite{Heesen2021Ap&SS.366..117H} has investigated a sample of 16 galaxies with radio halos and found that advection dominates over diffusion for galaxies with $\Sigma_{\rm SFR}\leq 2 \times 10^{-3} \;\mathrm{M_{\odot}}\,{\rm yr}^{-1 } {\rm kpc}^{-2}$. This is the case of most galaxies in our sample, whose surface density of star formation is given in Table~\ref{table:sigma_SFR-advection}. For these kinds of galaxies, \cite{Heesen2021Ap&SS.366..117H} obtained a velocity for the advected cosmic rays of
\begin{equation}
    v_{\rm adv}=10^{3.23\pm0.25} (\Sigma_{\rm SFR}\leq 2 \times 10^{-3} \;\mathrm{M_{\odot}}\, {\rm yr}^{-1 } {\rm kpc}^{-2})^{0.41\pm 0.13}.
    \label{eq:adv}
    \end{equation}

In our sample, the average value of $\Sigma_{\rm SFR}$ is $\sim 0.0074$, which implies transport velocities of $v_{\rm adv}\sim 220$~km\,s$^{-1}$. For a halo with a scale height of $\sim 1$~kpc, this translates to a dynamical age of $t_{\rm dyn}\approx 4.4$~Myr.

\begin{table}
    \caption{Surface density of star formation for the sample galaxies with detected halos and the advection velocity of cosmic ray electrons produced in the disk.}
    \centering
    \setlength\tabcolsep{12pt}
        \begin{tabular}{l c c}
            \toprule
            Name & $\Sigma_{\rm SFR}$ & $v_{\rm adv}$\\
                 & [$\mathrm{M_{\odot}}\,\mathrm{yr^{-1}\,kpc^{-2}}$] & [$\mathrm{km\,s^{-1}}$] \\
            \midrule
            ESO 005-G004 & 0.0036 & 170 \\
            ESO 079-G003 & 0.0096 & 253 \\
            ESO 163-G011 & 0.01 & 257 \\
            ESO 209-G009 & 0.0042 & 180 \\
            IC 3908      & 0.0117 & 274 \\
            NGC 1406     & 0.0086 & 241 \\
            NGC 1421     & 0.0053 & 198 \\
            NGC 1532     & 0.0020 & 132 \\
            NGC 2221     & 0.0159 & 311 \\
            NGC 4666     & 0.0095 & 252 \\
            NGC 7090     & 0.0027 & 150 \\
            \bottomrule
        \end{tabular}
        \tablefoot{The surface density of star formation was obtained as $\Sigma_{\rm SFR}=\mathrm{SFR} / A_{\rm SF}$, where the SFR are those listed in Table~\ref{table:general} and and $A_{\rm SF}$ was obtained from the radio diameters listed in Table~\ref{table:results-halo}. The advection velocity of the cosmic rays was obtained using Eq.~\ref{eq:adv} \citep{Heesen2021Ap&SS.366..117H}.}
    \label{table:sigma_SFR-advection}
\end{table}

By comparison, the timescale of synchrotron losses for electrons of 10~TeV in a magnetic field of 5~$\mu$G (see \citet{Heesen2009b} for estimates for the halo of NGC~253) is $t_{\mathrm{synch}}\sim 3.3 \times 10^4$~yr. It is clear then that some reacceleration process must be taking place in the halo, if high-energy gamma-rays are going to be produced in this region by the electrons through inverse Compton scattering. The most likely process is diffusive shock acceleration at the reverse shocks generated by the collision of the outflow with the external medium \citep{Romero2018}. 

Yet, for the production of the radio halo itself reacceleration is not strictly necessary. Synchrotron radiation at 1.28~GHz in a field of 5~$\mu$G can be produced by electrons with energies of $\sim 4$~GeV. The synchrotron cooling time at such energies in that field is much longer: $t_{\mathrm{synch}}=8.3\times 10^9 E_{\rm e}({\rm GeV})^{-1} B(\mu{\rm G})\sim 83$~Myr $>>t_{\rm dyn}$.
 
Several star-forming galaxies have been detected in $\gamma$-rays by the Large Area Telescope (LAT) on the \textit{Fermi} Gamma-ray Space Telescope (\textit{Fermi}). Notable examples are NGC~253 and M82, which are also suspected sources of ultrahigh-energy cosmic rays \citep{Veritas2009,Acero2009,Abdo2010,Anchordoqui1999,Anchordoqui2018}. Gamma-ray emission from starbursts is thought to be produced mainly by the interaction of hadronic cosmic rays, accelerated by supernova remnant shocks, with gas in the interstellar medium \citep{Paglione1996,Romero2003,Domingo-Santamaria2005,Ohm2016,Roth2021Natur.597..341R}. Some contribution may also come from the halos, where the relativistic electrons \citep{Romero2018,Peretti2019,Muller2020} can upscatter CMB or ambient IR photons. Hadronic interactions should also occur in the halo between relativistic protons, accelerated along with the electrons, and matter from clumps and wind-swept material. 

A correlation between the radio emission (of synchrotron origin) and the gamma-ray emission (produced by either inverse Compton interactions of the same electron population that emits in radio or by protons transported along with them) is expected. Such a correlation, between the $0.1-100~\mathrm{GeV}$ integrated $\gamma$-ray luminosity $L_{\gamma}$ and the 1.4~GHz radio luminosity $L_{\mathrm{1.4}}$, has indeed been found by several authors \citep{Ackermann2012,Kornecki2022}, although not under any specific assumption about the mechanism producing the $\gamma$-rays. The expression for this relation obtained by \cite{Kornecki2022} is

\begin{equation}
    \log \left( \frac{L_{\gamma}}{\mathrm{erg\,s^{-1}}} \right) = 1.26 \, \log \left( \frac{L_{\mathrm{1.4}}}{10^{21}\,\mathrm{W\,Hz^{-1}}} \right) +39.04.
\end{equation}

\begin{table}
    \caption{Predicted $0.1-100~\mathrm{GeV}$ $\gamma$-ray luminosities and fluxes of the sample galaxies with detected halos.}
    \centering
    \setlength\tabcolsep{10pt}
        \begin{tabular}{l c c}
            \toprule
            Name & $\log (L_{\gamma})$ & $\log (F_{\gamma})$\\
                 & [$\mathrm{erg \, s^{-1}}$] & [$10^{-14}\,\mathrm{erg \, s^{-1}\,cm^{-2}}$] \\
            \midrule
            ESO 005-G004 & 40.01 & 1.23 \\
            ESO 079-G003 & 40.26 & 1.09 \\
            ESO 163-G011 & 40.24 & 1.01 \\
            ESO 209-G009 & 39.43 & 1.21 \\
            IC 3908      & 39.73 & 1.04 \\
            NGC 1406     & 40.04 & 1.40 \\
            NGC 1421     & 40.01 & 1.28 \\
            NGC 1532     & 39.73 & 1.27 \\
            NGC 2221     & 40.08 & 0.94 \\
            NGC 4666     & 40.26 & 1.97 \\
            NGC 7090     & 38.52 & 0.67 \\
            \bottomrule
        \end{tabular}
        % \tablefoot{}
    \label{table:gamma}
\end{table}

\noindent We can use this relation to estimate the expected $\gamma$-ray luminosity from the radio luminosity of the galaxies in our sample. The results are shown in Table~\ref{table:gamma} along with the $\gamma$-ray fluxes obtained using the distances listed in Table~\ref{table:general}. Although the gamma-ray luminosities are in the range inferred from their observed fluxes for the nearby galaxies NGC~253, M82, NGC~4945, and NGC~1068, most of our galaxies are still undetectable. 

\begin{figure}[b]
    \centering
    \includegraphics[width=\linewidth]{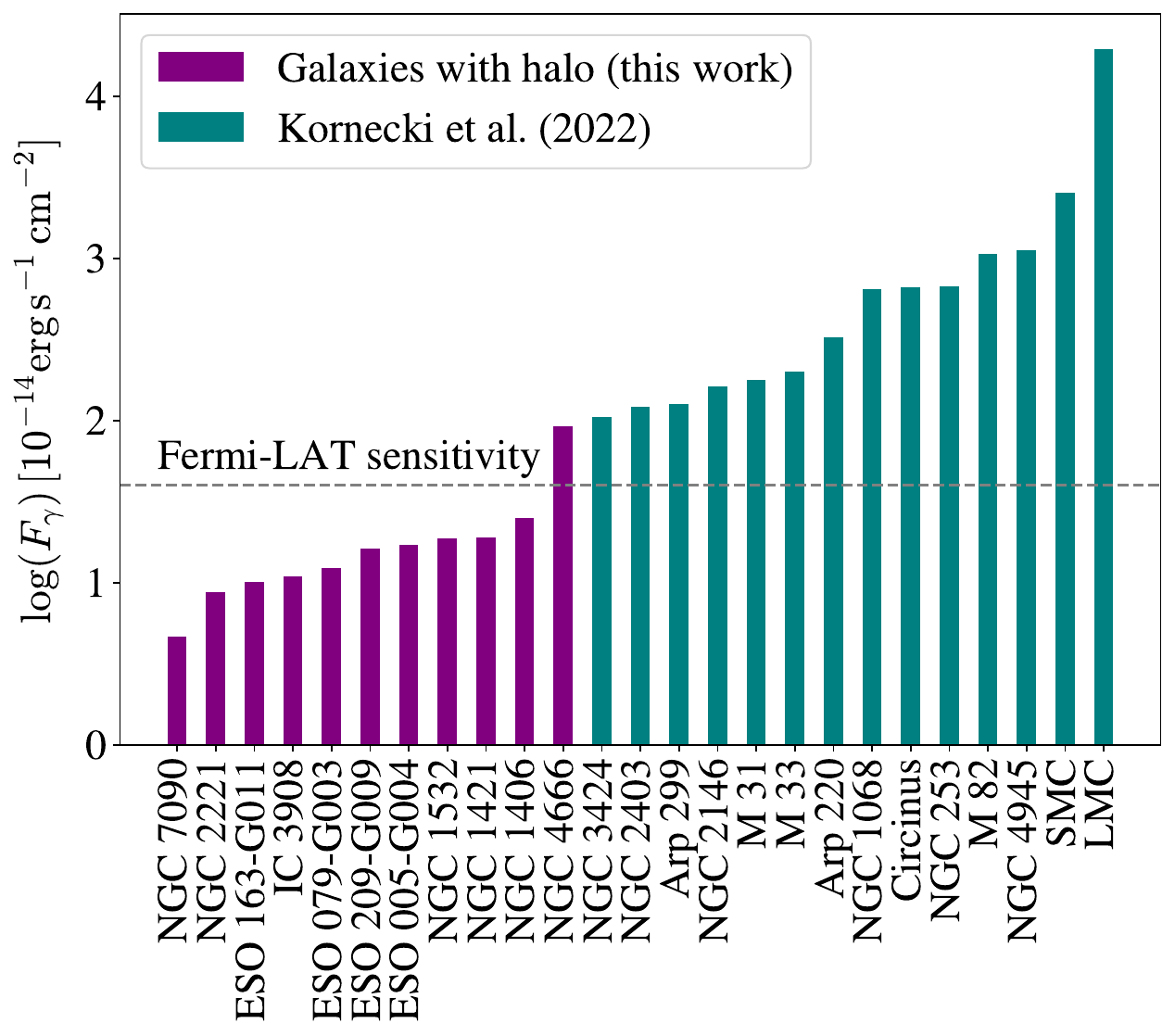}
    \caption{$0.1-100~\mathrm{GeV}$ gamma-ray fluxes predicted for our sample galaxies compared with the \cite{Kornecki2022} galaxy sample. The horizontal dashed line corresponds to an estimation of the 10-year sensitivity \citep{Funk2013}. Only NGC~4666 has a flux high enough to potentially be detected by Fermi in the near future.}
    \label{figure:gamma-flux}
\end{figure}

In Fig.~\ref{figure:gamma-flux}, we show the estimated $\gamma$-ray flux of the haloed galaxies in our sample, along with an estimate of the \textit{Fermi} 10-year sensitivity. The sample of galaxies from \cite{Kornecki2022} is also shown for comparison. NGC~4666 and ESO~079-G003 are the two galaxies with the expected higher $\gamma$-ray luminosity in our sample. However, ESO~079-G003 is at a much greater distance than NGC~4666. As a result, only NGC~4666 has a $\gamma$-ray flux high enough to potentially be detected by \textit{Fermi} in the near future. In the \textit{LAT 12-year Source Catalog (4FGL-DR3)} no source is reported at the position of the galaxy. A dedicated reanalysis of the available data may be worthwhile. In any case, NGC~4666 with its large halo remains as a potential $\gamma$-ray source for future observations, either with \textit{Fermi} or with the Cherenkov Telescope Array (CTA).

\section{Summary and conclusions} \label{sec:conclusions}

In this paper, we present the analysis of the extended radio emission for 25 galaxies with inclinations $\gtrsim80^{\circ}$ gathered from a sample published in the MeerKAT 1.28 GHz Atlas of Southern Sources in the IRAS Revised Bright Galaxy Sample. The resulting sample, described in Table~\ref{table:general}, contains galaxies at distances nearer than 46~Mpc, IR luminosities between $10^{9}-10^{11}~\mathrm{L_{\odot}}$, SFRs between $0.25-8~\mathrm{M_{\odot}\,yr^{-1}}$, and 1.4~GHz radio luminosities in the range $0.3-26 \times 10^{21}~\mathrm{W\,Hz^{-1}}$. The angular resolution of the radio images is $\sim 7.5''$ and the rms is $\sim 20~\mu\mathrm{Jy\,beam^{-1}}$, except for particular cases. 

We constructed averaged vertical intensity profiles and modeled them with one or two component(s). When two components were needed to accurately fit the profiles, we interpreted it as the presence of a radio halo. We found halos in 11 galaxies distributed along the entire range of distances and inclinations, seven of which were detected in this work for the first time. Our results agree with other works for the four galaxies with previously reported halos. All the resulting parameters are listed in Tables~\ref{table:results-halo} and \ref{table:results-nohalo}. The images and the obtained profiles are shown in Fig. \ref{figure:galaxies-example} and Appendix~\ref{appendixA}. Our main results can be summarized as follows:

\begin{itemize}
    \item The disks were generally better fitted with exponential rather than Gaussian functions. The disk scale heights ranged between 0.1 to 2.5~kpc with errors lower than 20\%.
    
    \item The halo scale heights ranged from 0.7 to 1.5~kpc, with a mean value of $\sim1~\mathrm{kpc}$ and errors lower than 19\%, and they covered an order of magnitude with respect to the disk scale heights, that is, $1<z_{0}^{\mathrm{h}}/z_{0}^{\mathrm{d}}<10$.
    
    \item We found no correlation between the halo scale heights and the SFR. However, we did find a direct relation between them and the radio diameter of the galaxies. The halo luminosities, on the other hand, showed a direct relation with the SFR and the galaxy IR luminosity.
    
    \item All of the sample galaxies followed the radio-IR relation between the radio and total IR luminosities, with a $q$ parameter of $2.5\pm0.1$. A similar trend was found for the radio luminosity of the halos.
    
    \item All of the galaxies with detected halos but one presented high values of FIR 60 to 100~$\mu$m flux ratios, $\log(S_{60}/S_{100})>-0.4$, and IR luminosity per unit of star formation area, $\log(L_{\mathrm{IR}}/A_{\mathrm{SF}})>0.5$. These values are in agreement with those expected for galaxies with starburst activity.
    
    \item We did not find any relation between the halo scale heights or luminosity with the optical properties of the galaxies.
\end{itemize}

The direct relation of the halo luminosities with the SFR and IR luminosity supports the hypothesis that radio halos result from synchrotron radiation produced by relativistic electrons and this points toward the fact that star formation activity plays a crucial role in the formation of the halos. The absence of a relation between the halo scale heights and SFR can be explained by the fact that they depend on many other factors, such as the age of the system and the density of the surrounding medium. However, the origin of the direct relation between the halo scale heights and the radio diameters remains uncertain. 

As \textit{Fermi} detected several star-forming galaxies in gamma rays, we investigated the possibility that any of our sample galaxies with halos could be observed with this telescope in the future. A correlation between the radio  and the gamma-ray emission is expected in starbursts. We used the correlation found by \cite{Kornecki2022} to estimate the gamma-ray luminosity and flux of the galaxies in our sample. Taking the 10-year \textit{Fermi} sensitivity into account, we found that only NGC~4666 has a gamma-ray flux high enough to potentially be detected by \textit{Fermi} or with other instruments, such as the CTA, in the near future.

Based on the results presented in this paper, we we have started to carry out new dedicated low-frequency observations of selected haloed galaxies to study the spectral index and polarization distributions. Such an investigation will help to shed light on the magnetic field and particle spectrum in the superwind region, and on the coupling between star formation activity and cosmic ray injection in the intergalactic medium.

\begin{acknowledgements}
We would like to thank an anonymous reviewer for constructive comments.  GER acknowledges financial support from the State Agency for Research of the Spanish Ministry of Science and Innovation
under grant PID2022-136828NB-C41/AEI/10.13039/501100011033/. Additional support came from PIP 0554 (CONICET). 
\end{acknowledgements}

\bibliographystyle{aa} % style aa.bst
\bibliography{bibliography}

\begin{appendix}

\section{Contour maps and \textit{z} profiles of all the sample galaxies}\label{appendixA} %First appendix

In this appendix we list the radio contour maps superimposed on Digital Sky Survey (DSS) images (\textit{top panels}) and $z$ profiles (\textit{bottom panels}) of all the sample galaxies, except for ESO~428-G028 and ESO~209-G009 which were already shown in Fig.~\ref{figure:galaxies-example} of Section~\ref{sec:results-general}. The horizontal axis of the profiles is the offset in arcseconds from the galactic plane in $z$ direction, and the vertical axis is the logarithm of the average flux density in $\mu \mathrm{Jy}$. The data is shown in dots, the fitted disk in dashed line for the galaxies with no halo, while for the galaxies with halos, the fitted disk component is shown in dashed line, the halo component in dotted line and the total fit in solid line.

\begin{figure*}[!b]
    \centering 
    \includegraphics[width=0.4\textwidth]{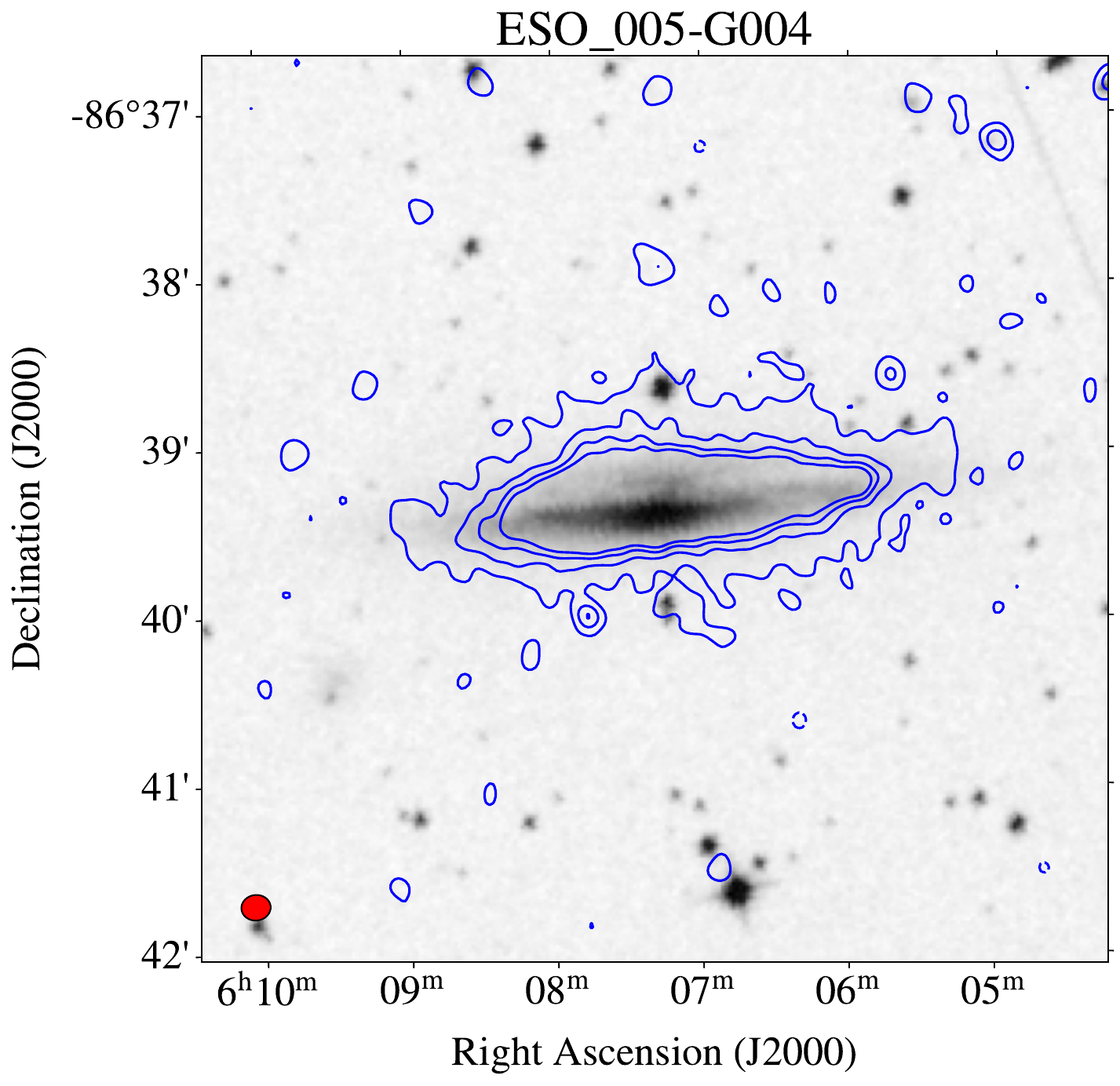}
    \includegraphics[width=0.4\textwidth]{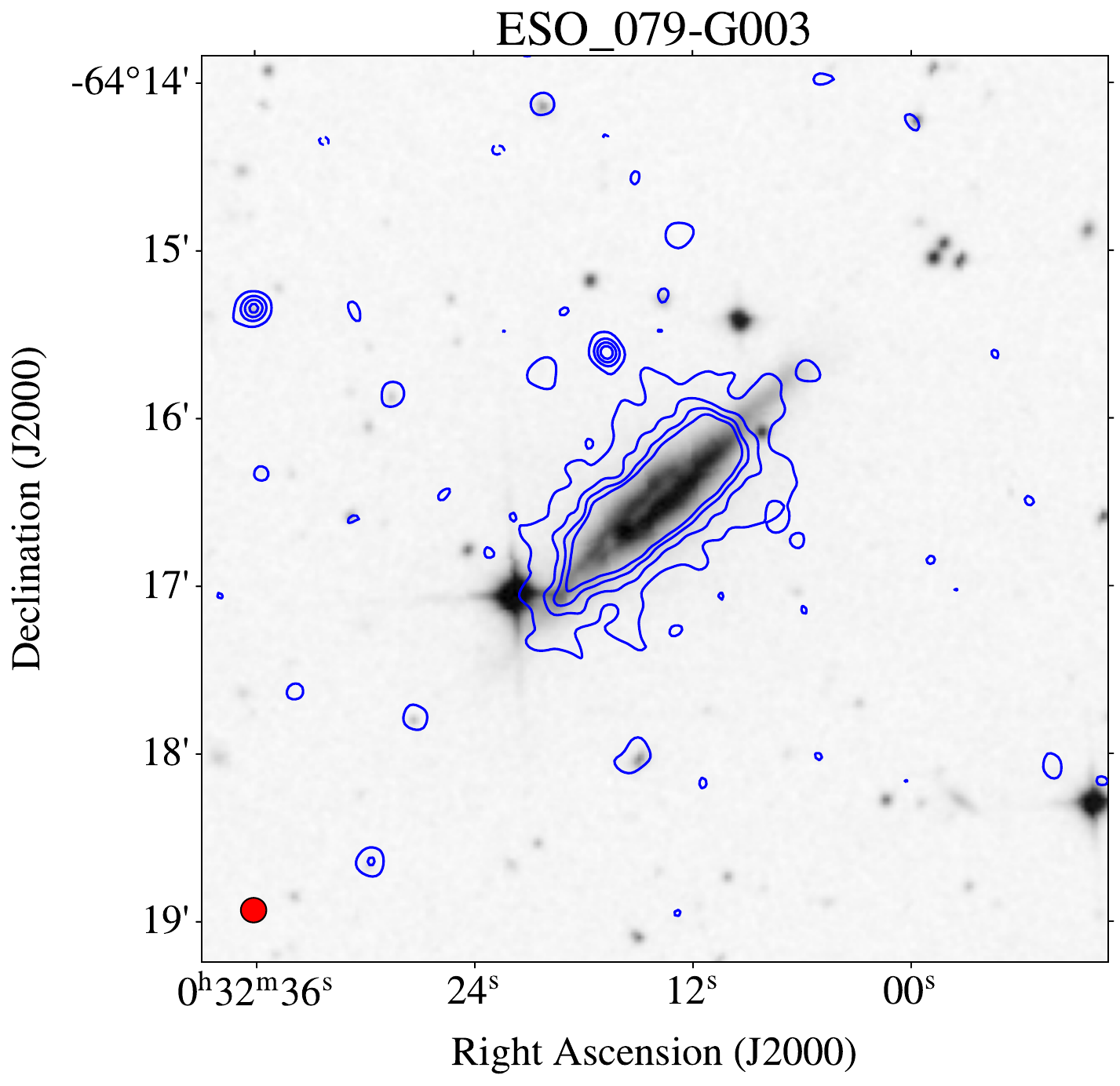}
    
    \includegraphics[width=0.4\textwidth]{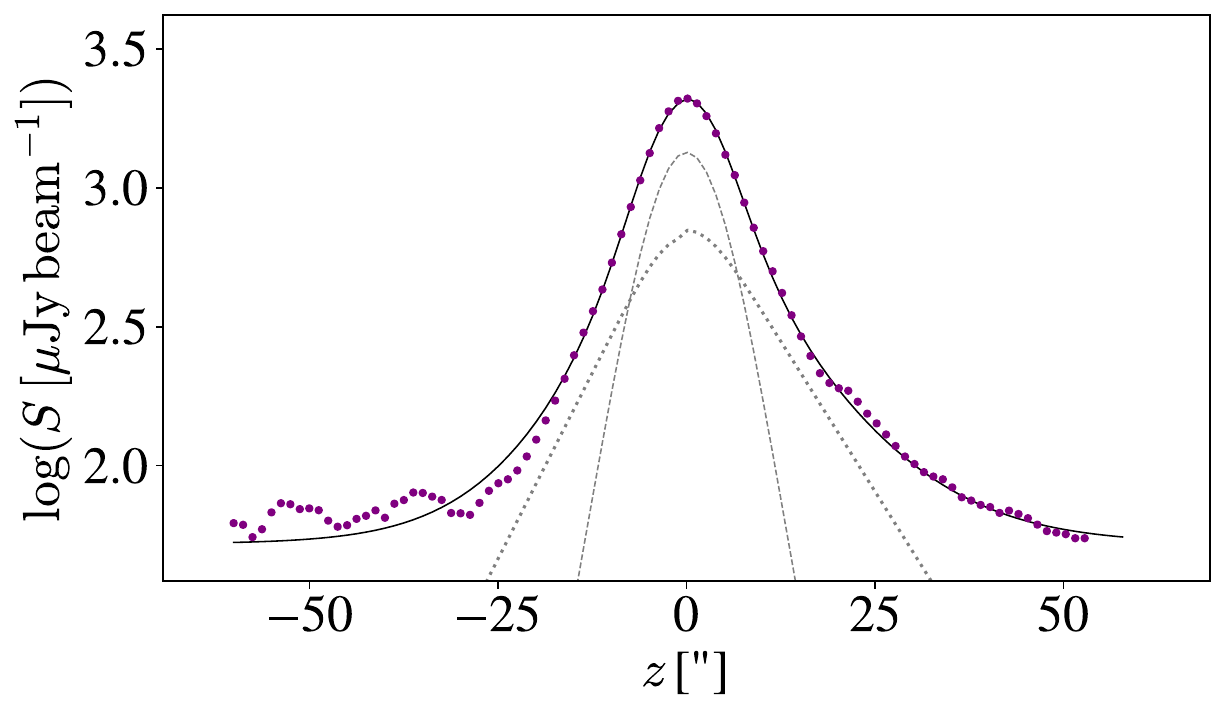}
    \includegraphics[width=0.4\textwidth]{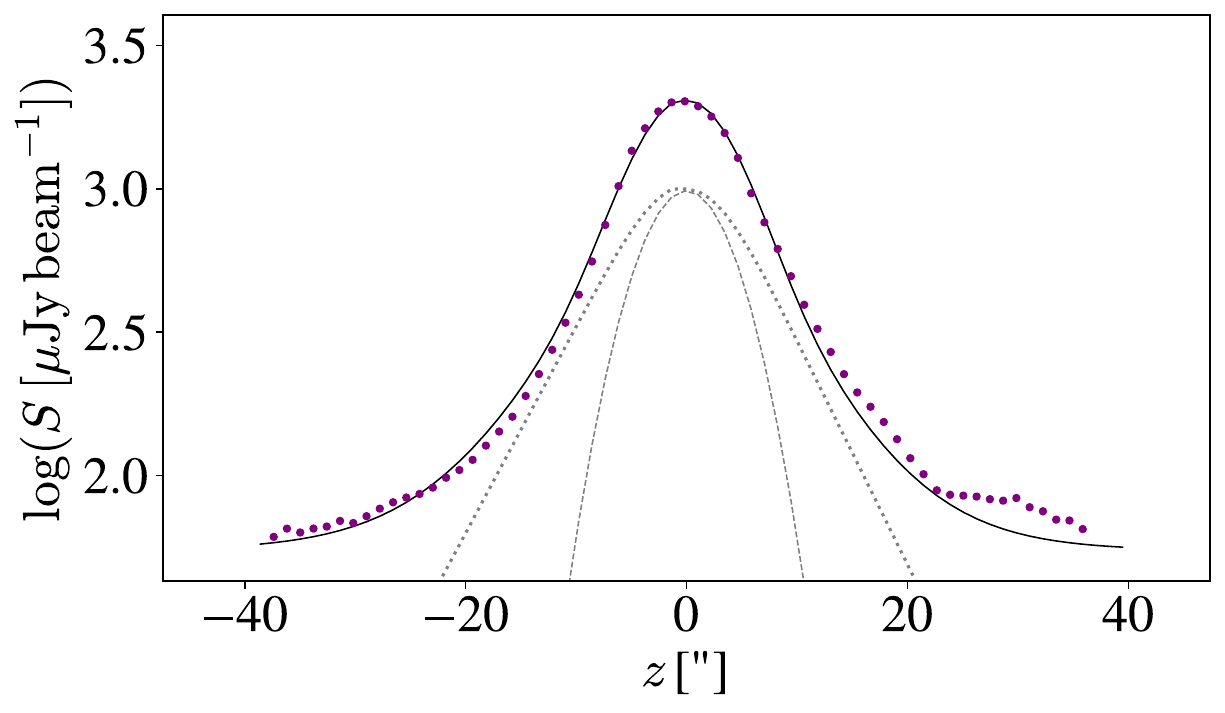}
    \includegraphics[width=0.4\textwidth]{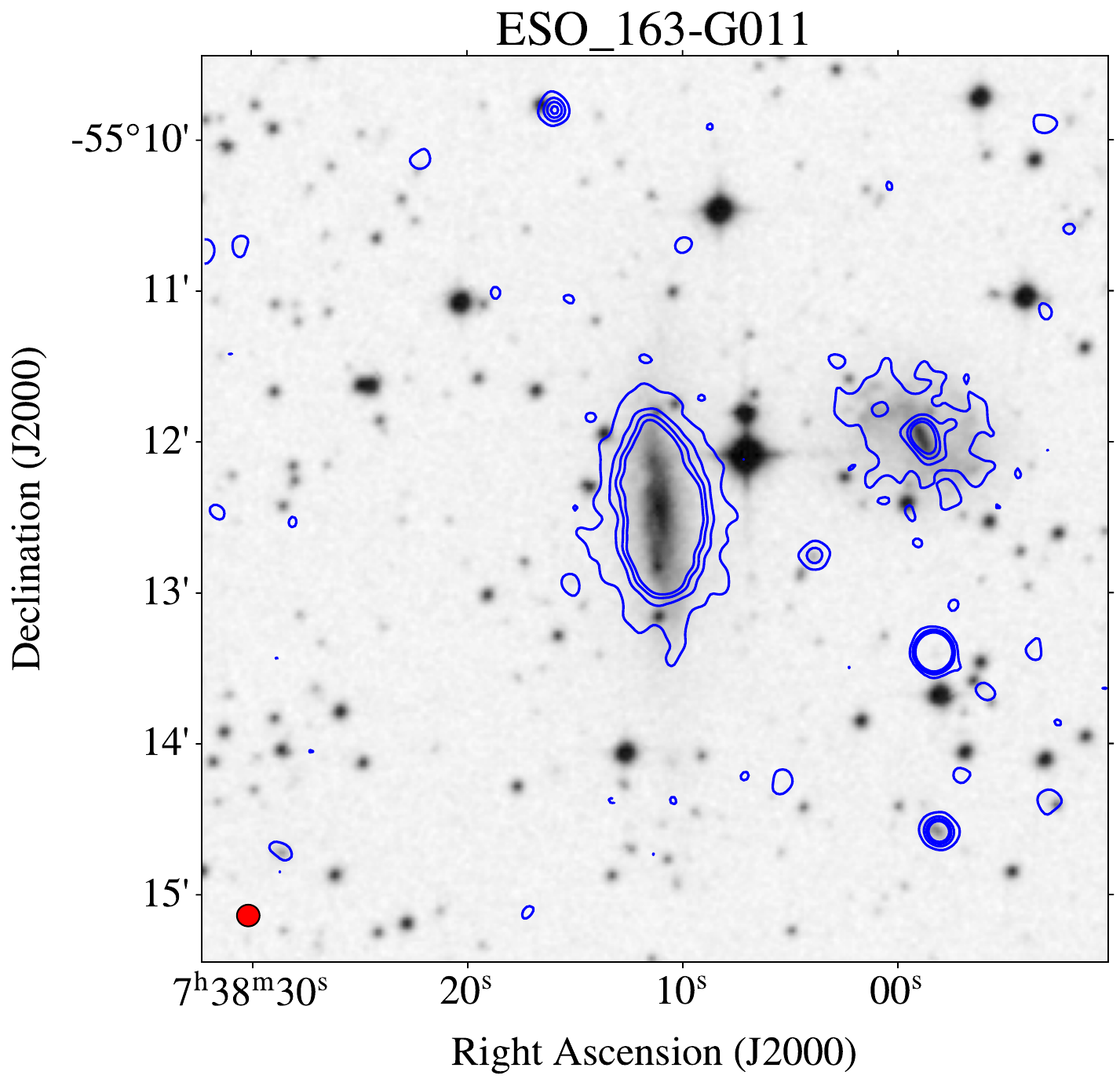}
    \includegraphics[width=0.4\textwidth]{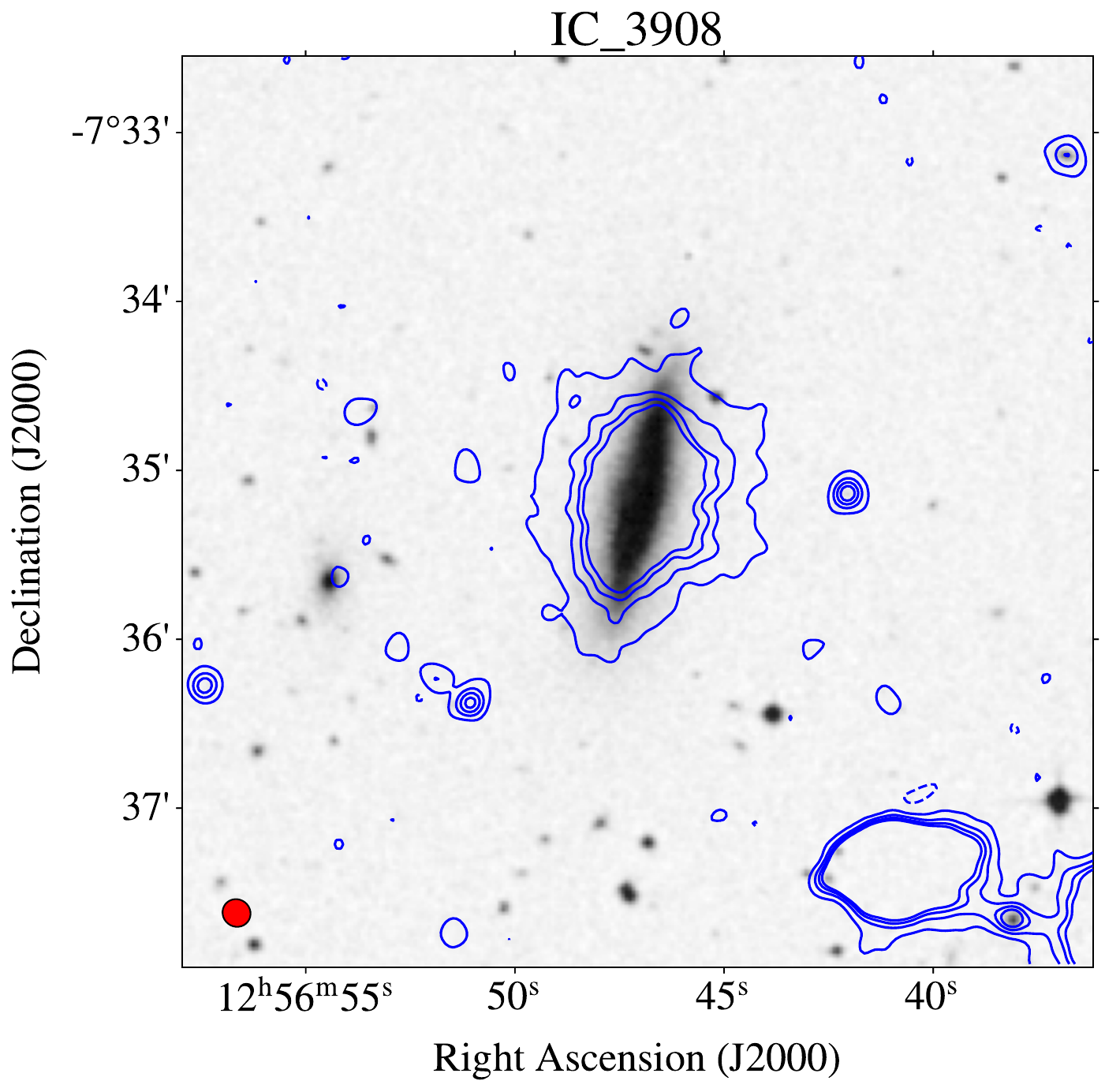}
    \includegraphics[width=0.4\textwidth]{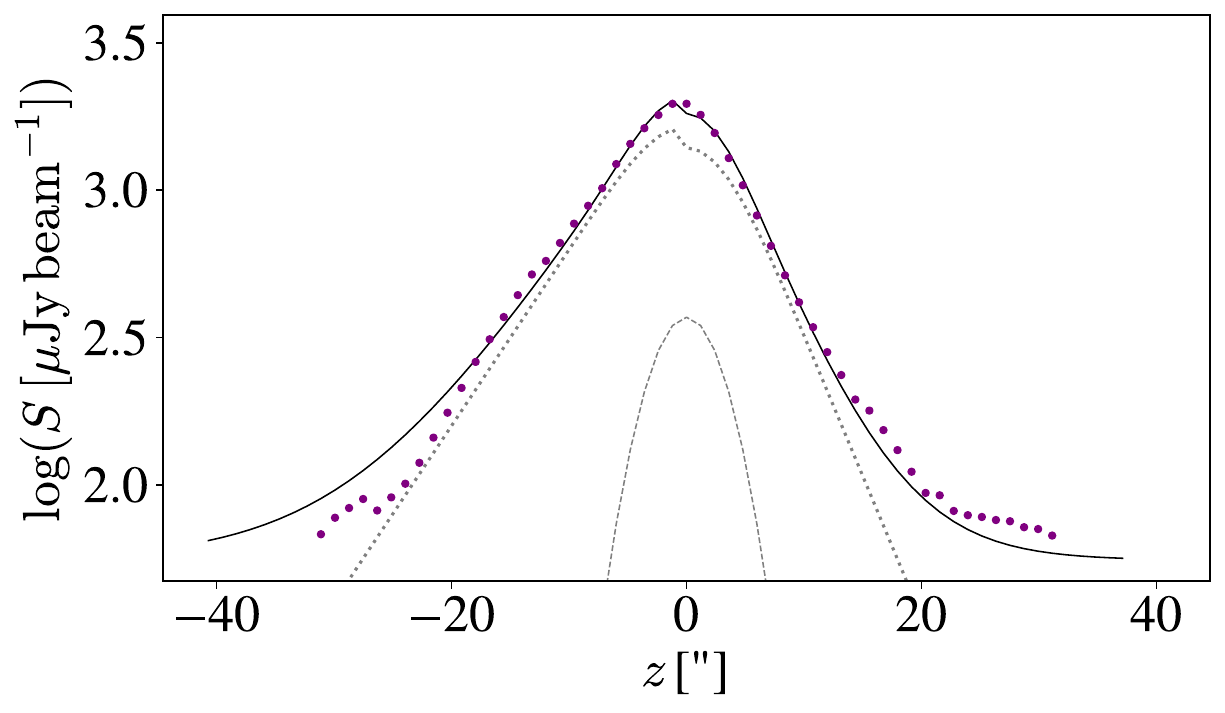}
    \includegraphics[width=0.4\textwidth]{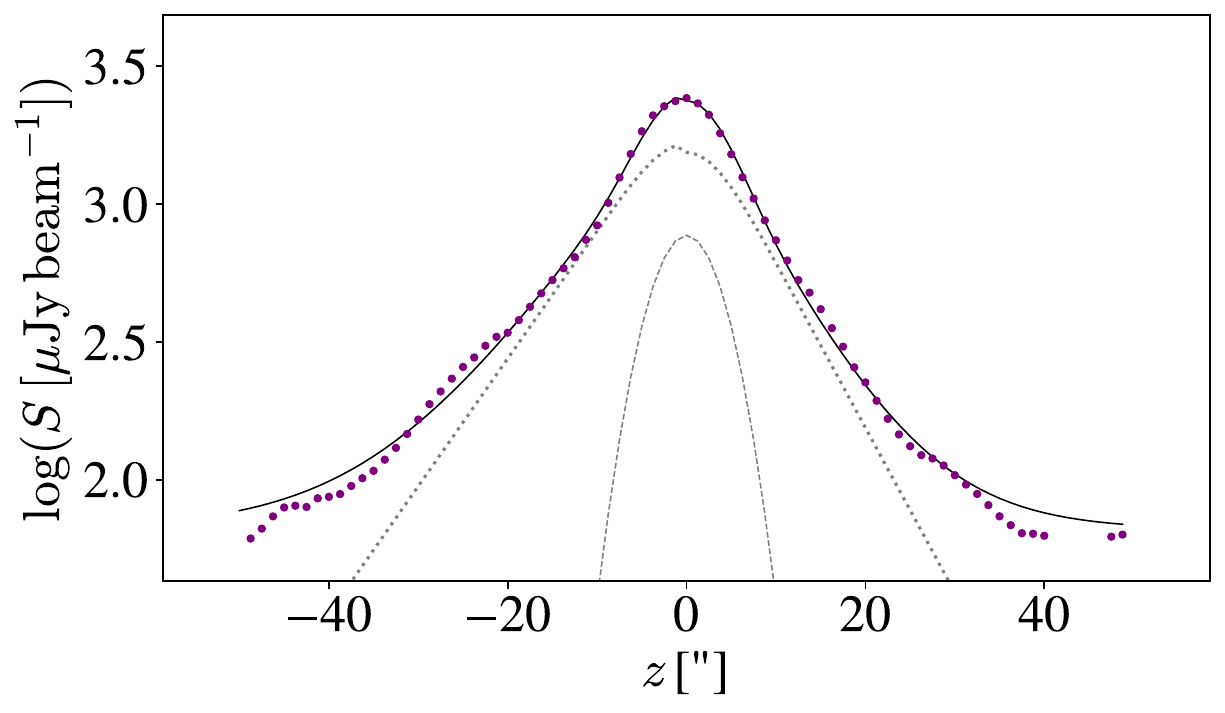}
    \caption{Radio contour maps superimposed on a DSS image. The contour levels are -3, 3, 9, 15,  and 21 times the rms of each radio continuum image, in units of mJy\,beam$^{-1}$. The rms are listed in Tables~\ref{table:results-halo} and \ref{table:results-nohalo}. The synthesized beam of $7.5''$ is shown in the bottom left corner of each image.}
\end{figure*}

%------------------------------

\begin{figure*} \ContinuedFloat
\centering   
    \includegraphics[width=0.4\textwidth]{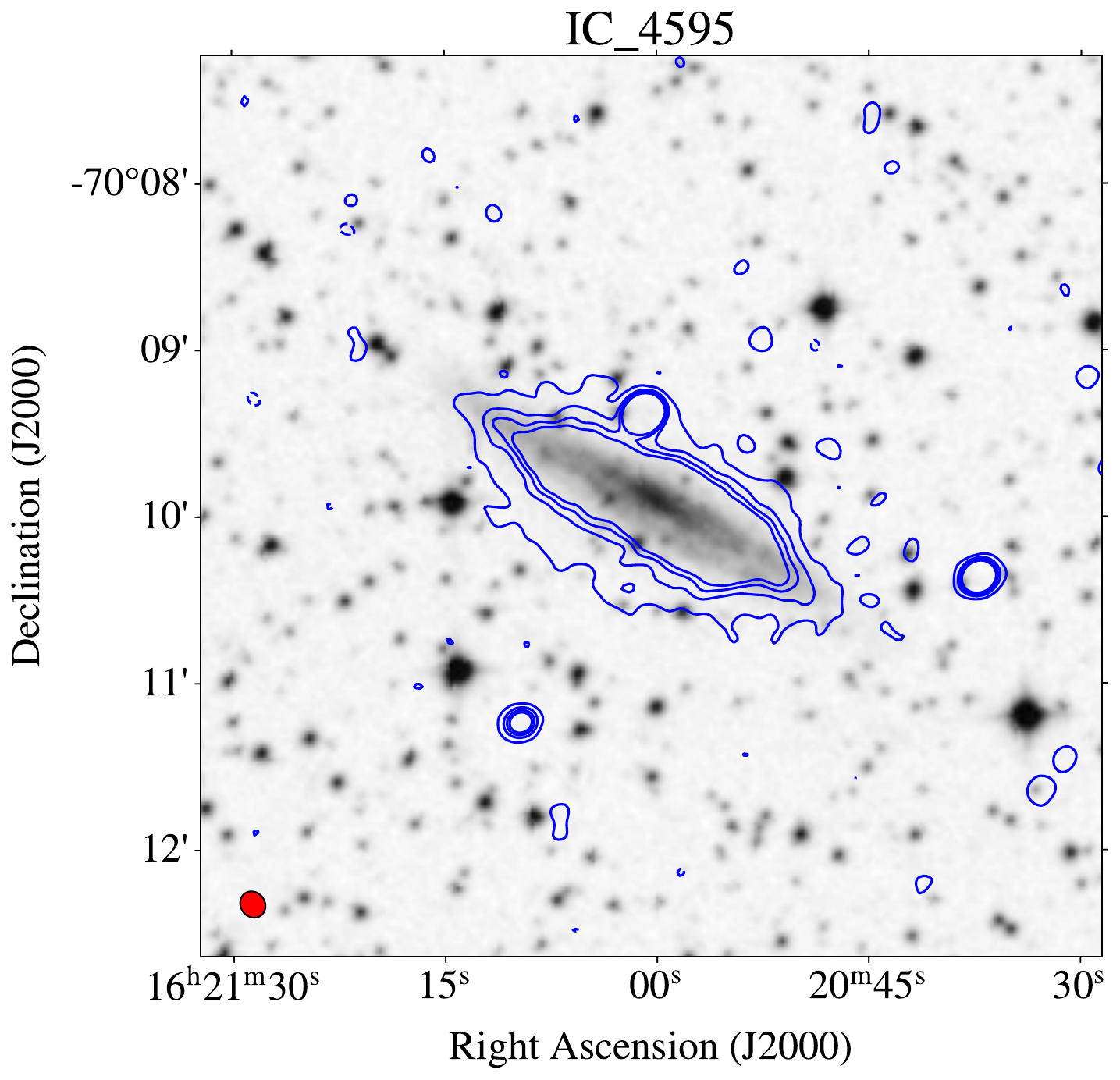}
    \includegraphics[width=0.4\textwidth]{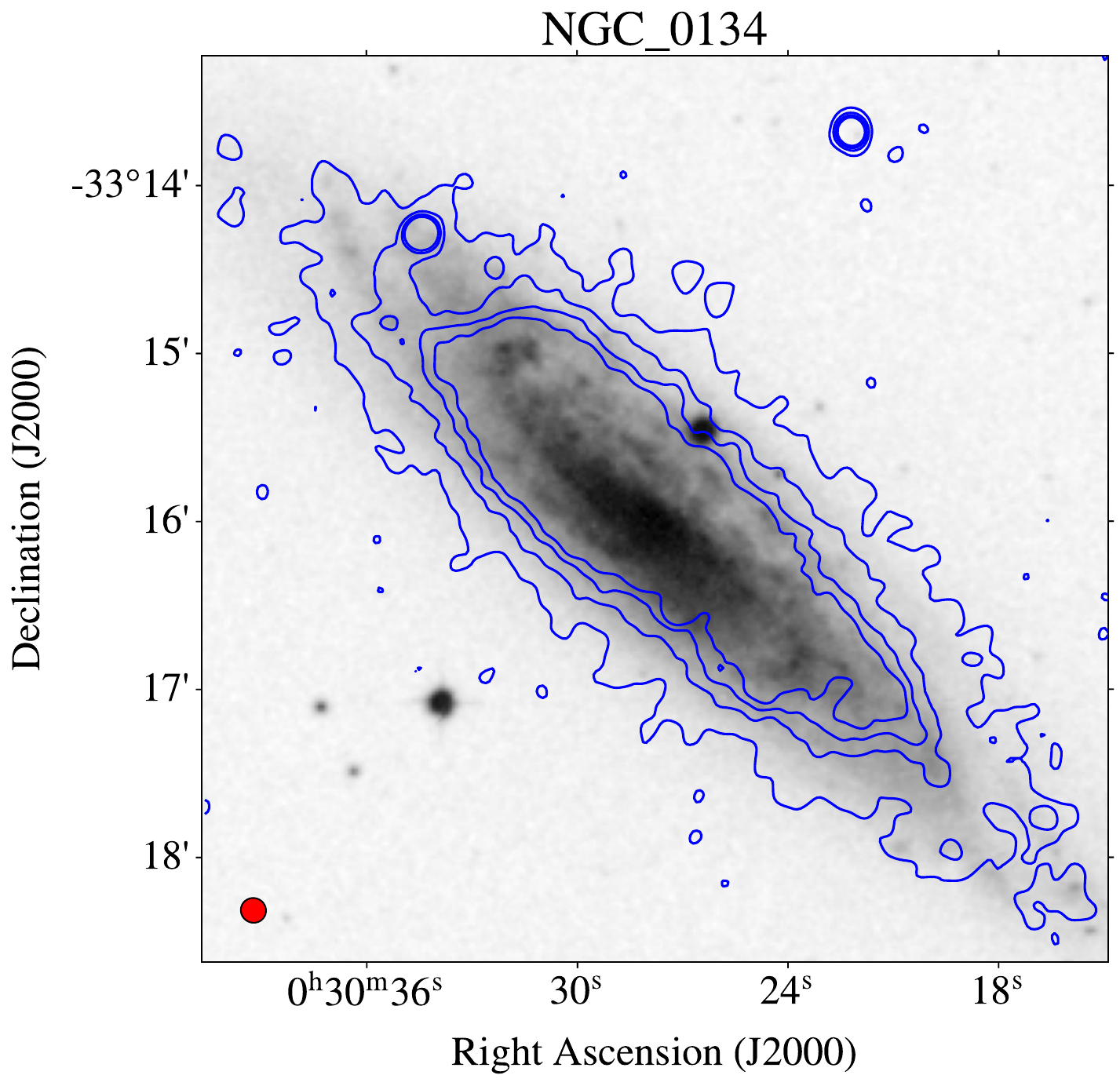}
    \includegraphics[width=0.4\textwidth]{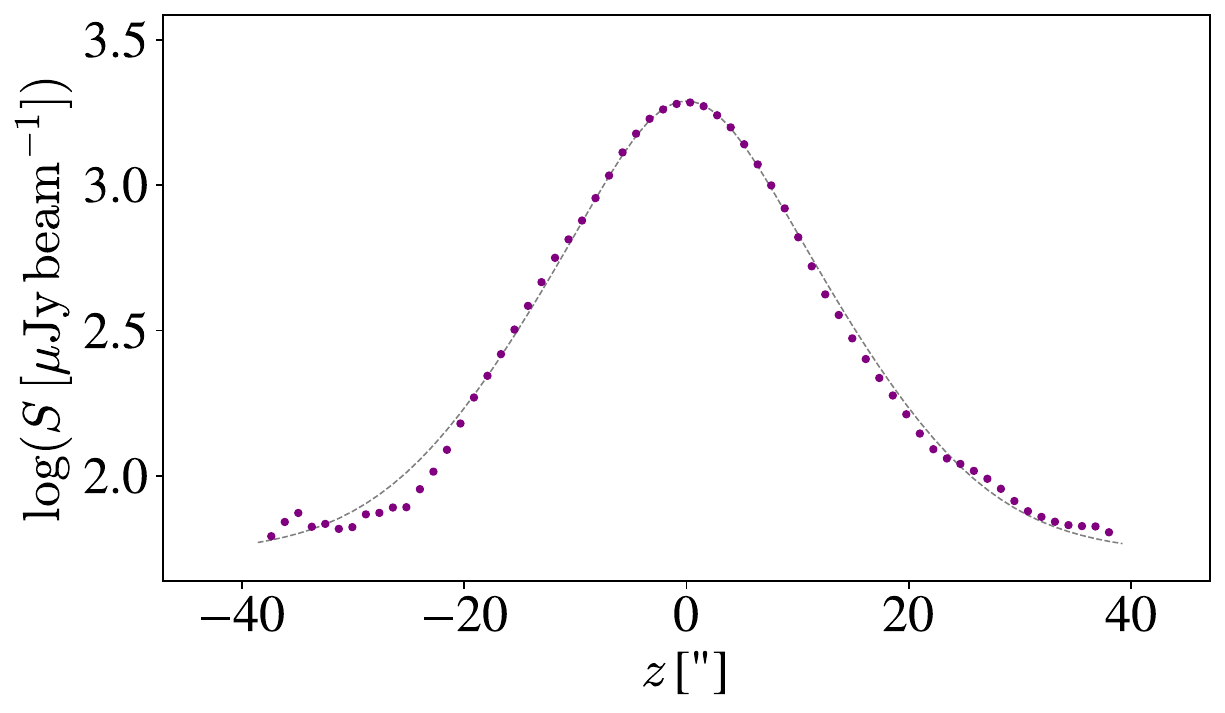}
    \includegraphics[width=0.4\textwidth]{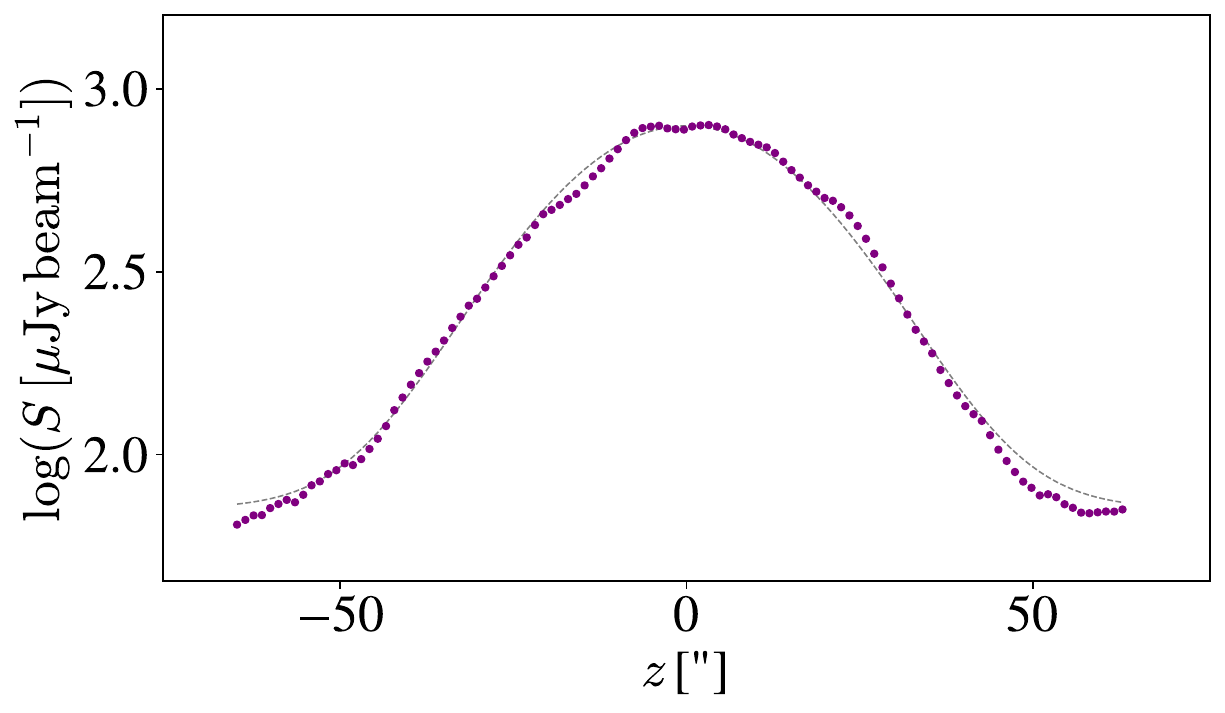}
    \includegraphics[width=0.4\textwidth]{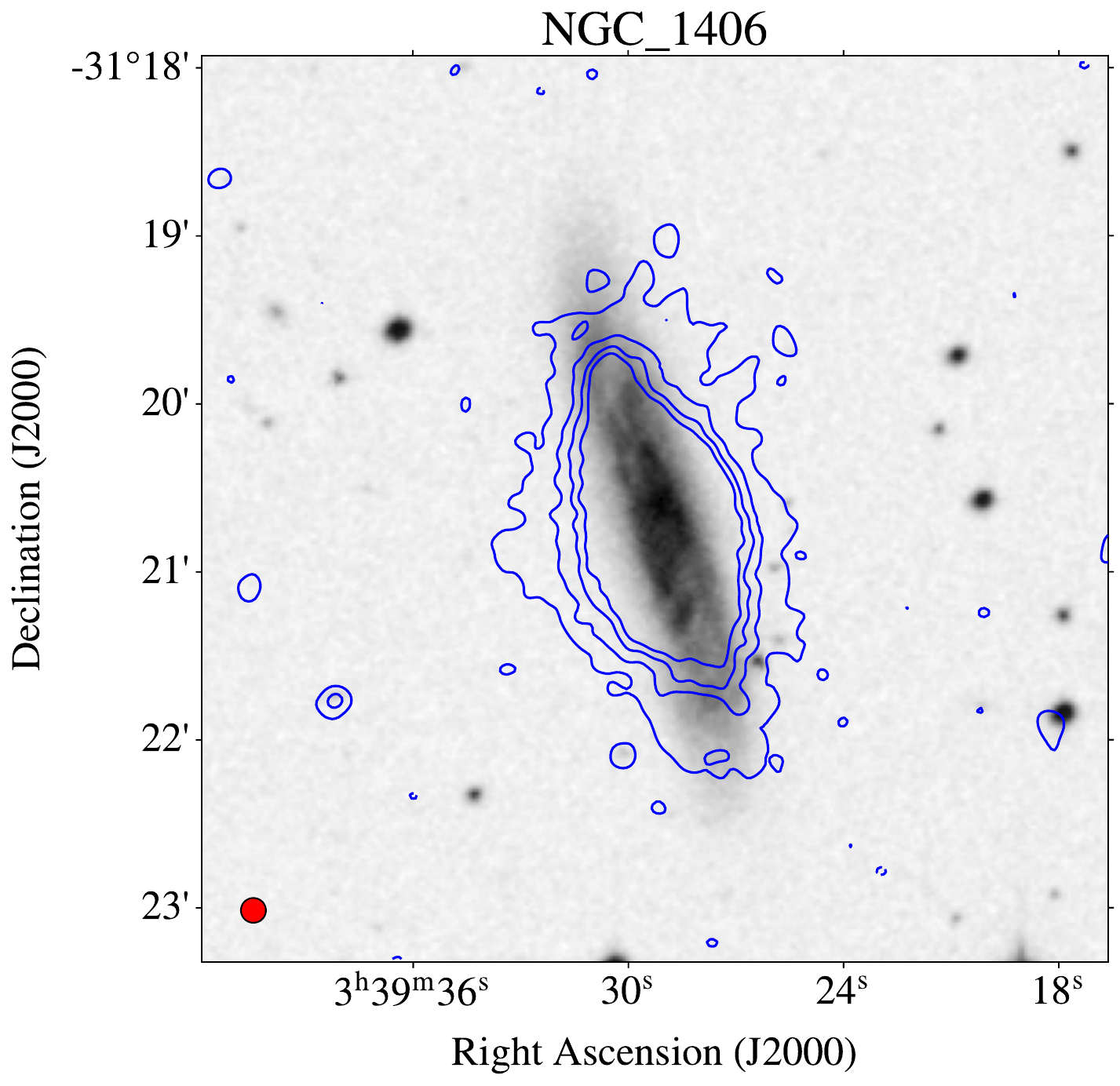}
    \includegraphics[width=0.4\textwidth]{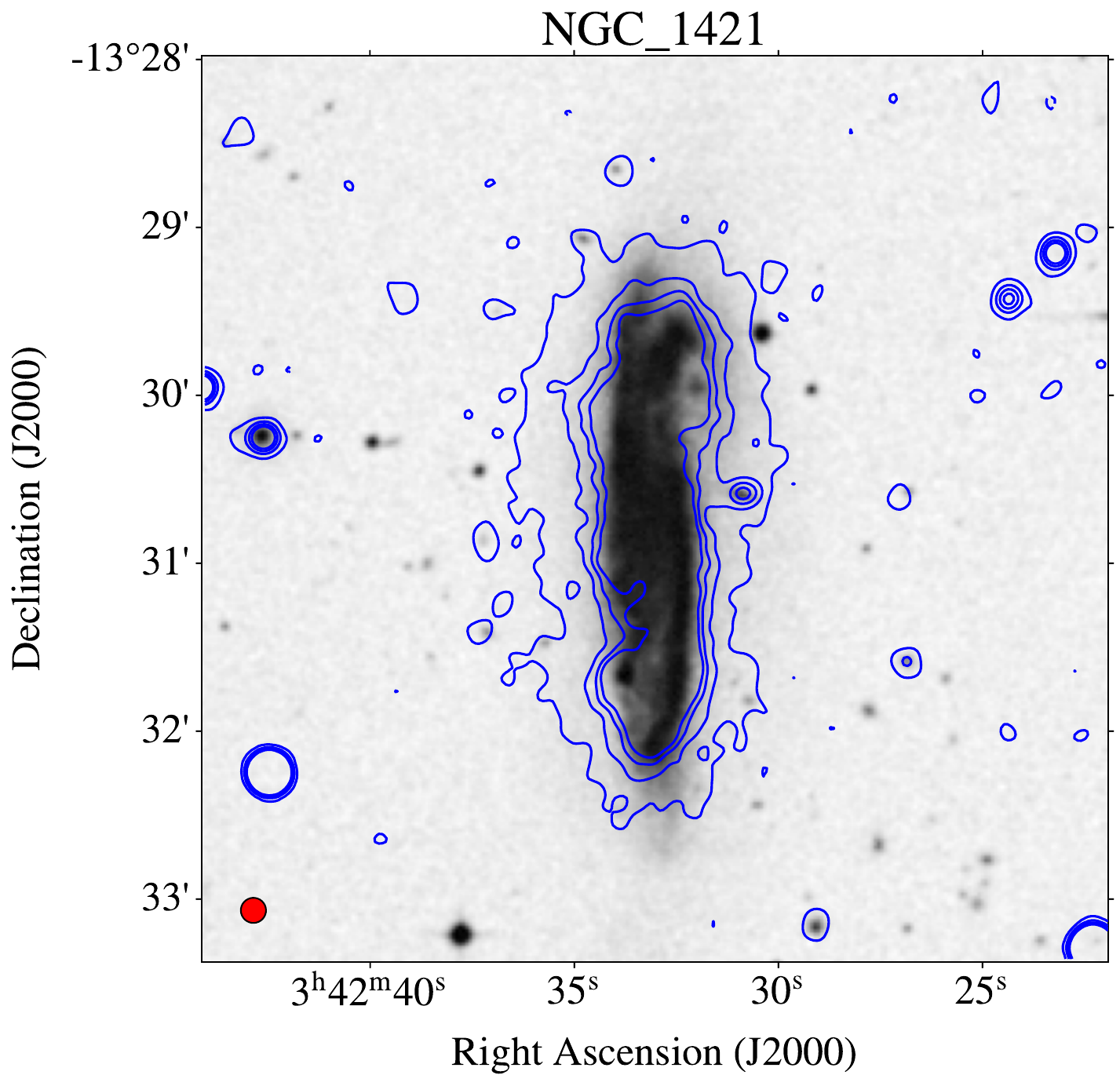} 
    \includegraphics[width=0.4\textwidth]{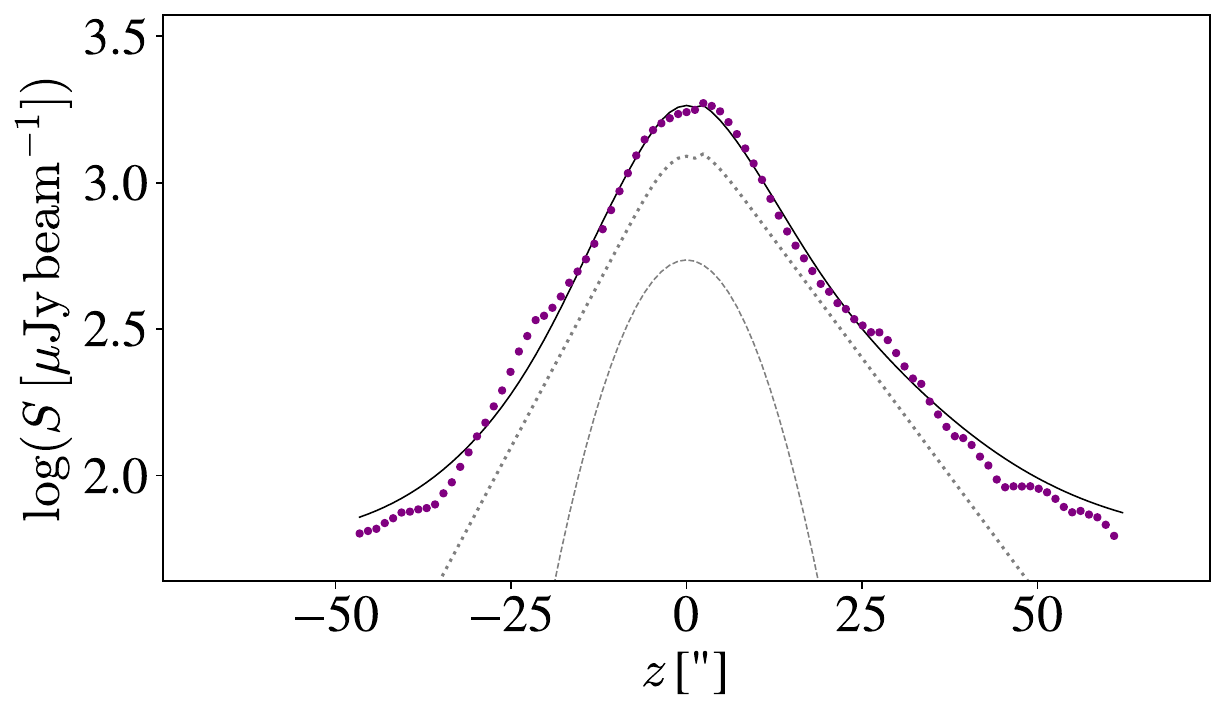}
    \includegraphics[width=0.4\textwidth]{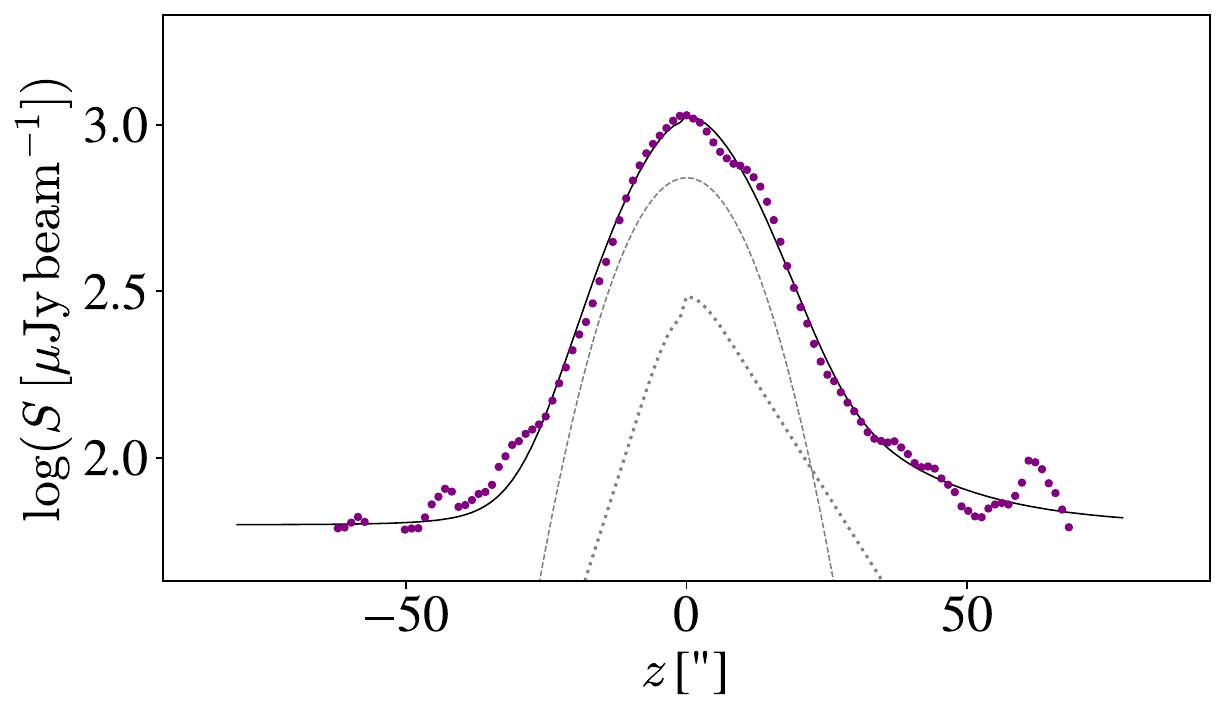}
    \caption{continued.}
\end{figure*}

\begin{figure*} \ContinuedFloat
\centering
    \includegraphics[width=0.4\textwidth]{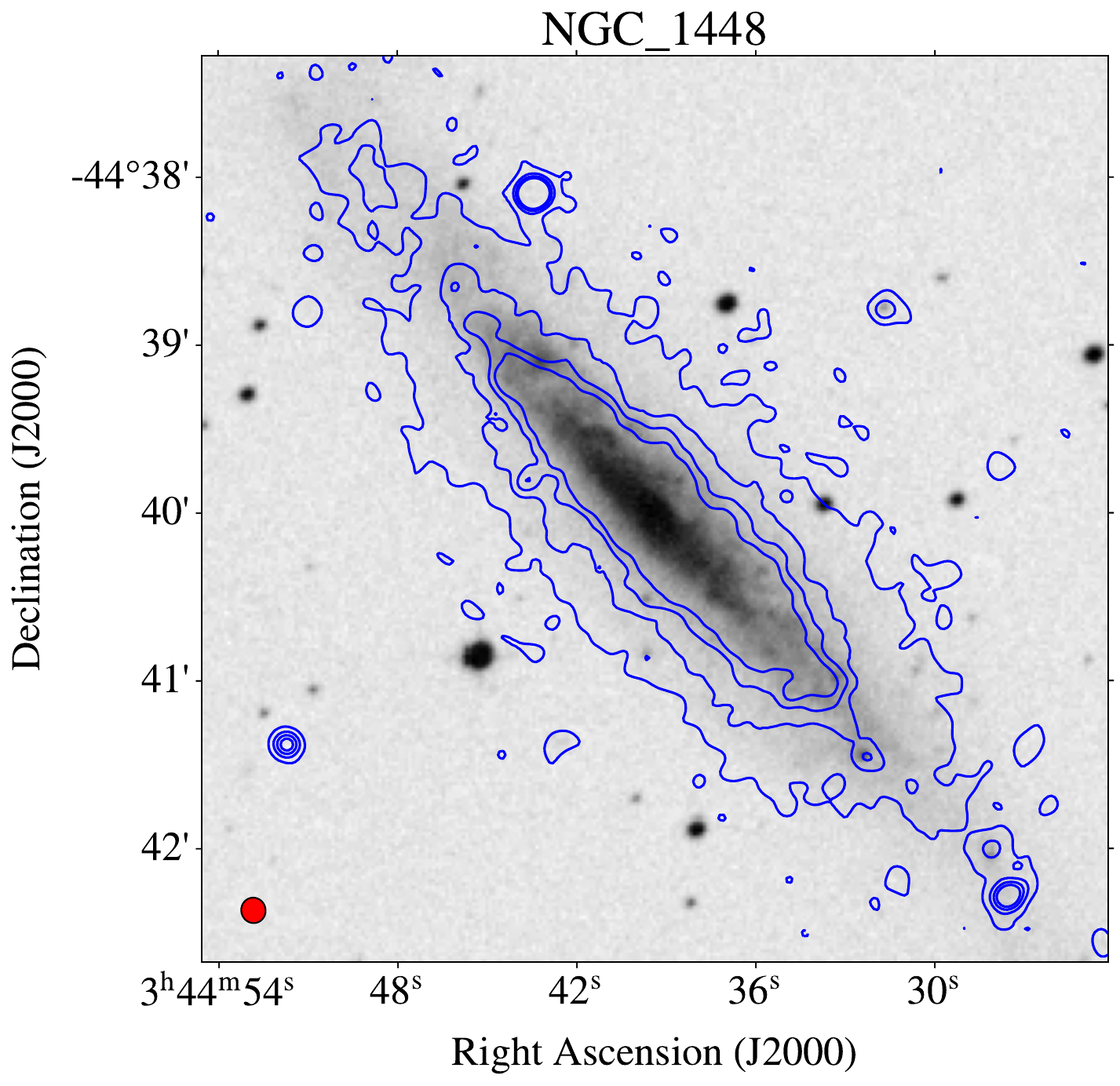}
    \includegraphics[width=0.4\textwidth]{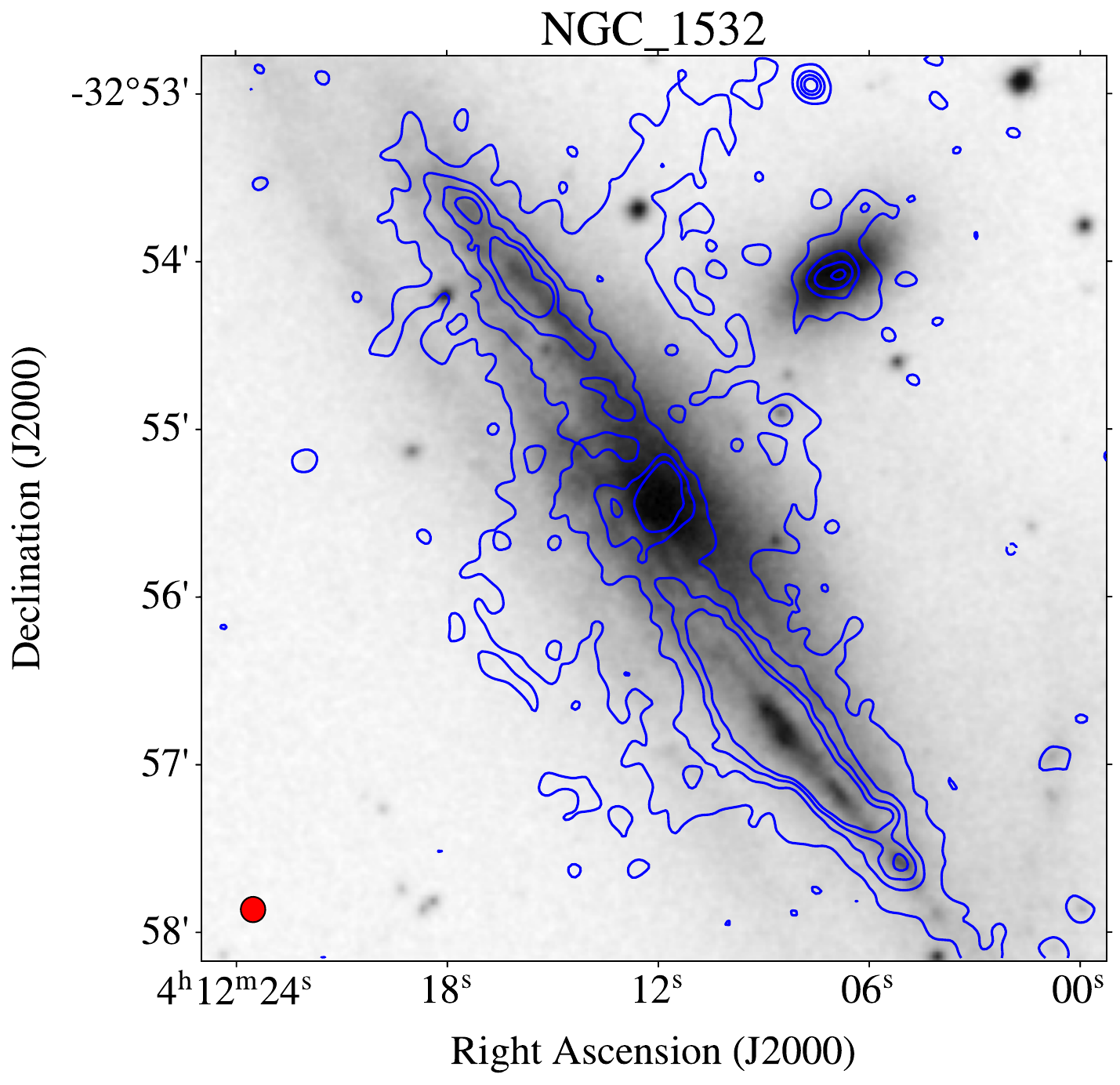}
    \includegraphics[width=0.4\textwidth]{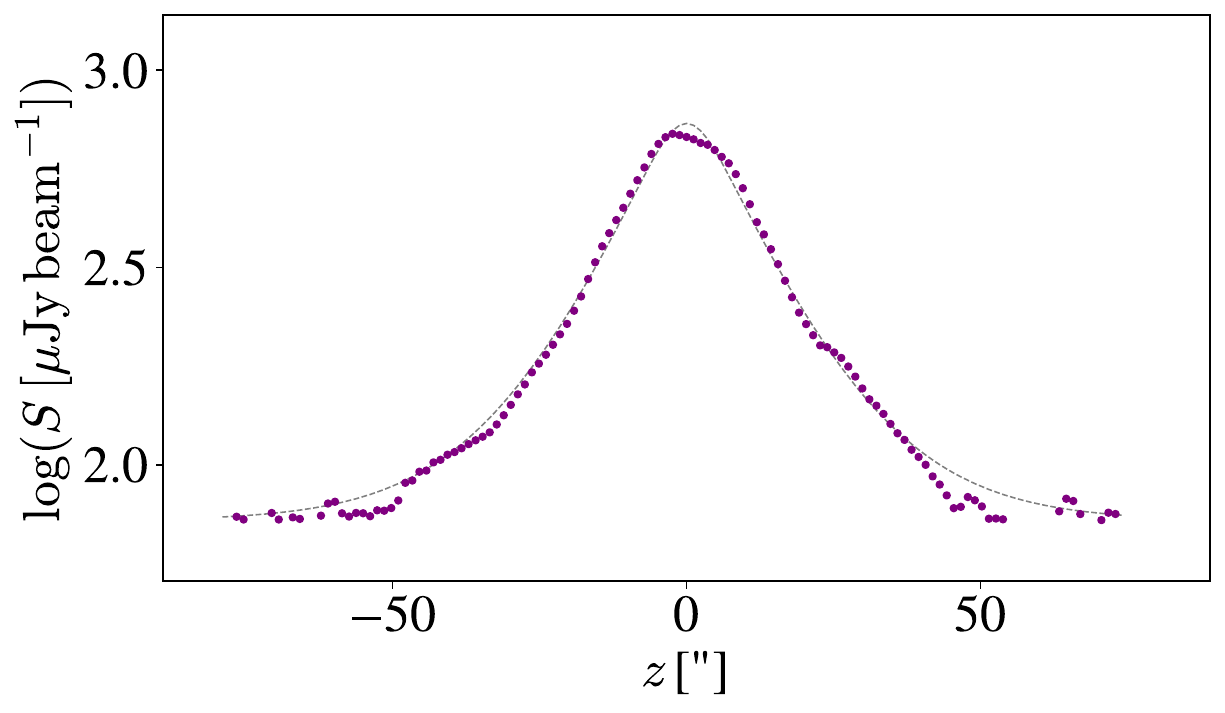}
    \includegraphics[width=0.4\textwidth]{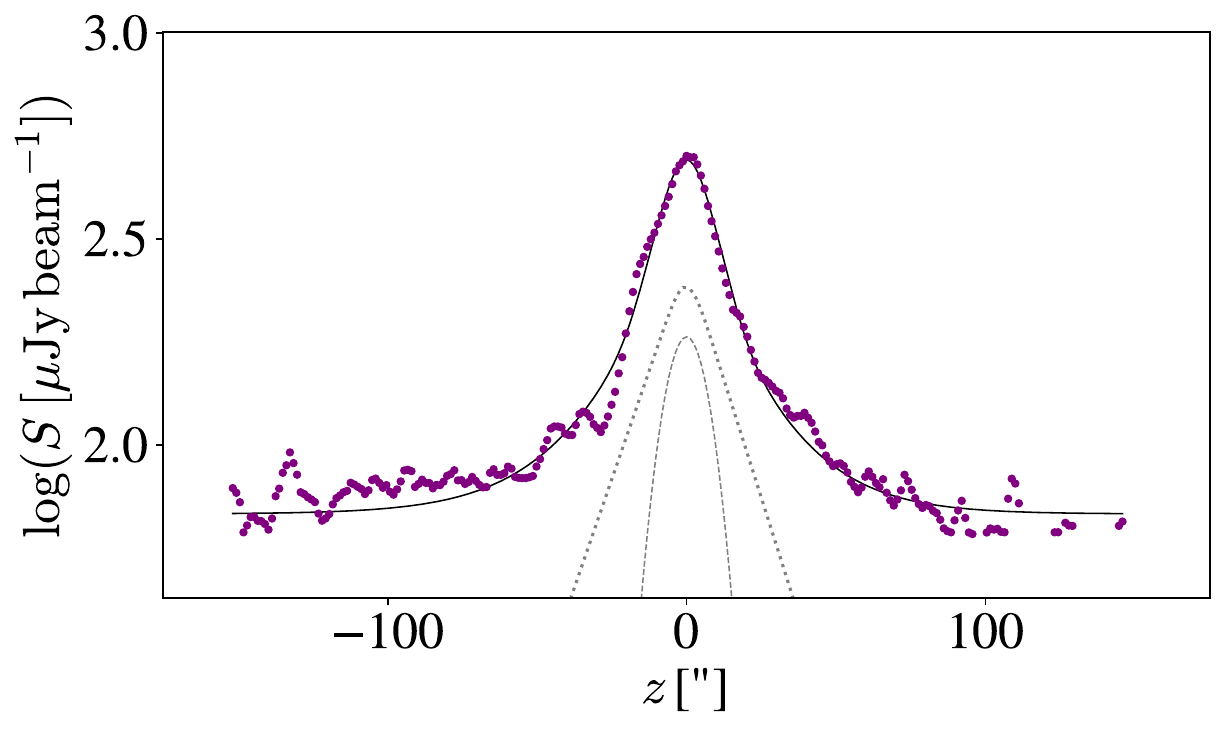}
    \includegraphics[width=0.4\textwidth]{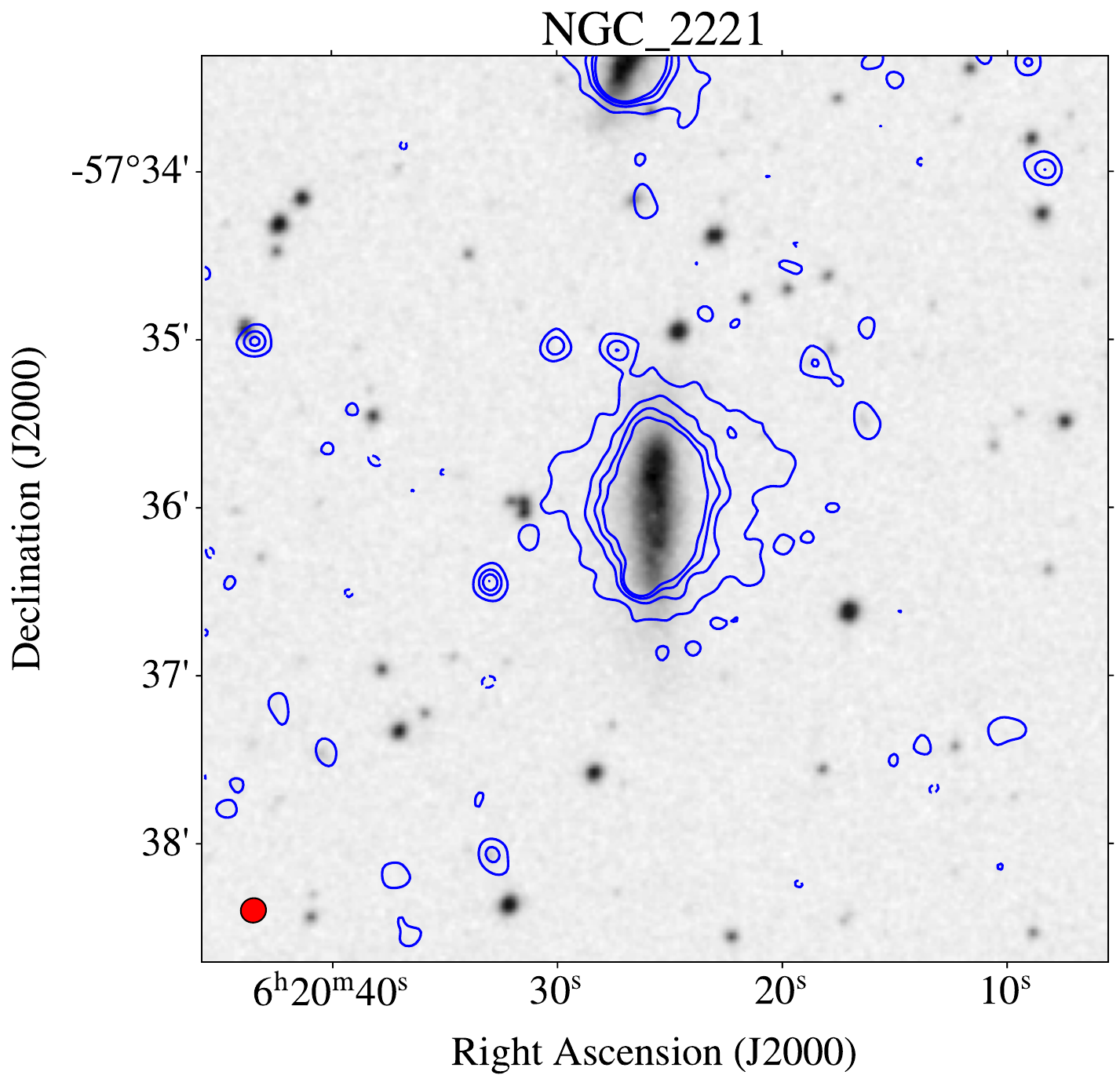}
    \includegraphics[width=0.4\textwidth]{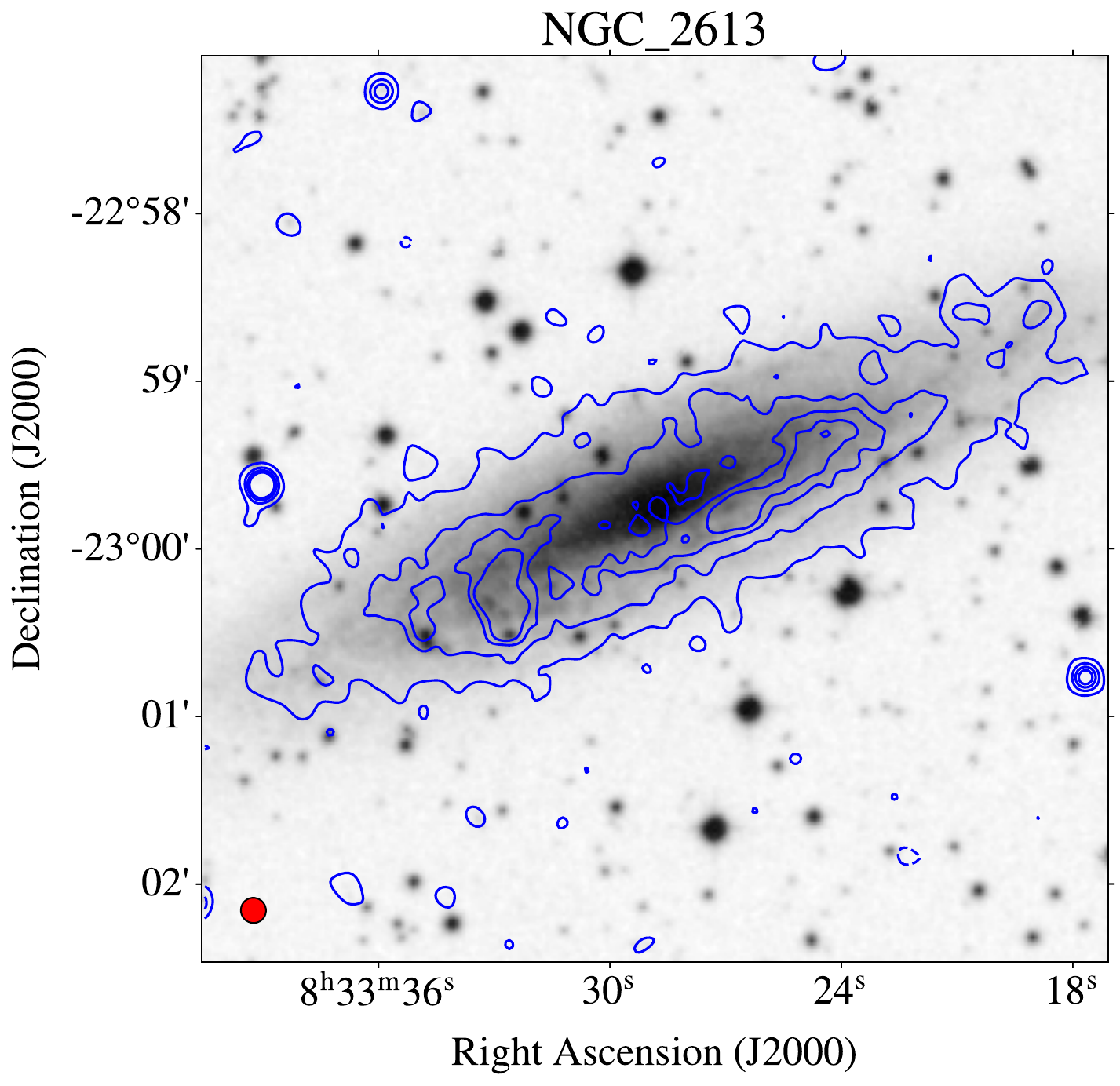}
    \includegraphics[width=0.4\textwidth]{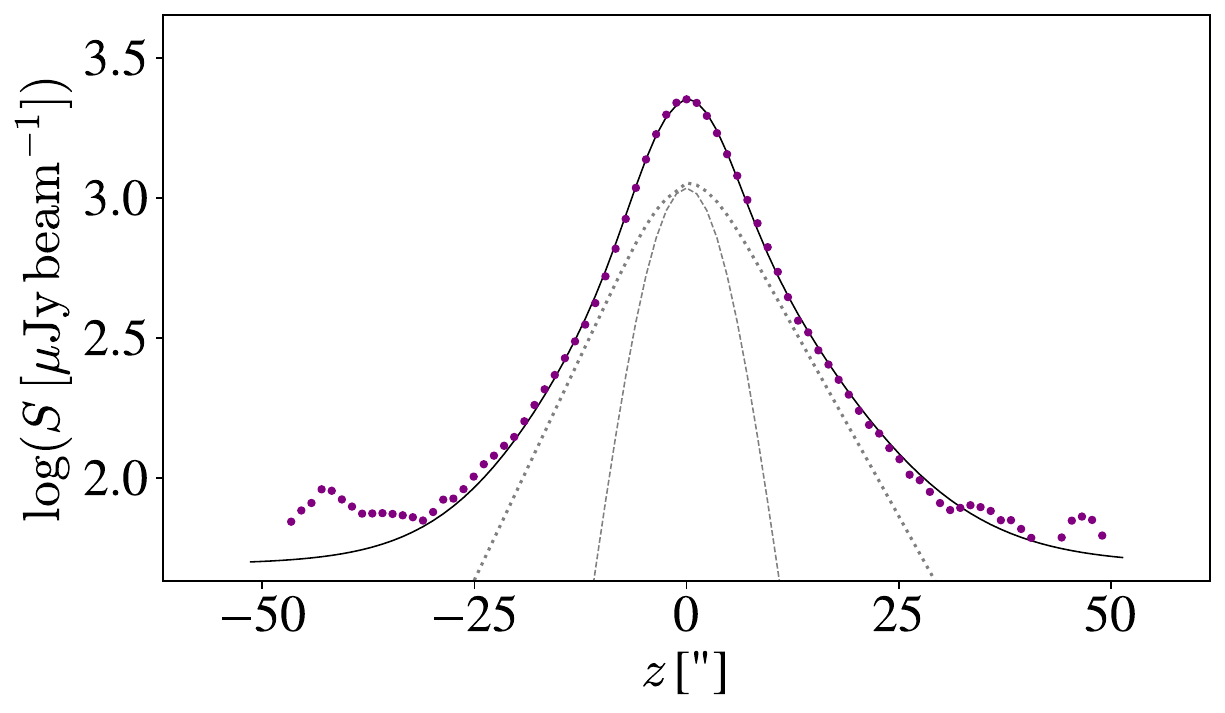}
    \includegraphics[width=0.4\textwidth]{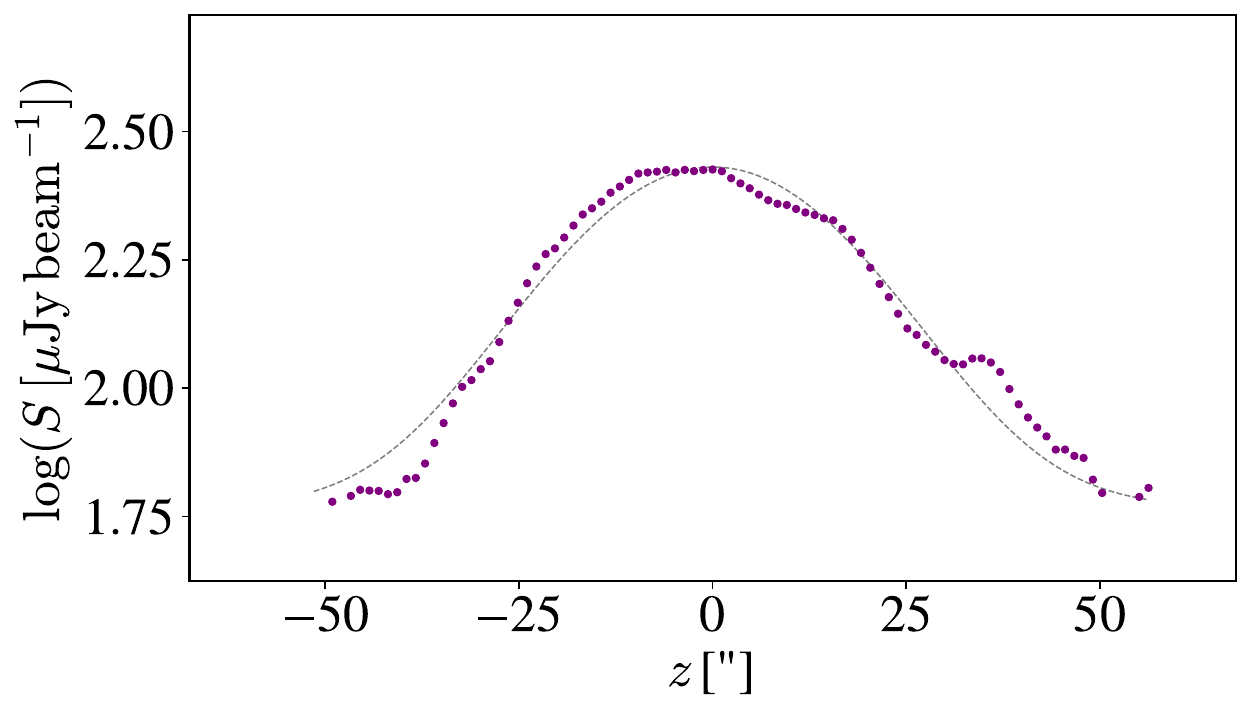} 
    \caption{continued.}
\end{figure*}

\begin{figure*} \ContinuedFloat
\centering
    \includegraphics[width=0.4\textwidth]{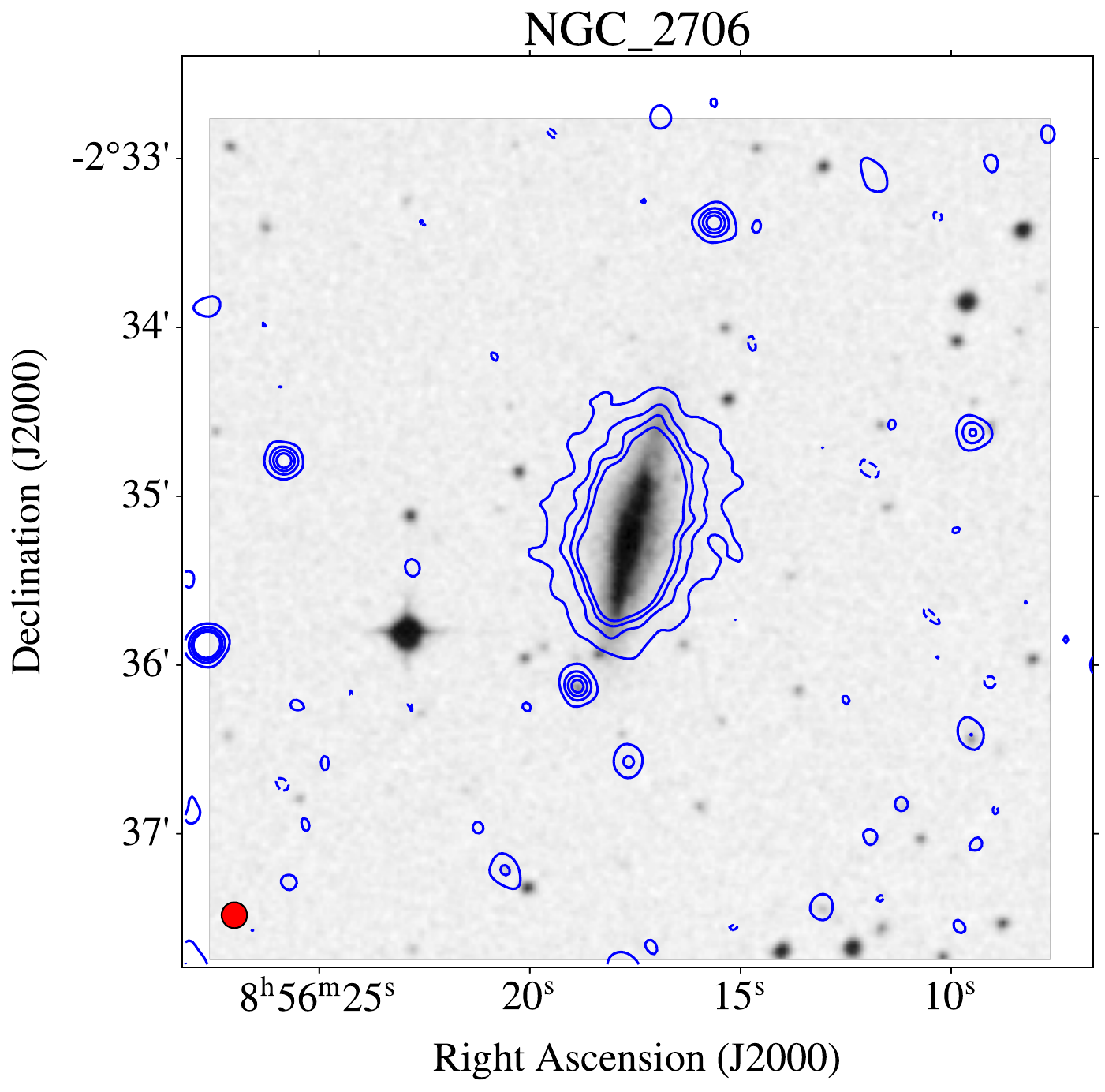}
    \includegraphics[width=0.4\textwidth]{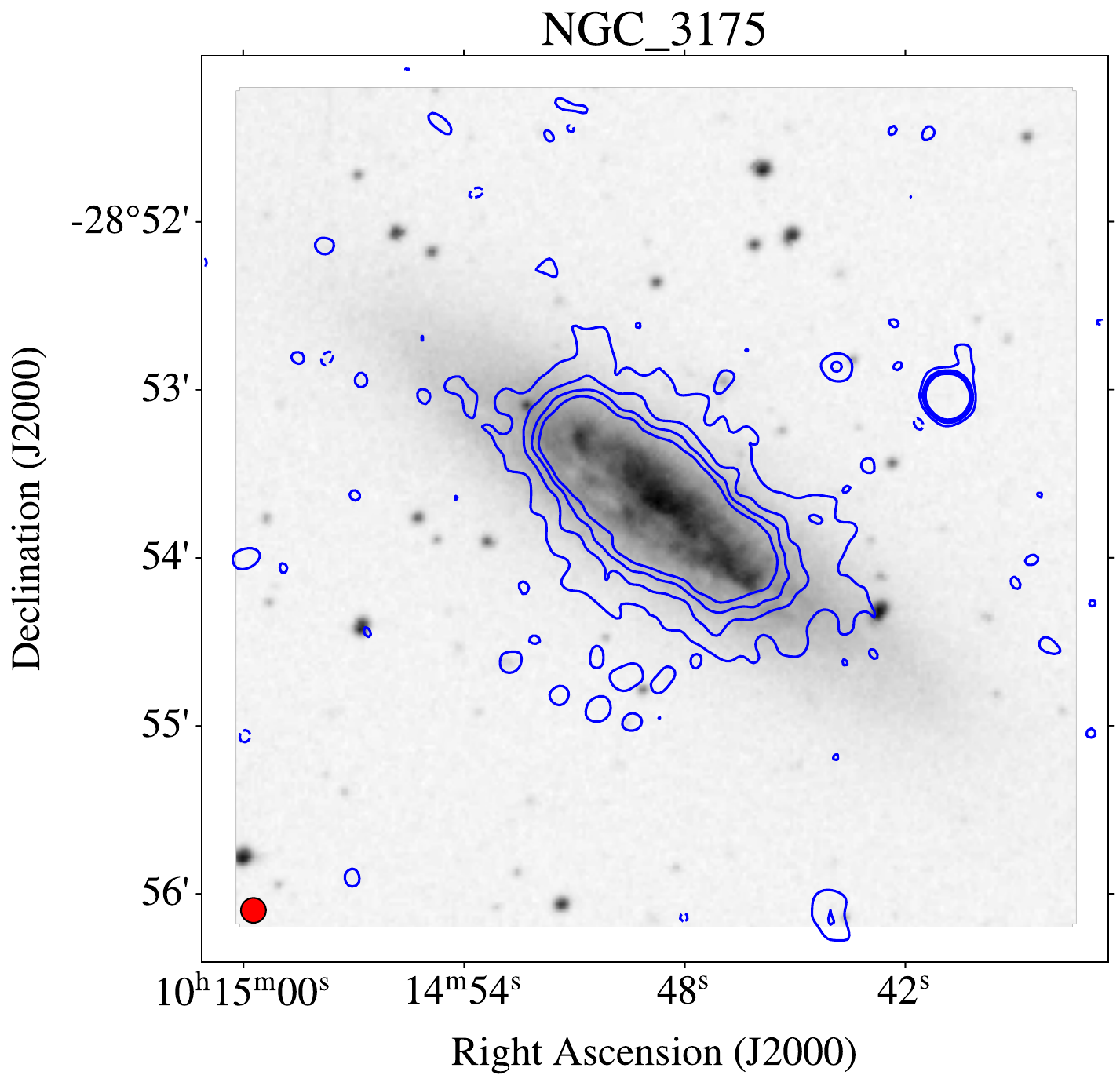}   
    \includegraphics[width=0.4\textwidth]{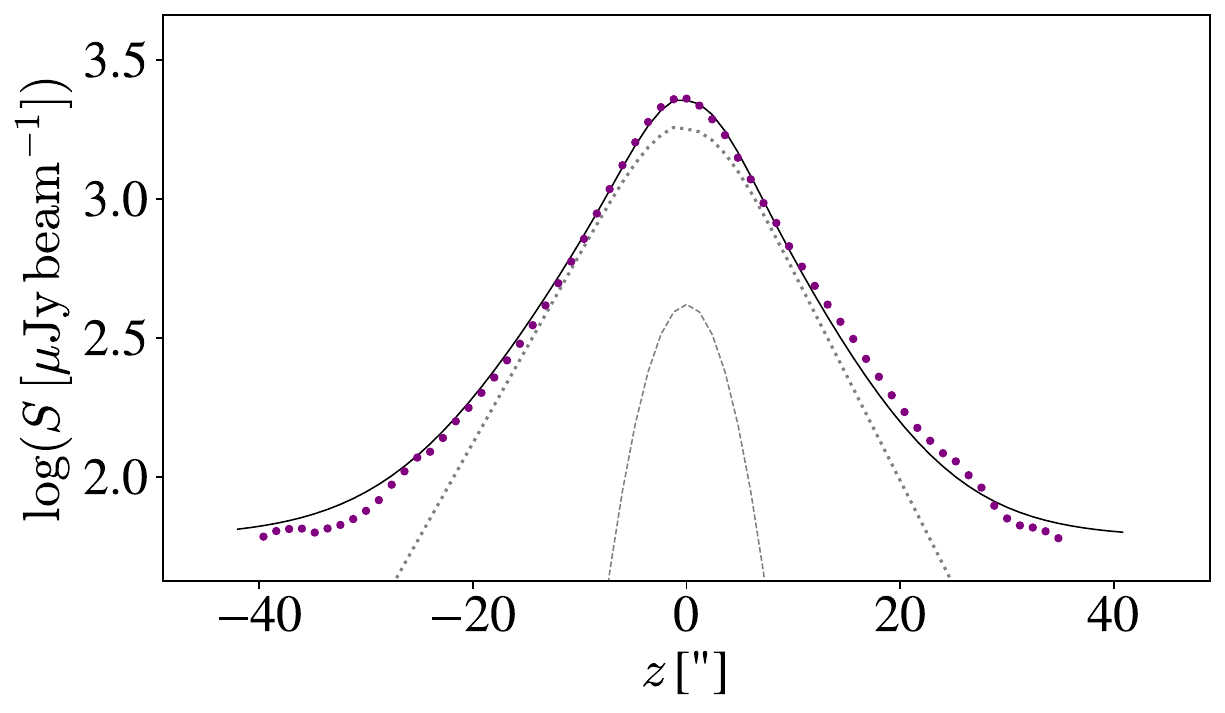} 
    \includegraphics[width=0.4\textwidth]{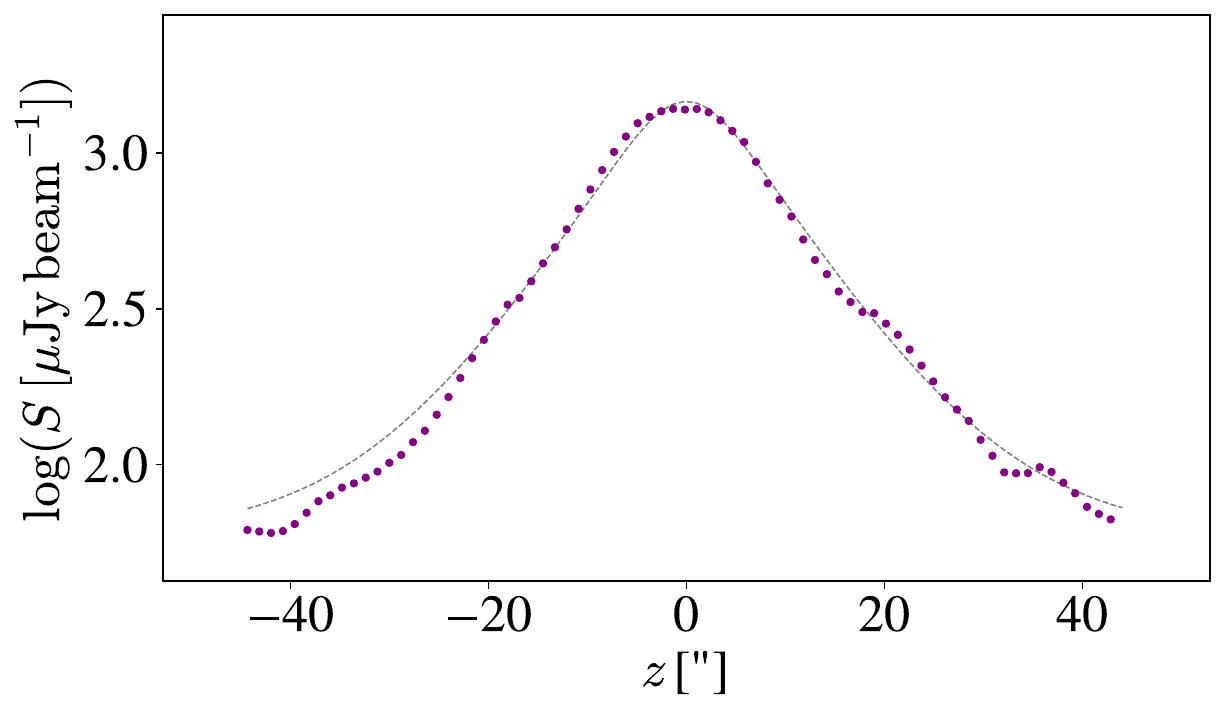} 
    \includegraphics[width=0.4\textwidth]{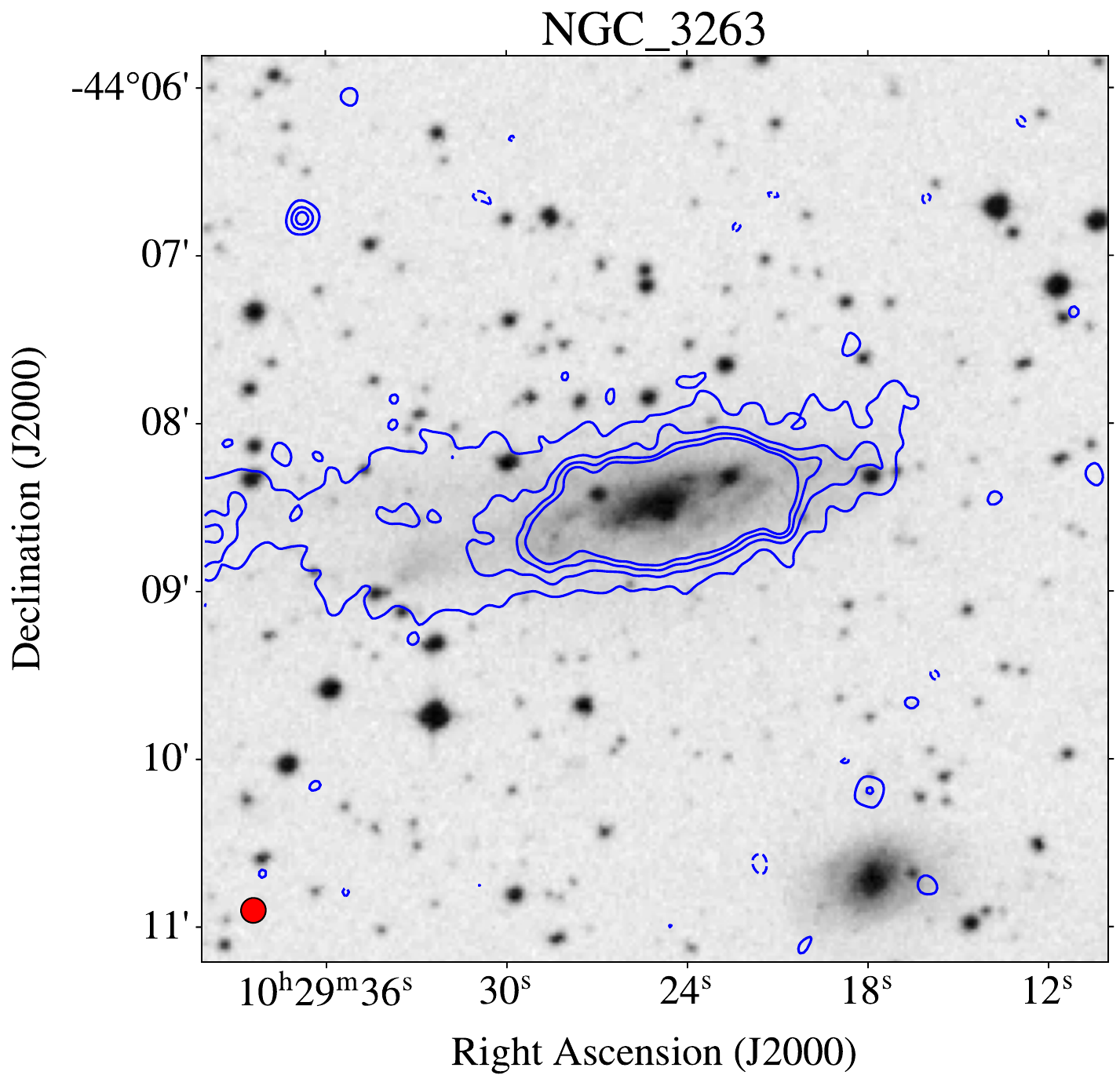} 
    \includegraphics[width=0.4\textwidth]{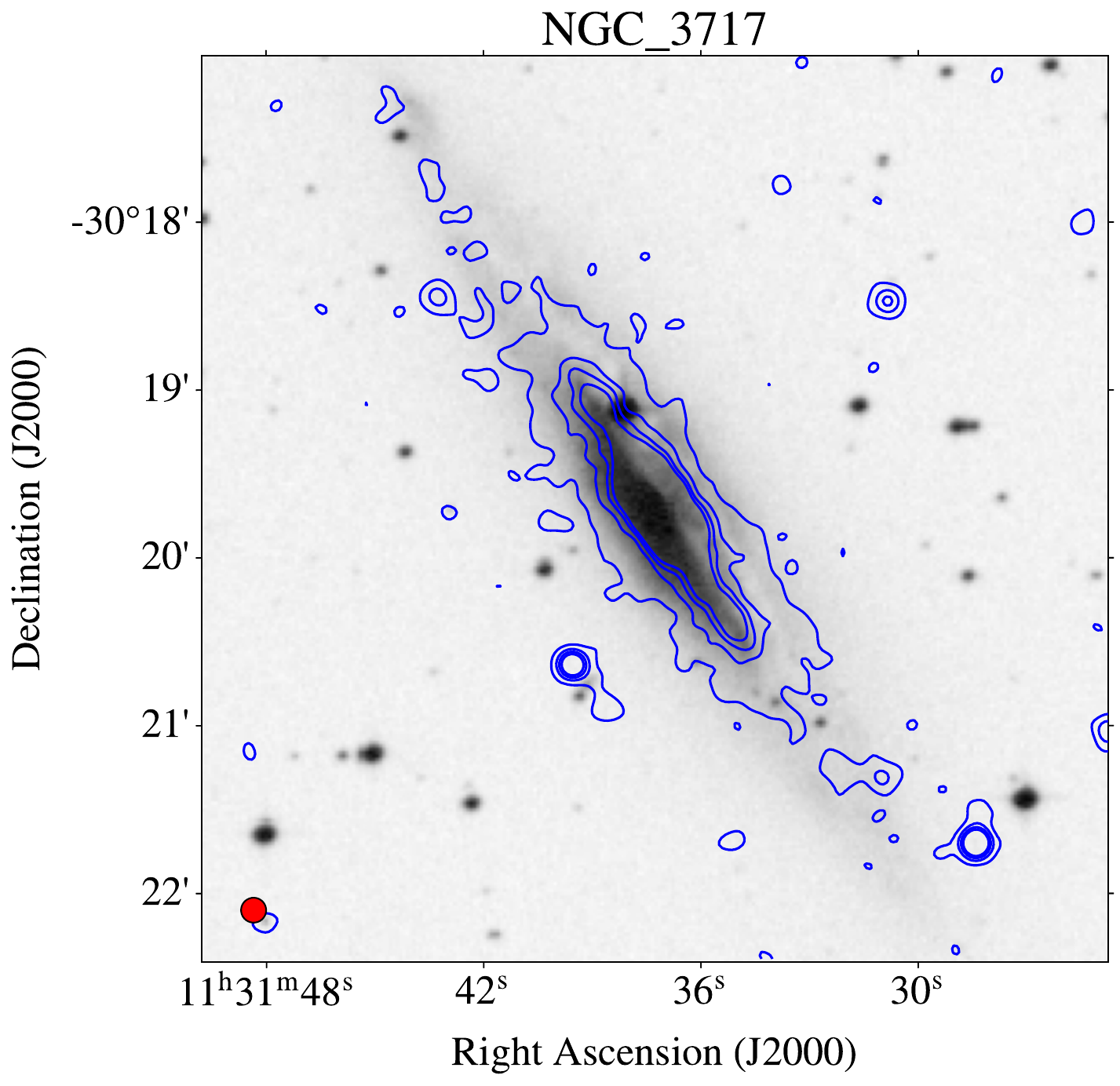}
    \includegraphics[width=0.4\textwidth]{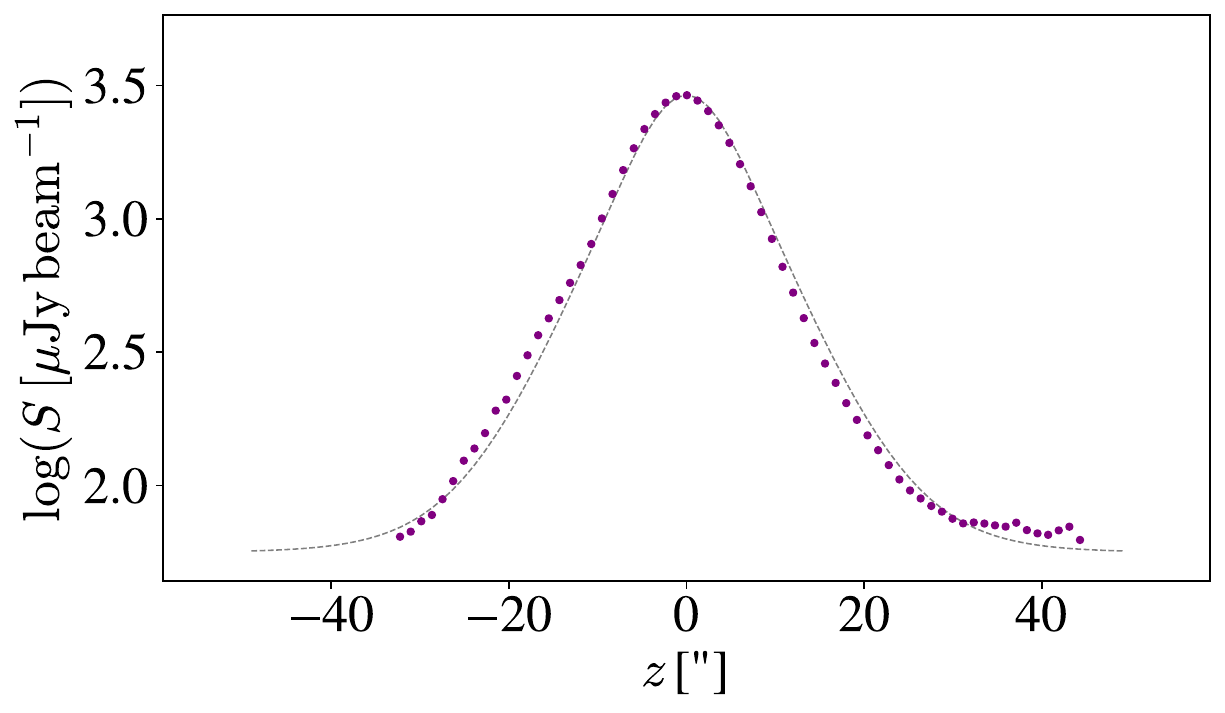}
    \includegraphics[width=0.4\textwidth]{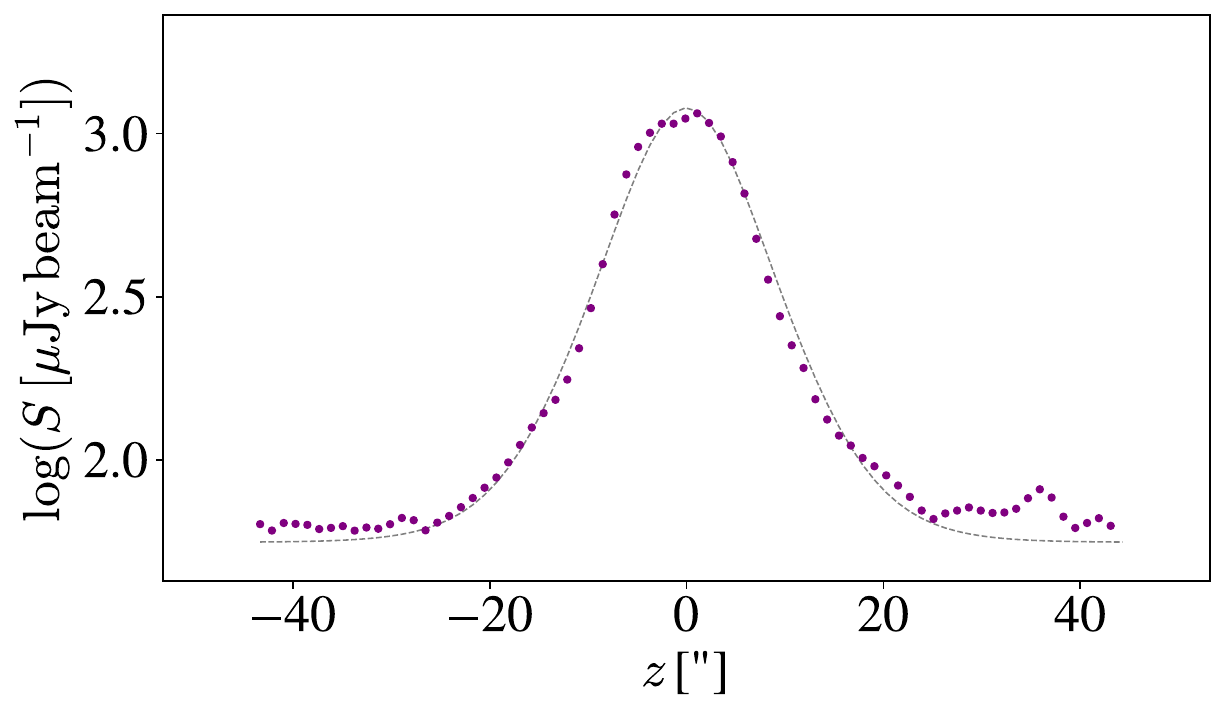}
\caption{continued.}
\end{figure*}

%  %------------------------    

\begin{figure*} \ContinuedFloat
\centering    
        \includegraphics[width=0.4\textwidth]{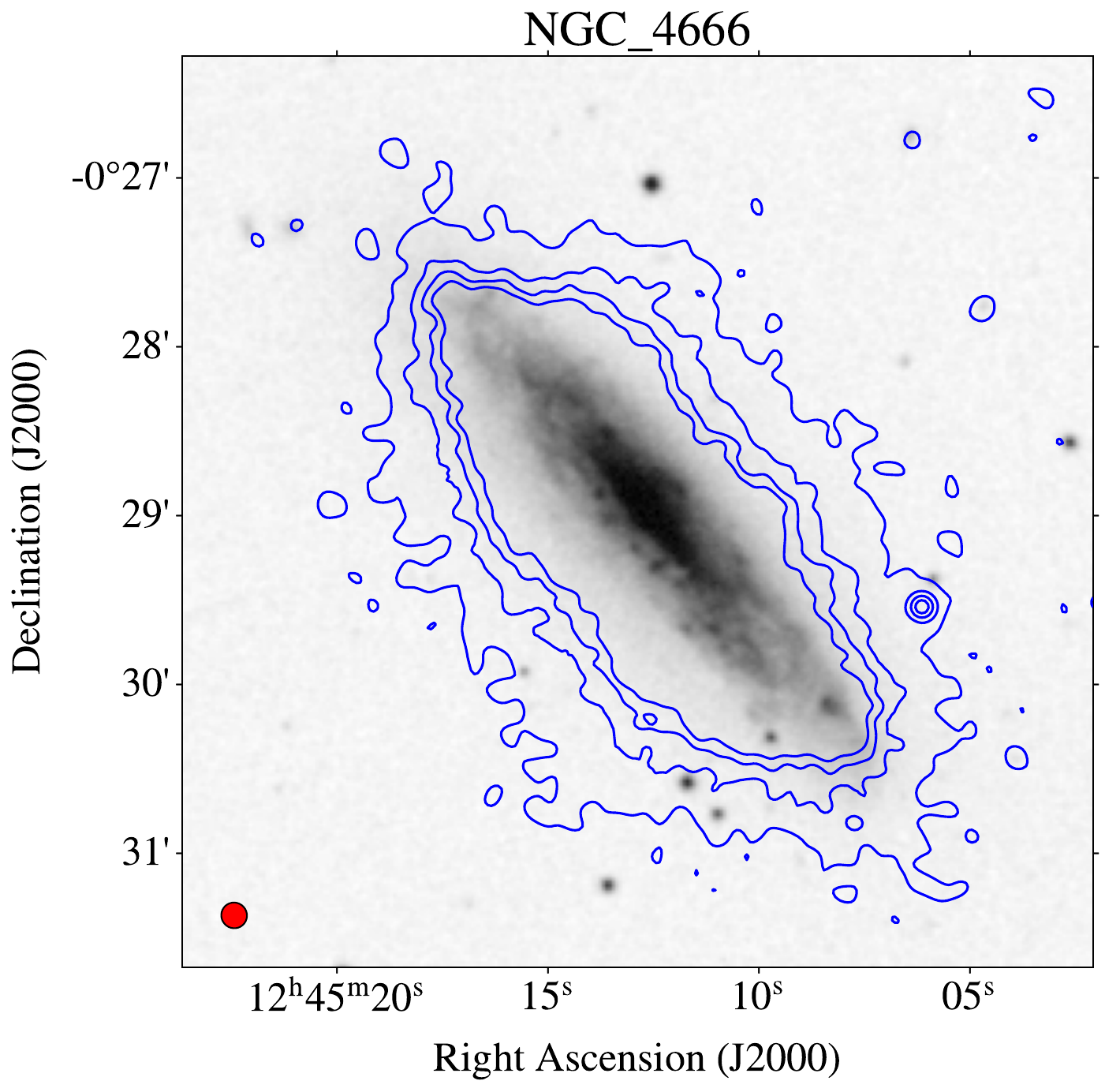} 
        \includegraphics[width=0.4\textwidth]{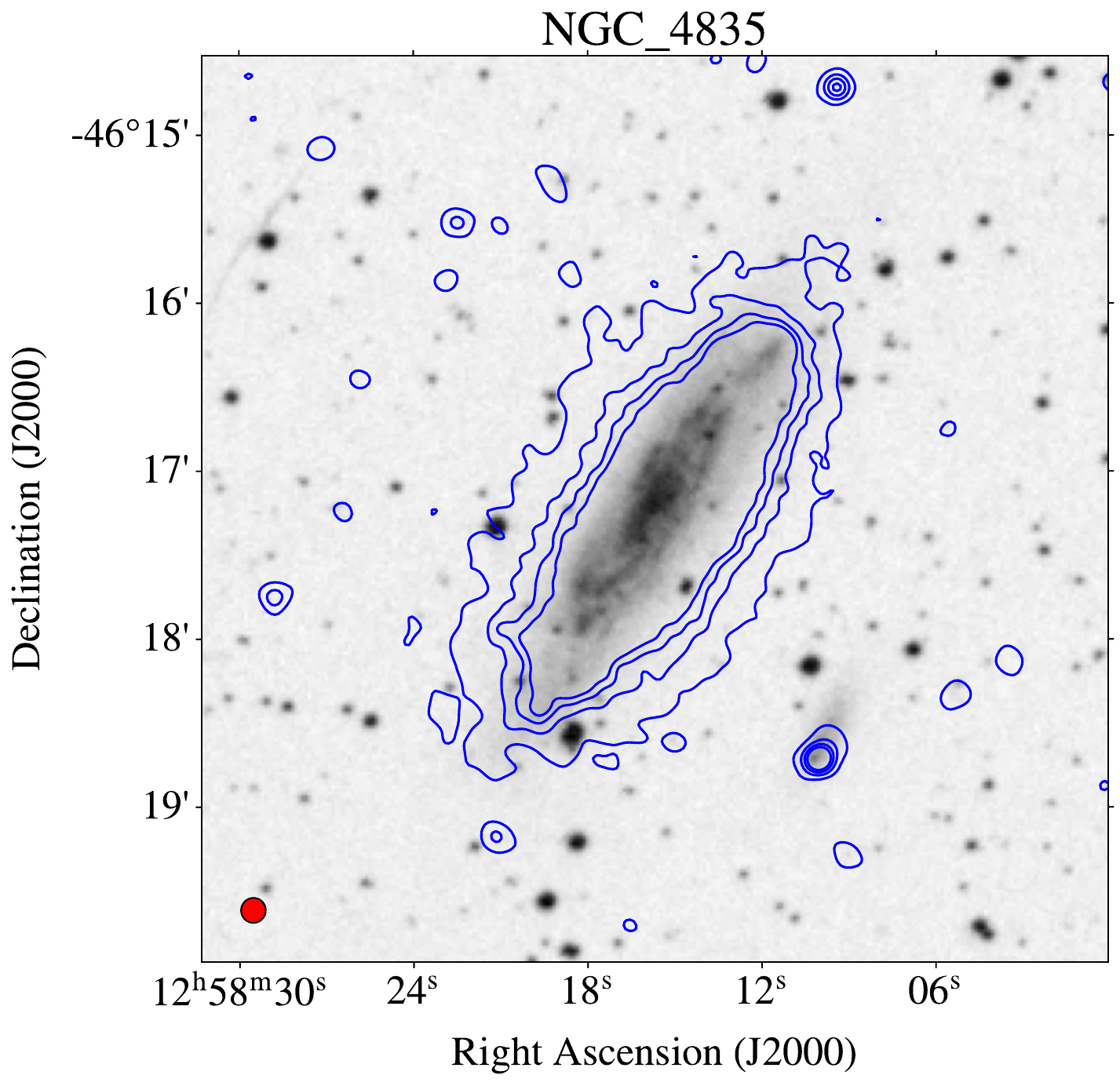} 
        \includegraphics[width=0.4\textwidth]{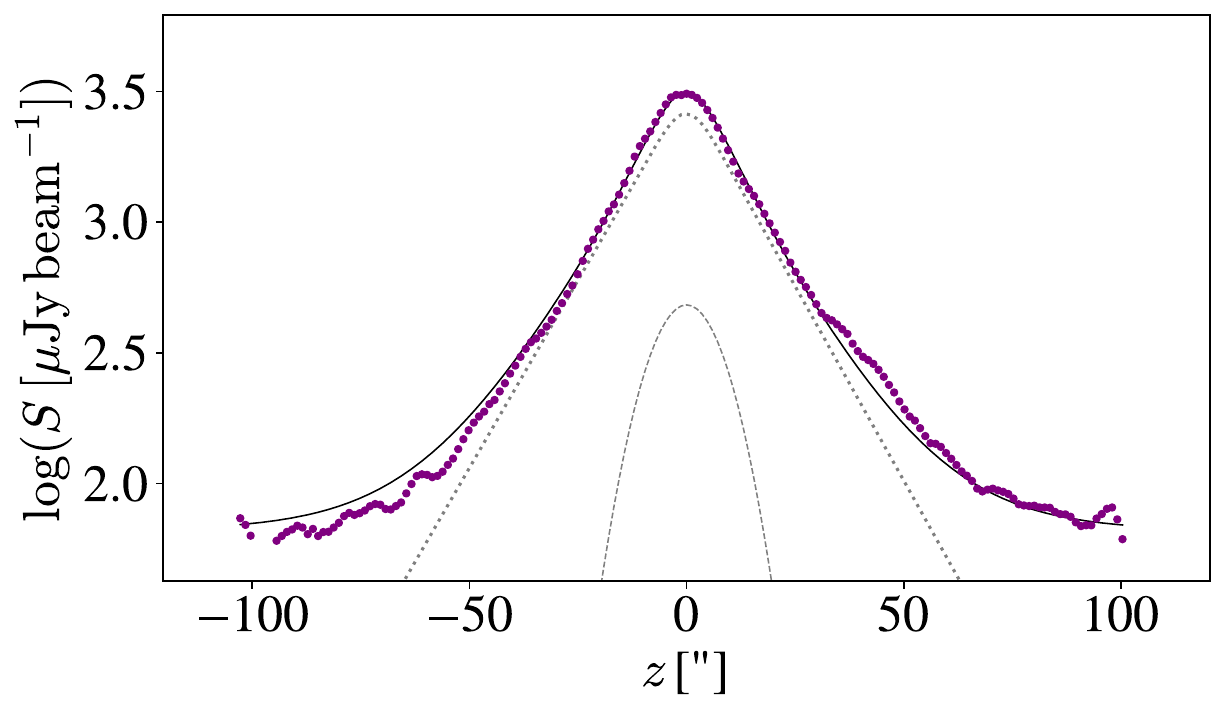} 
        \includegraphics[width=0.4\textwidth]{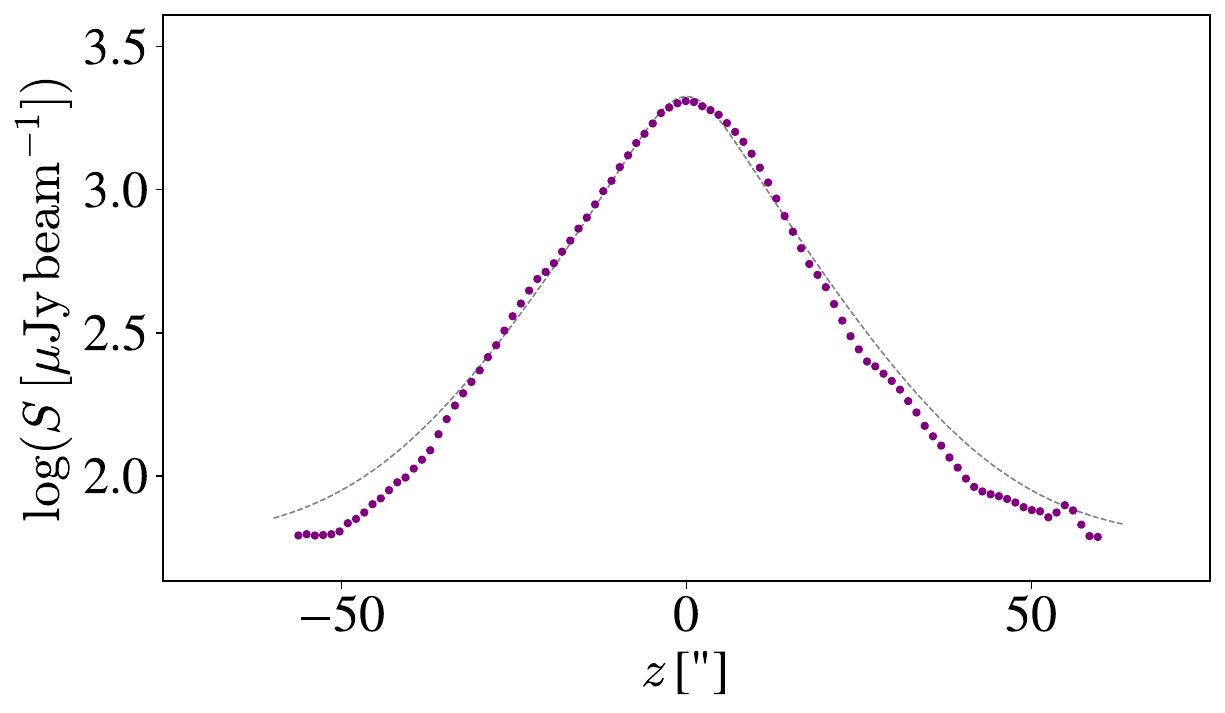} 
        \includegraphics[width=0.4\textwidth]{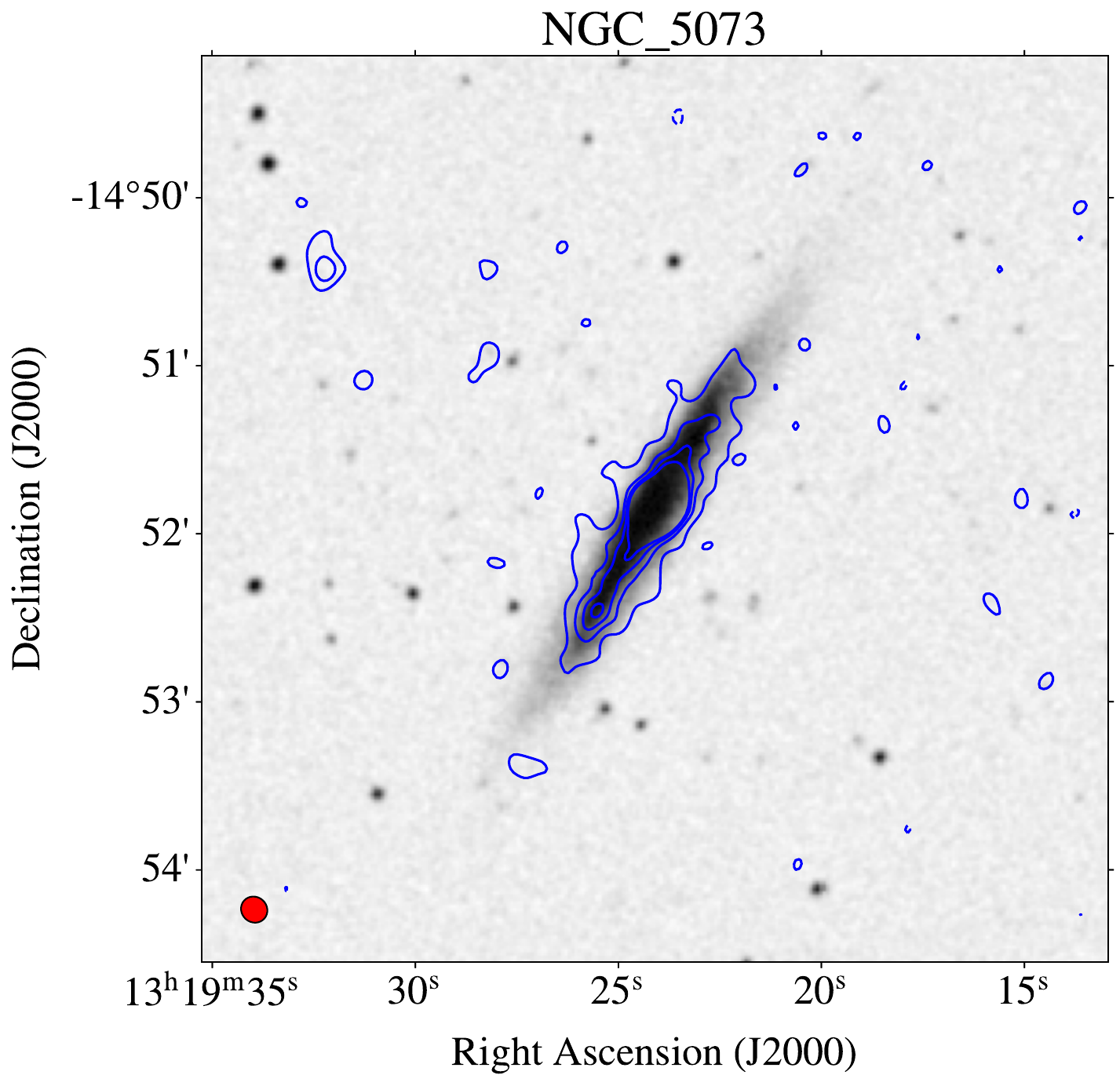} 
        \includegraphics[width=0.4\textwidth]{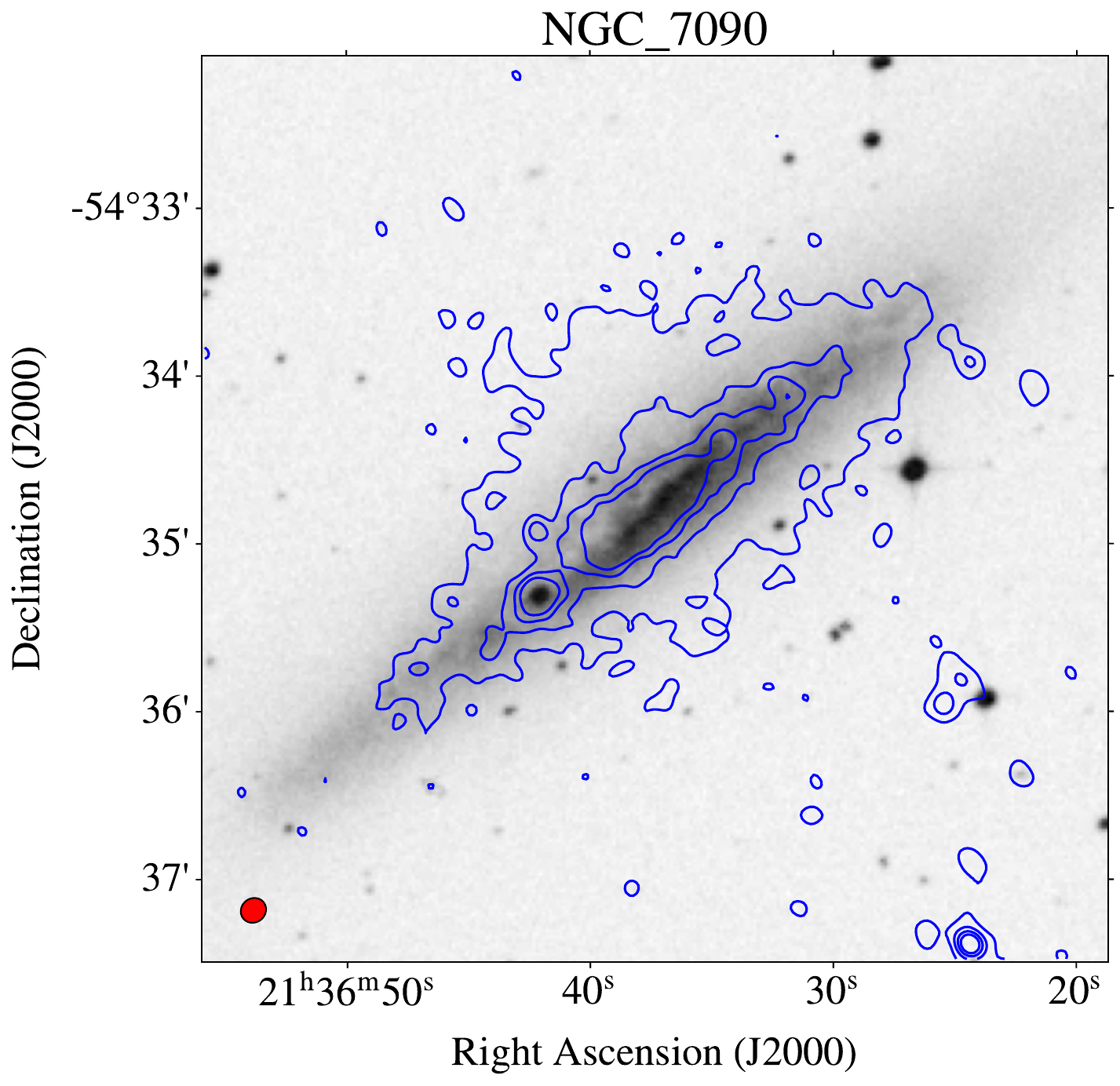} 
        \includegraphics[width=0.4\textwidth]{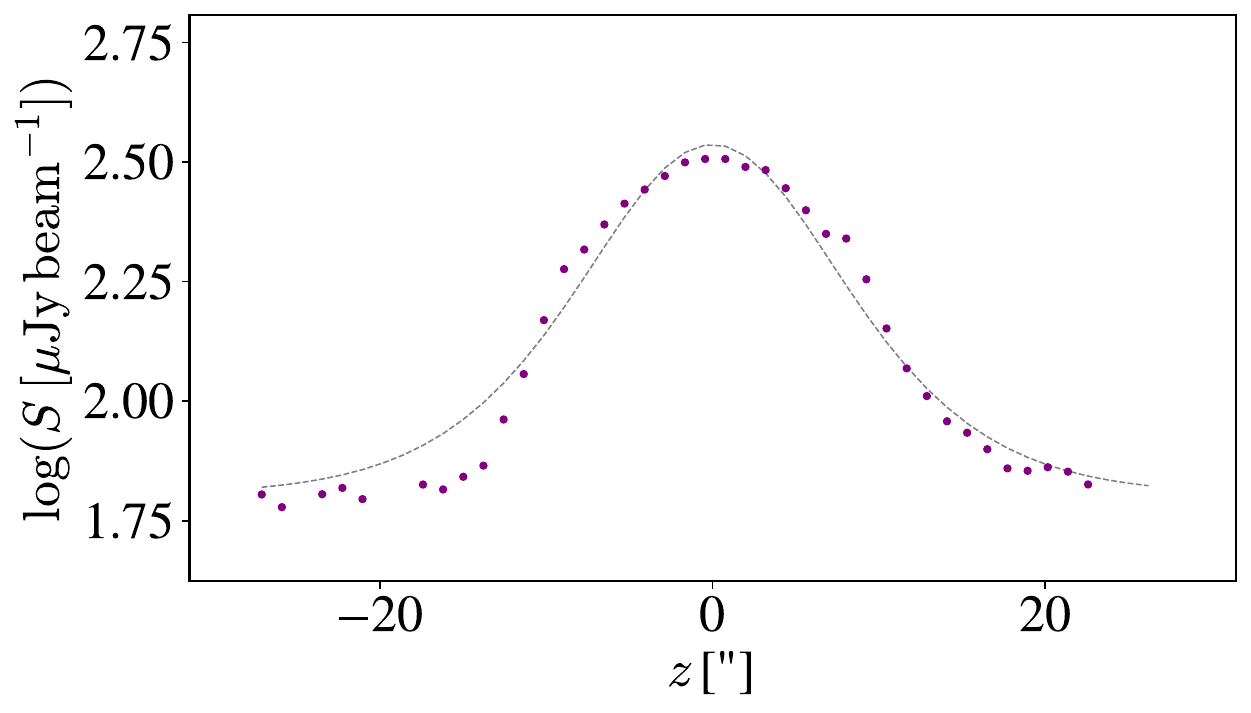} 
        \includegraphics[width=0.4\textwidth]{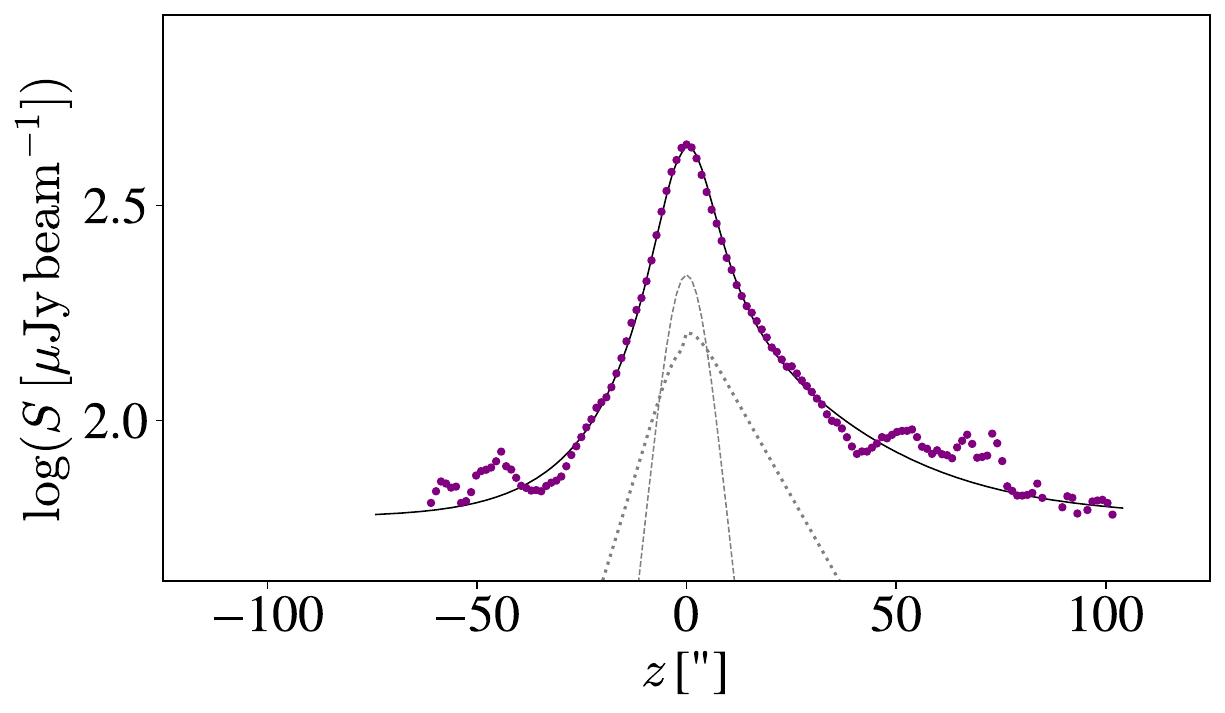}
\caption{continued.}
\end{figure*}

\begin{figure*} \ContinuedFloat
\centering    
        \includegraphics[width=0.4\textwidth]{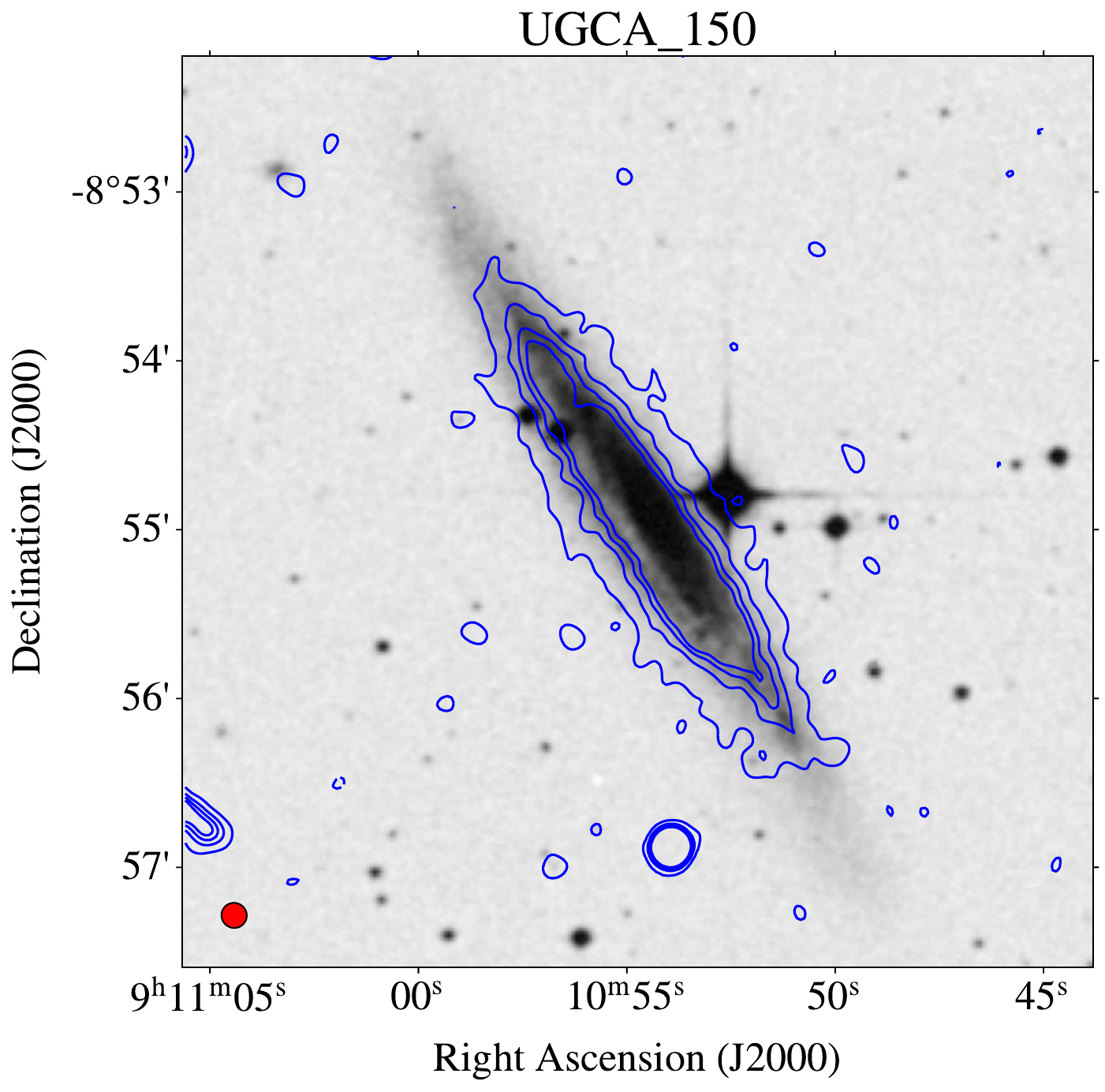}
        \includegraphics[width=0.4\textwidth]{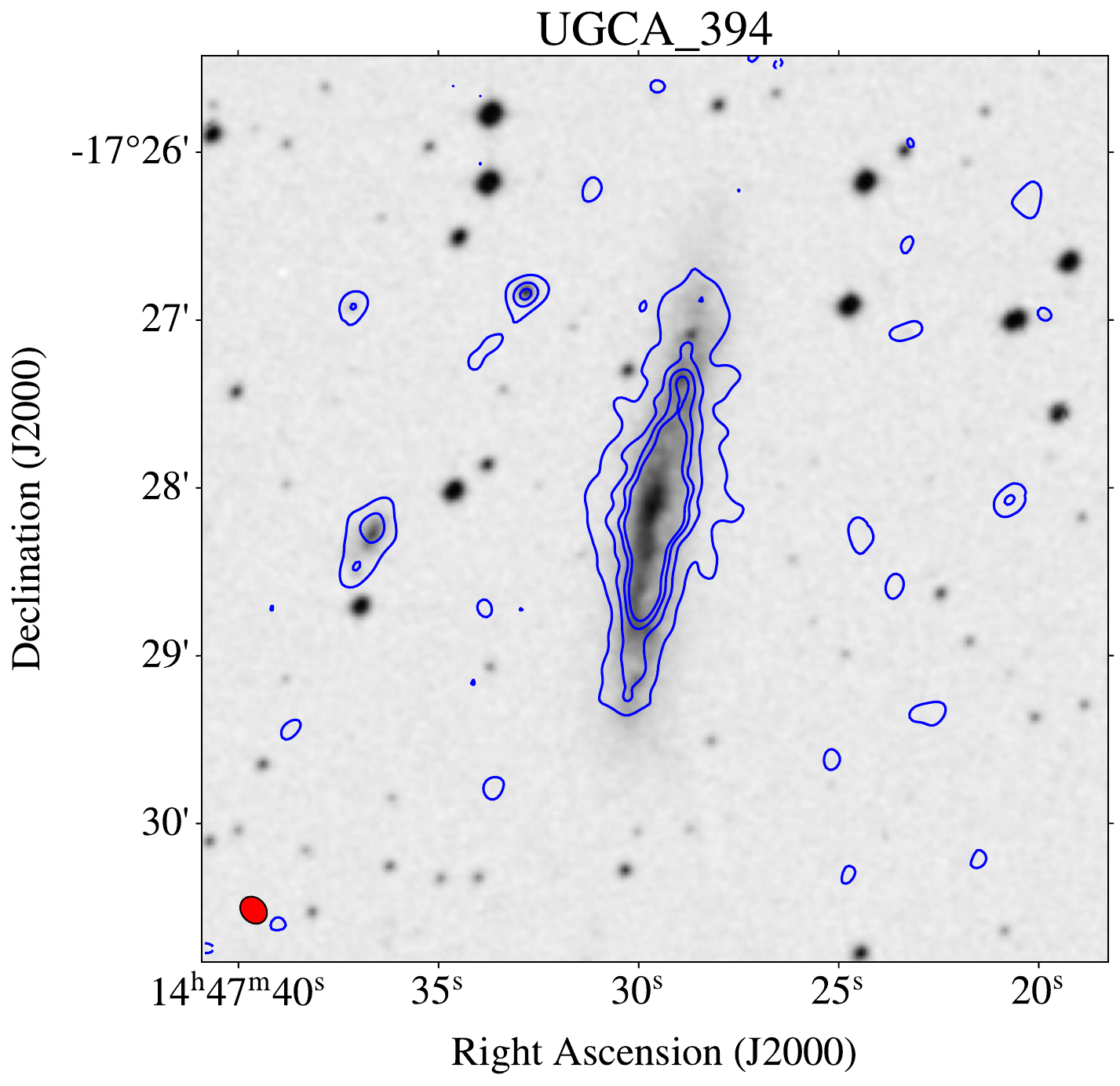}
        \includegraphics[width=0.4\textwidth]{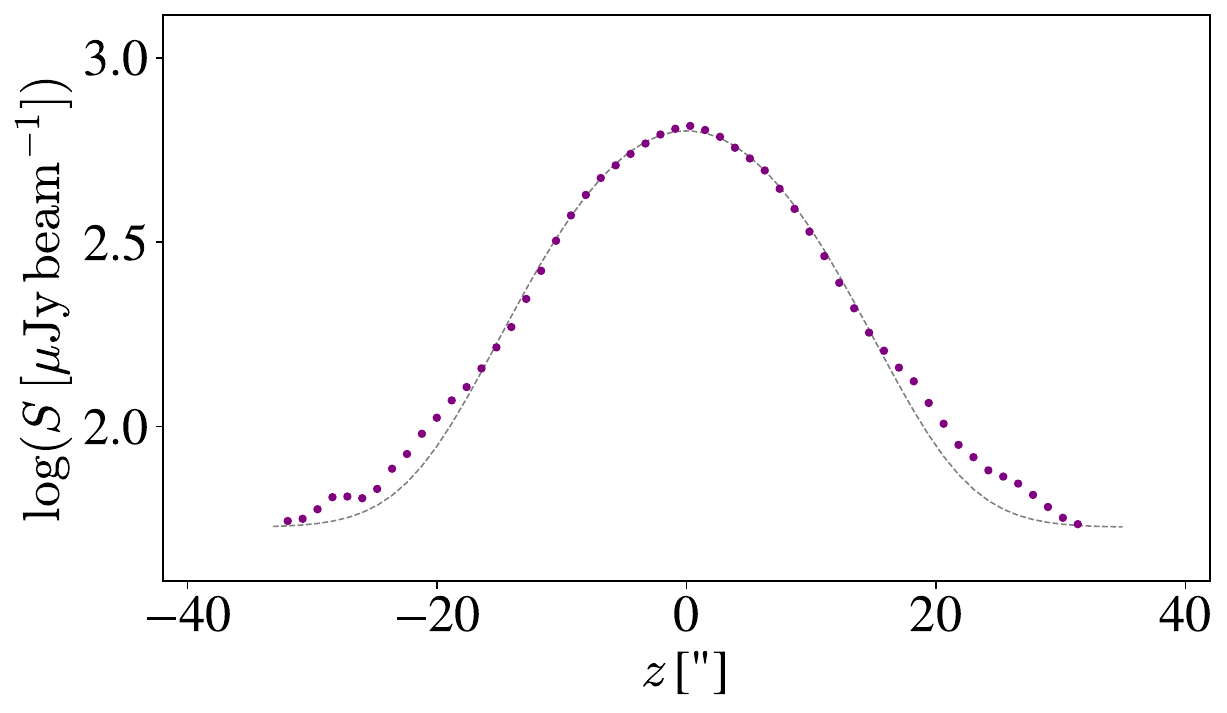}
        \includegraphics[width=0.4\textwidth]{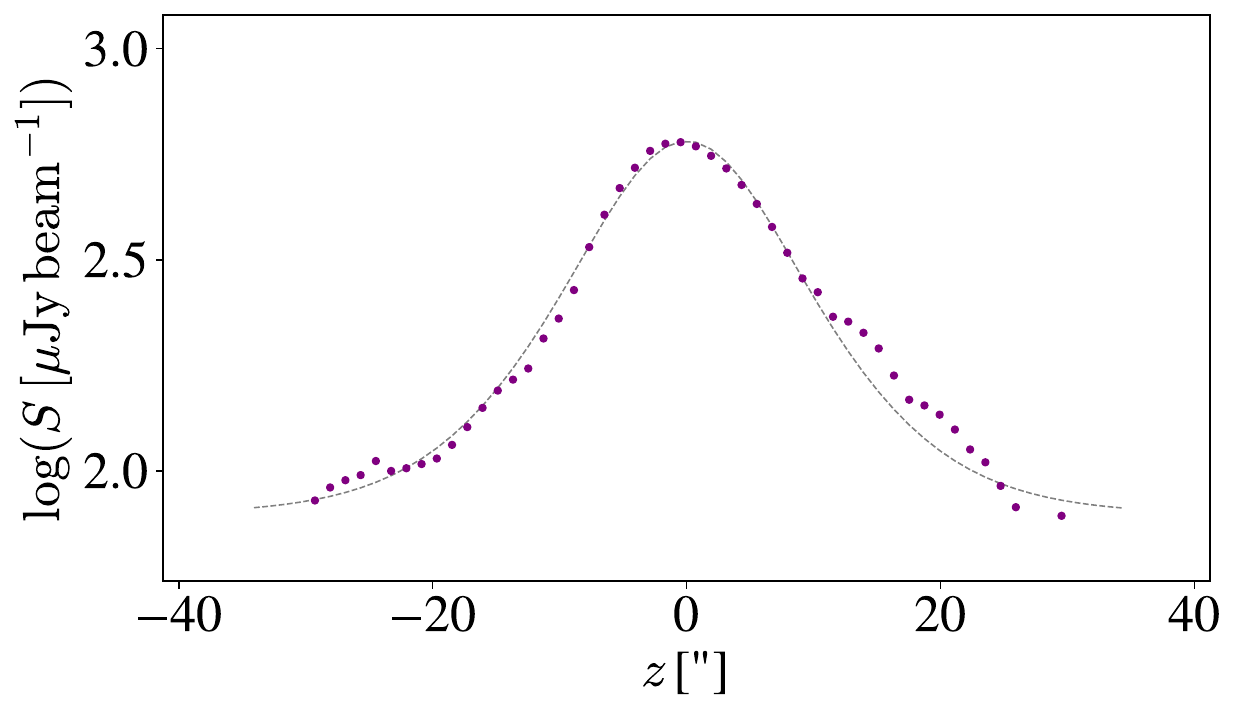}
        \includegraphics[width=0.4\textwidth]{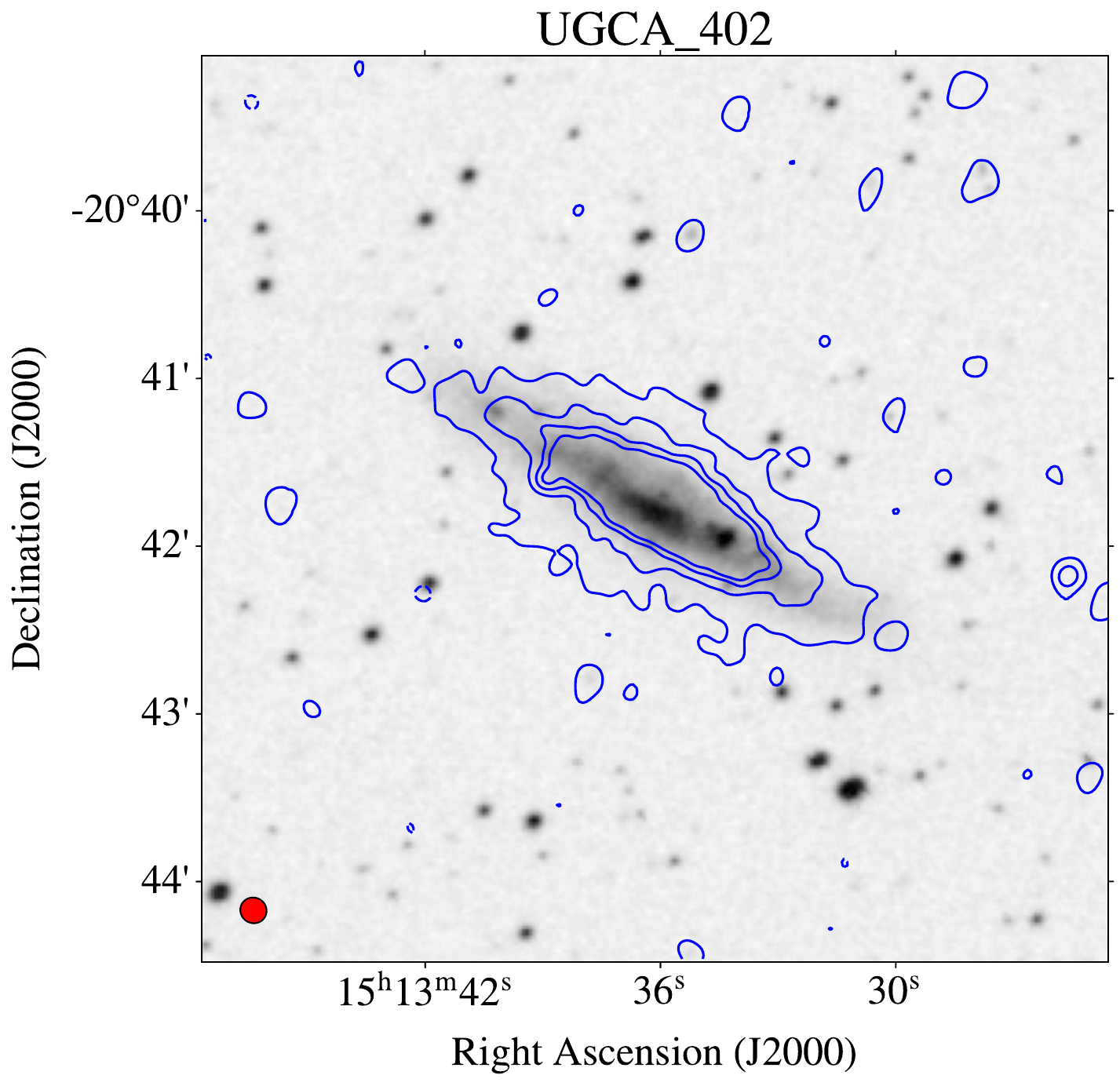}
	
        \includegraphics[width=0.4\textwidth]{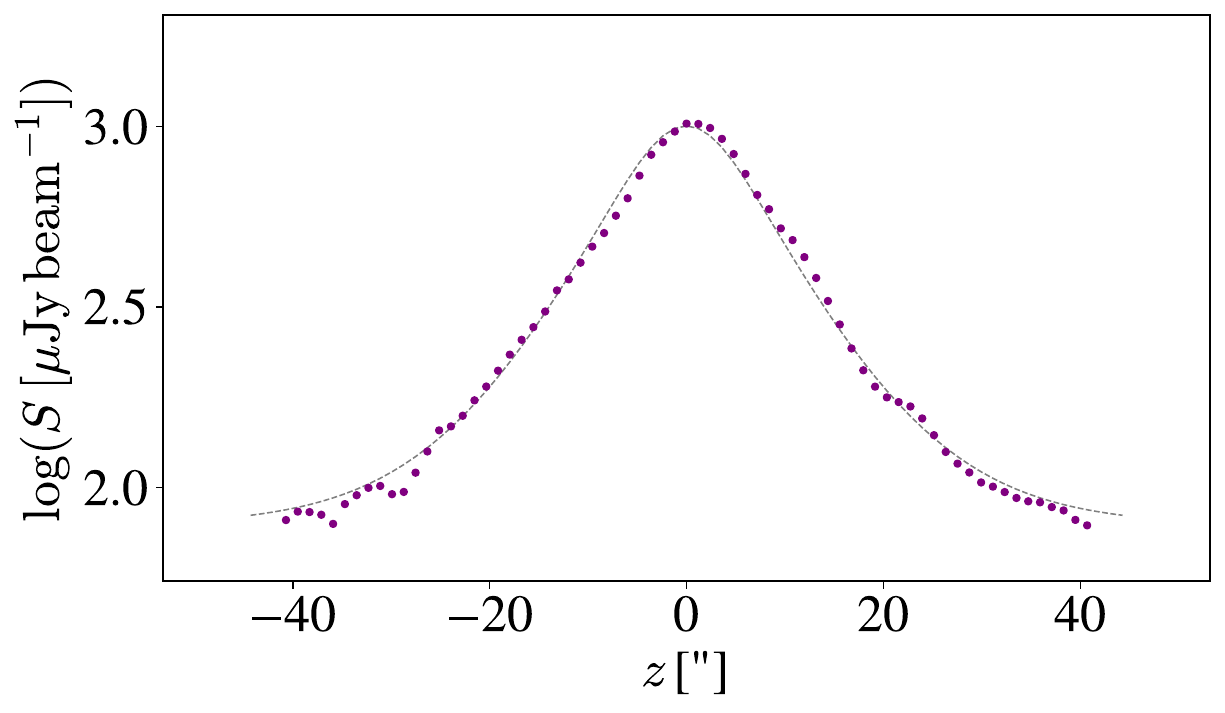}
\caption{continued.}
\end{figure*}

\end{appendix}

\end{document}